\documentclass[nofootinbib,twocolumn,preprintnumbers,floatfix]{revtex4-1}
\pdfoutput=1
\usepackage{amsmath,amsthm,amssymb,multirow,psfrag}
\usepackage{epsfig}
\usepackage{color}
\usepackage{soul}
\usepackage{appendix}

\usepackage{verbatim}
\usepackage{epsfig}
\usepackage{color}
\usepackage{comment}
\usepackage{slashed}
\usepackage{float}
\usepackage{bm}
\usepackage{commath}
\usepackage{multirow}
\usepackage{physics}
\usepackage{mathtools}
\usepackage{enumitem}

\usepackage{subcaption} 

\usepackage{graphicx}
\graphicspath{{Plots/}}

\usepackage[compat=1.1.0]{tikz-feynman} 
\usepackage{tikz} 
\usetikzlibrary{shapes,arrows,positioning,automata,backgrounds,calc,er,patterns}
\usepackage{feynmf}
\usepackage{tikz-network}
\usepackage{soul}
\usepackage[normalem]{ulem}
\setstcolor{red}
\tikzfeynmanset{warn luatex=false}

\begin{document}

\newcommand{\fref}[1]{Fig.~\ref{#1}} 
\newcommand{\eref}[1]{~\ref{#1}} 
\newcommand{\eqnref}[1]{Eq.~\eqref{#1}} 
\newcommand{\eqnSref}[1]{Eqs.~\eqref{#1}} 
\newcommand{\appenref}[1]{Appendix~\ref{#1}}
\newcommand{\tabref}[1]{Table~\ref{#1}}
\newcommand{\sectref}[1]{Section~\ref{#1}}
\newcommand{\sectSref}[1]{Sections~\ref{#1}}

\newcommand{\cc}[1]{\textcolor{magenta}{[CC: #1]}}
\newcommand{\comments}[1]{\textcolor{blue}{Note: #1}}
\newcommand{\he}[1]{\textcolor{red}{[HE: #1]}}
\newcommand{\ds}[1]{\textcolor{blue}{[DS: #1]}}
\newcommand{\tg}[1]{\textcolor{violet}{[TG: #1]}}

\title{\LARGE{{\bf{Baryogenesis and Dark Matter in Multiple Hidden Sectors} \\
}}}

\author{ {\bf{
Hassan Easa$^1$,~~
Thomas Gregoire$^1$,~~
Daniel Stolarski$^1$,~~
Catarina Cosme$^{1,2}$~~
\smallskip 
\\
} }
$^1$Ottawa-Carleton Institute for Physics, Carleton University, 1125 Colonel By Drive, Ottawa, Ontario K1S 5B6, Canada
\\ 
$^2$Instituto de F\'{i}sica Corpuscular (IFIC), Universitat de Val\`{e}ncia-CSIC, Parc Cient\'{i}fic UV, C/ Catedr\'{a}tico Jos\'{e} Beltr\'{a}n 2,
E-46980 Paterna, Spain
}

\email{
Hassaneasa@cmail.carleton.ca \\
gregoire@physics.carleton.ca \\
stolar@physics.carleton.ca  \\
catarina.cosme@ific.uv.es
}

\begin{abstract}
We explore a mechanism for producing the baryon asymmetry and dark matter in models with multiple hidden sectors that are Standard-Model-like but with varying Higgs mass parameters. If the field responsible for reheating the Standard Model and the exotic sectors carries an asymmetry, it can be converted into a baryon asymmetry using the standard sphaleron process. A hidden sector with positive Higgs mass squared can accommodate dark matter with its baryon asymmetry, and the larger abundance of dark matter relative to baryons is due to dark sphalerons being active all the way down the hidden sector QCD scale.
This scenario predicts that dark matter is clustered in large dark nuclei and gives a lower bound on the effective relativistic degrees of freedom, $\Delta N_{\rm eff} \gtrsim 0.05$, which may be observable in the next-generation cosmic microwave background experiment CMB-S4.
\end{abstract}
\maketitle


\newcommand{\GutBaryo}{Ignatiev:1978uf, PhysRevLett.42.746, PhysRevD.19.1036, PhysRevD.18.4500, Ellis:1978xg, PhysRevLett.42.850, Yoshimura:1979gy, PhysRevD.20.2494, Nanopoulos:1979gx, Yildiz:1979gx, PhysRevD.22.2977}

\newcommand{\EWBaryo}{Rubakov:1996vz, Riotto:1999yt}
\newcommand{\DMref}{Cirelli:2010xx, Bernal:2017kxu, Roszkowski:2017nbc, Boveia:2018yeb, Schumann:2019eaa, Profumo:2019ujg}
\newcommand{\DMevidence}{Persic:1995ru, Rubin:1985ze, Bosma:1981zz, Clowe:2006eq, Planck:2018vyg}

\section{Introduction}
\label{introduction}

The existence of a large number of Standard-Model-like hidden sectors~\cite{PhysRevD.80.055001}, and the N-naturalness~\cite{Arkani-Hamed:2016rle} paradigm in particular provide a novel and interesting solution to the hierarchy problem. If there are a large number $N$ of sectors where the Higgs mass parameter takes on random values, then one expects one sector to have a value of order $\Lambda^2/N$ where $\Lambda$ is the cutoff of the theory. If such a sector is identified with the Standard Model (SM), this setup can explain why the Higgs mass is parametrically smaller than the cutoff as long as $N\gg 1$.

In order for such a scenario to describe our Universe, a novel cosmological history is required that naturally allows most of the energy of the Universe to be in the SM sector with relatively little in the others. In~\cite{Arkani-Hamed:2016rle}, this was accomplished by a ``reheaton'' field which carries all of the energy of the Universe at early times. The reheaton field has a weak scale mass and democratic coupling to the Higgs in all the sectors. Because of its weak scale mass, it can easily decay into the SM sector, but decays to sectors with a heavier Higgs are kinematically suppressed naturally giving the SM the dominant fraction of the energy of the Universe after the reheaton decay. The coupling of the reheaton to all sectors is required to be very small in order to prevent loop-level interactions across the sectors, and this small coupling ensures that the reheaton width is orders of magnitude smaller than its mass. 

Because of the unique cosmology, it remains an open question of how the baryon asymmetry of the Universe (BAU) and the dark matter abundance are generated in such a framework (see \cite{Bertone:2004pz} for a dark matter review and \cite{Boucenna:2013wba} for a review of models that relate dark matter to the BAU).
Obviously, BAU and dark matter are necessary ingredients in any realistic model, and in this work, we build a model that addresses both of these problems. The reheaton field is a fermion that carries a lepton number. At very early times, it lives in a thermal bath of a ``reheaton sector'' that does not contain any SM-like fields. Such a sector has similar matter content as traditional leptogenesis models~\cite{Fukugita:1986hr}, and the dynamics of that sector create an asymmetric population of reheatons. 

When the reheaton decays, it transfers its asymmetry to the lepton number in each of the various sectors. By the same kinematic mechanism described in~\cite{Arkani-Hamed:2016rle}, the dominant decay of the reheaton is to the SM sector, and thus most of the asymmetry is transferred to the lepton number of the SM sector. The lepton asymmetry can then be transferred to a baryon asymmetry by the SM sphaleron process~\cite{Manton:1983nd,Klinkhamer:1984di,Kuzmin:1985mm,Khlebnikov:1988sr}. In order for the reheaton to dominantly decay its energy into the SM sector, the reheating temperature of the reheaton decay must be of order the weak scale, which in turn sets an upper bound of the width of the reheaton~\cite{Arkani-Hamed:2016rle}. If we naively assume that reheating is instantaneous, then the temperature of the SM is always around or below the weak scale, and the sphaleron rate is exponentially suppressed for nearly all time. Doing a more careful calculation of the thermal history, taking into account the early decays of the reheaton, shows that the SM sector in fact reaches temperatures much higher than the naive reheating temperature~\cite{PhysRevD.31.681,Chung:1998rq, Giudice:2000ex, Maldonado:2019qmp, Allahverdi_2021}, and the SM sphaleron can be unsuppressed for a sufficiently long time to generate the observed BAU. 

The subdominant decays of the reheaton will populate the other sectors, with the temperature of the other sectors decreasing as the Higgs mass increases. The lepton asymmetry of the reheaton will also be transferred to other sectors. If the sphaleron in other sectors is active, then the dark baryon asymmetry can serve as an asymmetric dark matter candidate~\cite{Nussinov:1985xr,Kaplan:2009ag,Petraki:2013wwa, Zurek:2013wia}. 
Because constraints on extra relativistic degrees of freedom~\cite{Planck:2018vyg} require that the vast majority of the energy of the reheaton decay to the SM sector, one also expects that the dark baryon asymmetry is significantly smaller than the SM baryon asymmetry. Furthermore, because the Higgs mass in the dark sectors is larger than in the SM, the sphaleron decouples earlier further suppressing the dark baryon asymmetry.

The above analysis, however, changes qualitatively if one considers dark sectors with \textit{positive} Higgs mass squared parameter. Such sectors were dubbed ``exotic'' in~\cite{PhysRevD.101.095016} and shown to produce interesting gravitational wave phenomenology. Because the Higgs mass squared is positive, the Higgs does not break electroweak symmetry, and $SU(2)\times U(1)$ is a good symmetry until the QCD phase transition at much lower temperatures. Therefore, even though the lepton asymmetry transferred to exotic sectors is smaller than that of the SM sector, the sphaleron process is significantly more efficient, and this can accommodate the observed dark matter abundance of $\Omega_{\rm DM} \sim 5 \Omega_{B}$. 

We note that it was previously argued that sectors with a positive Higgs mass parameter cannot have a baryon asymmetry~\cite{arkani2005predictive}. The argument was that after QCD and electroweak phase transition, even though the sphaleron is exponentially suppressed, it is still faster than Hubble (which is Planck suppressed), and thus all the baryon asymmetry is washed out. Because of the exponential suppression of the sphaleron, this conclusion depends very sensitively on the phase transition temperature compared to the electroweak boson mass. While that ratio can be computed on the lattice in an SM-like setup, QCD in the exotic sector has six massless flavors, so the phase transition is expected to be first order~\cite{PhysRevD.29.338} and qualitatively different than the SM. Furthermore, the electroweak boson mass depends on the $SU(2)$ gauge coupling, $g$. While the minimal N-naturalness setup has $g$ being the same in the exotic and SM sectors, this need not be the case to solve the hierarchy problem. Therefore, we find that a baryon asymmetry in the dark sector is indeed possible, and asymmetric dark matter can be accommodated.

The dark matter will be dominated by the neutrons in the exotic sector with the lightest Higgs mass. Furthermore, the pions in these sectors are much lighter than the SM pions, so these neutrons can form large nuclei~\cite{Krnjaic:2014xza,Hardy:2014mqa}. In fact,~\cite{Hardy:2014mqa} showed that such an agglomeration of neutrons is generic and the exotic nuclei can be expected to grow quite large. This means that dark matter self-interaction is suppressed and the bounds from the Bullet Cluster~\cite{Randall_2008} can be evaded.

It was emphasized in~\cite{Arkani-Hamed:2016rle} that $N$ decoupled sectors, each with light photons and neutrinos, gives rise to both bounds and possible signatures in cosmological observations. The setup we present here is no different: one important signature is that if the dark matter is comprised of exotic sector baryons, this requires a minimum contribution to effective relativistic degrees of freedom, $\Delta N_{\rm eff} \gtrsim 0.05$, which could be observable at CMB-S4~\cite{abazajian2019cmb}. A significant constraint comes from the neutrinos of SM-like sectors which are heavier than the SM neutrinos. We focus on the scenario where the neutrinos are Dirac, and we find $N \lesssim 10^5 - 10^8$ depending on the assumptions about what fraction of the dark matter those neutrinos make up. In particular, at the high end of that range, the neutrinos make up a sizable fraction of the cold dark matter. We focus on the dark neutrons being the dominant component of dark matter, but that need not be the case. 

We note that the constraint on the number of sectors means that this model cannot fully solve the large hierarchy, which would require $N=10^{16}$ sectors. It does generically solve the little hierarchy with $N\gtrsim 10^4$, and can ameliorate the large one. Models with different scaling of the Higgs vacuum expectation value (VEV)  with sector number (such as the one in~\cite{Arkani-Hamed:2016rle}) would have different neutrino phenomenology and can weaken the bound on the number of sectors. The phenomenology of the baryon asymmetry and hidden neutron dark matter would remain qualitatively similar. 

This paper is organized as follows. In~\sectref{sec: Model setup}, we give a qualitative overview of the model including the field content and the mechanisms employed to generate the BAU and asymmetric dark matter.~\sectref{sec: Lepton asymmetry} quantitatively presents the details of the reheaton sector and how its asymmetry is generated. The reheating of the various sectors via the decays of the reheaton field and the constraints imposed by cosmological observations are analyzed in~\sectref{sec: Reheating Dynamics}. In~\sectref{sec: Baryon Asymm}, we discuss the baryon asymmetry across the various sectors and the relevant parameter space consistent with observations, while~\sectref{sec: DM} focuses on dark matter phenomenology.~\sectref{conclusions} contains our conclusion and a summary of the baryon asymmetry and dark matter within the N-naturalness framework. Various technical details are given in the appendices.

\section{Model Overview}
\label{sec: Model setup}

\begin{figure}[tb]
\begin{centering}
\includegraphics[scale=0.3]{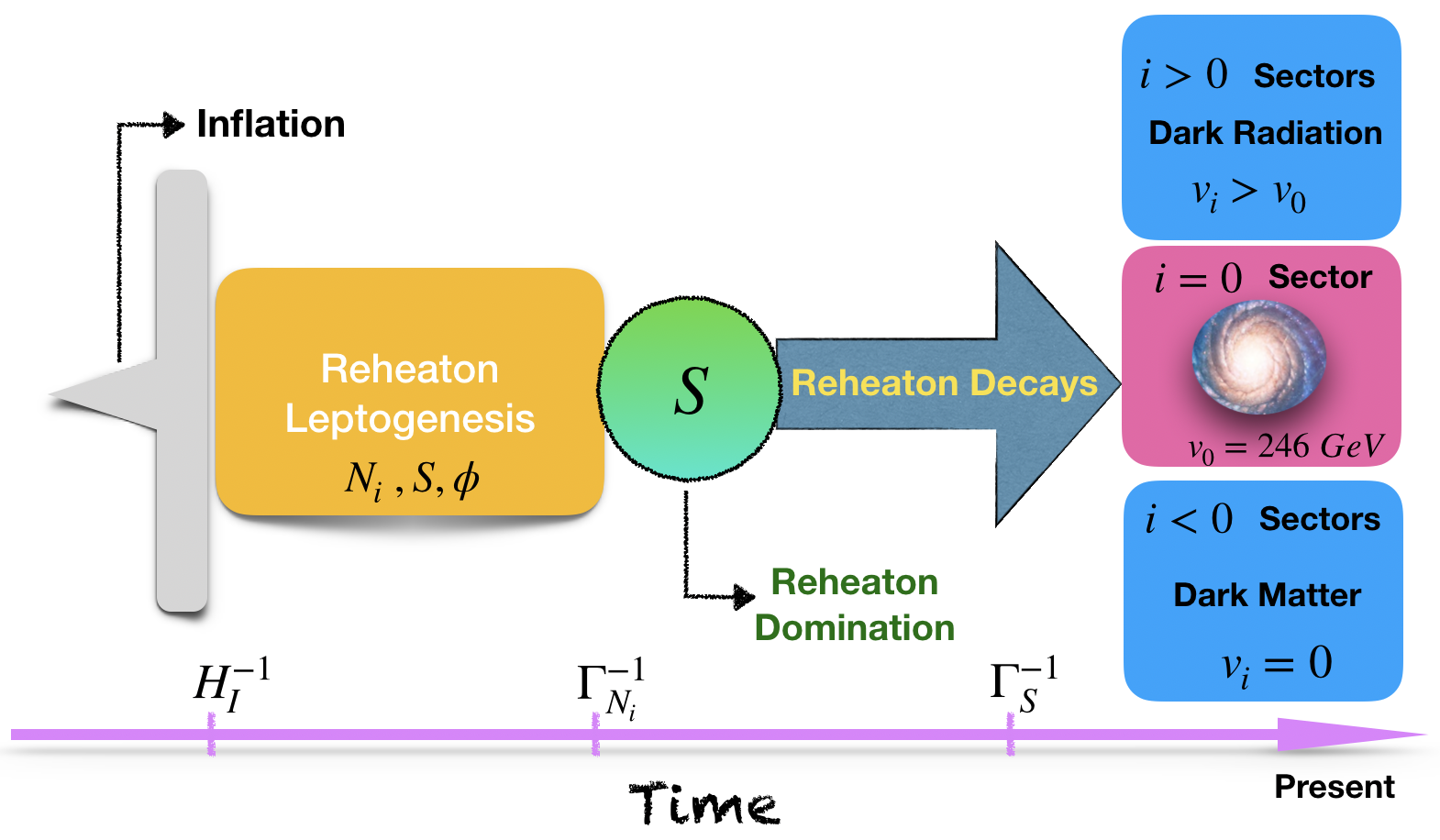}
\par\end{centering}
\caption{A schematic of the cosmological evolution of our model. After inflation, there is a thermal bath composed of Majorana fermions, $N_i$, scalars $\phi$ and reheatons, $S$. The decays of the heavy $N_i$ generate the lepton number asymmetry, which will be carried by the reheaton. Since $\phi$ can only decay into $S$, at some point, the energy density of the Universe is dominated by $S$. Finally, the reheaton decays into the various sectors and populates them. This process leads to a baryon asymmetry in the SM $i=0$ sector, as well as an asymmetry in the first exotic $i=-1$ sector, which is the dominant source of dark matter in the Universe.} 
\label{fig:HistoryDiagram}
\end{figure}

A schematic diagram of the cosmological history of our model is shown in \fref{fig:HistoryDiagram}. 
The full Lagrangian of our model contains a reheaton fermionic field $S$, $N$ copies of the Standard Model with varying Higgs masses, a sector responsible for generating a lepton asymmetry initially stored in the reheaton,  and finally a sector that allows the reheaton to decay to the various Standard Model copies through a vectorlike lepton portal denoted ${L_4}_i$: 
\begin{equation}
    \mathcal{L} = \mathcal{L}_\text{lepto} + \mathcal{L}_S + \sum_{i} \mathcal{L}_{{L_4}_i}+ \sum_{i} \mathcal{L}_{\text{SM},i} \; . \;
\end{equation}
The precise nature of each of these sectors is covered in \sectref{sec: Lepton asymmetry} and in the discussion below.

We begin our story after inflation when the inflaton decays and deposits all of its energy into a reheating sector that contains a thermal bath that includes the following:
\begin{itemize}
    \item At least two heavy Majorana fermions $N_i$.
    \item A complex scalar $\phi$.
    \item A weak-scale Dirac fermion $S$ which we refer to as the reheaton.
\end{itemize}
The reheaton carries an approximately conserved lepton number and the purpose of this sector is to generate an asymmetry for it. It is constructed to mimic the standard leptogenesis setup~\cite{Fukugita:1986hr}. The dynamics encoded in $\mathcal{L}_\text{lepto}$ are detailed in \sectref{subsec: Toy Model}. In short, out of equilibrium decays of the heavy $N_i$ fields will eventually impart a lepton number asymmetry on the thermal bath, and the $\phi$ scalar will decay away. There will thus be an era where the Universe is dominated by an asymmetric population of reheatons. The asymmetry of the reheatons will be partially transferred to SM baryons, explaining the baryon asymmetry of the Universe, and partially to an exotic sector, giving rise to asymmetric dark matter.

Below the scale of the heavy $N_i$ and $\phi$ fields, the content of our model is very similar to the original N-naturalness model, with $N$ nearly decoupled SM-like sectors each containing a copy of the SM field content plus a vectorlike lepton doublet $(L_{4},L_{4}^{c})$ with charges of $\mathbf{2}_{\pm 1/2}$ under a given sector's $SU(2)\times U(1)$ gauge symmetry. These TeV-scale vectorlike leptons serve as a portal through which the reheaton $S$ decays to the different sectors. Without this portal, the reheaton could in principle couple directly to the leptons of each sector, but this would lead to a heavy reheaton and too much mixing between the reheaton and the neutrinos~\cite{Arkani-Hamed:2016rle}.
 In the $N$ SM-like sectors, the dimensionless parameters: gauge, Yukawa, and quartic couplings are (approximately) equal to those of the SM, but the dimensionful Higgs mass parameter takes a different value in each sector. Having a range of possible Higgs masses with the lowest one dynamically preferred by the reheaton decay  is how this setup addresses the hierarchy problem~\cite{Arkani-Hamed:2016rle}. For simplicity, we parameterize the Higgs mass squared parameter in the $i$th sector by
\begin{equation}
\left ( m_{H}^{2} \right )_{i} = - \frac{\Lambda^{2}}{N} \left ( 2 \ i + r \right ) \; , \quad  -\frac{N}{2} \leq i \leq \frac{N}{2} \; , \label{Higgs mass2 param}
\end{equation}
and we can identify the $i=0$ sector, the one with a negative mass parameter with the smallest magnitude, as the SM sector. Sectors with $i>0$ are dubbed ``standard sectors'' and have a similar structure to the SM but with a larger Higgs VEV. Those with $i<0$ are ``exotic sectors'' in which the Higgs does not acquire a VEV at all and can be integrated out in an $SU(3)\times SU(2)\times U(1)$ symmetric theory. These exotic sectors will be the dominant source of dark matter (DM) in the Universe. The parameter $r$ indicates the spacing between sectors, with $r=1$ corresponding to uniform spacing and $r<1$ corresponding to a large splitting between our sector and the next one~\cite{Arkani-Hamed:2016rle}. More details on the mass spectrum of each sector are given in \appenref{App: mass spectrum}.

Similar to~\cite{Arkani-Hamed:2016rle}, the vectorlike lepton doublet $L_4$ and $L_4^c$ serves as the portal to the reheaton. This is given by the following  Lagrangian:
\begin{equation} 
\begin{split} 
-\mathcal{L}_S-\sum_i\mathcal{L}_{{L_4}_i}  & \supset   \sum_{i} \left [ \lambda S^{c} \left (  L_{4} H \right )_{i}  + \mu_{L} \left (  l \;  L_{4}^{c}  \right )_{i} \right. \\
& \left. + M_{L} \left ( L_{4}^{c} L_{4}  \right )_{i} \right ] +  
 m_{S} S S^{c} + \text{ H.c. } \; , \\
\end{split}
\label{eq:lagl4}
\end{equation}
where $\mu_L$ is the bilinear coupling and $l_i$ denotes the SM left-handed lepton doublet, and as before the $i$ index labels the sectors. We are using Weyl notation for the fermion fields. For simplicity, we also take the dimensionful parameters $\mu_L$ and $M_L$ to be the same in all sectors. In the standard sectors, when the Higgses get VEVs and the heavy lepton doublets are integrated out, the reheaton mixes with the neutrinos of the various sectors. To give masses to all neutrinos, we need to add neutrino mass terms to this Lagrangian in all sectors; those mass terms can be either Majorana or Dirac. As we will discuss, in our scenario, bounds on warm dark matter lead us to consider neutrinos that are almost purely Dirac.

For this model to reproduce the observation that much of the structure of the Universe is governed by SM fields, the reheaton must preferentially decay into the SM sector. Since $S$ couples democratically (with the same $\lambda$) to all sectors, kinematic effects are required to achieve our observable Universe with most of the energy in the visible sector. 
This can be accomplished if the reheaton's mass is comparable to the SM Higgs mass, such that its decays into sectors with larger Higgs masses are suppressed. This way, the reheaton dynamically selects the sector with the smallest Higgs mass (defined as the Standard Model sector, with $i=0$) and populates it preferentially.
Since the population of $S$ is asymmetric, its decay will transfer the asymmetry into all sectors, with the quantum number being a linear combination of the lepton number of each sector.

In the SM sector, the lepton asymmetry transferred by the reheaton is converted into baryon asymmetry via SM sphalerons \cite{Manton:1983nd, Klinkhamer:1984di, Kuzmin:1985mm,Khlebnikov:1988sr}. The baryon asymmetry depends on the evolution of the temperature of the SM thermal bath, as the sphaleron rate becomes exponentially suppressed after the electroweak phase transition. 
A first pass estimate of this effect would be to assume an instantaneous
decay of the reheaton, that is, assuming that the energy density stored
in the reheaton is suddenly deposited in the other sectors when $\Gamma_{S}\sim H$,
where $\Gamma_{S}$ is the width of the reheaton and $H$ is the Hubble
parameter, which would imply a reheating temperature $T_{\rm RH}\sim\sqrt{\Gamma_{S}M_{P}}$, with $M_P$ the reduced Planck mass.
Nevertheless, this estimate does not tell us how the temperature of
the Universe evolves before reheating, as $T_{\rm RH}$ only depends on
the duration of the reheating period, $\tau_{S}=\Gamma_{S}^{-1}$.
Hence, we need to track the evolution of the temperature by solving
the Boltzmann equations for the evolution of the energy density of
each component of the Universe, assuming reheating is not instantaneous
(see~\sectref{sec: Reheating Dynamics}). As we will see in~\sectref{sec: Reheating Dynamics}, the temperature
of the Universe is larger than $T_{\rm RH}$ for $t<\Gamma_{S}^{-1}$ and
therefore, during that period, the sphaleron process can occur very
efficiently. Although the decays of the reheaton do not heat up
the Universe, they make it cool more slowly due to entropy release~\cite{PhysRevD.31.681,Chung:1998rq, Giudice:2000ex, Maldonado:2019qmp, Allahverdi_2021}. Thus, the temperature at the end of reheaton decays---the so-called
reheating temperature---is not the largest temperature achieved by
the thermal bath, but the one at which the entropy density levels off \cite{PhysRevD.31.681}.
The reheating process from the reheaton and its consequences for the BAU are explored in detail in~\sectSref{sec: Reheating Dynamics} and \ref{sec: Baryon Asymm}, respectively.

Qualitatively, the process for the generation of the baryon asymmetry in the other sectors is essentially the same as described above, but the different Higgs masses and VEVs of those sectors lead to different baryon asymmetries. In the case of the standard sectors ($i>0$), the corresponding Higgs VEV is larger than in the usual SM sector, which suppresses the associated baryon asymmetry in two different ways: (1) the reheaton decay is kinematically suppressed, implying that there is less total energy in the standard sector compared to the SM sector, which allows us to satisfy the bounds coming from $\Delta N_{\text{eff}}$, and (2) in these sectors, the electroweak phase transition happens at higher temperatures and, therefore, the sphaleron freezes out earlier.
In the exotic sectors ($i<0$), electroweak symmetry breaking is not due to the Higgs, but due to the confinement of the SU(3) color group instead~\cite{PhysRevD.20.2619}. Hence, the electroweak scale is $\sim100$ MeV, allowing the sphaleron to be active at much lower temperatures. Therefore, even though the energy density of the exotic sectors is lower than in the SM sector, the corresponding baryon asymmetry can be larger than the one of the SM. This way, the lightest baryon in the exotic sector, the neutron, can be a DM candidate, and one can easily accommodate $\Omega_\text{DM} \sim 5 \Omega_B$. This DM dynamics is described quantitatively in \sectref{sec: DM}.

\section{Origin of Lepton Number Asymmetry}
\label{sec: Lepton asymmetry}
In this section, we show that the lepton number asymmetry in our framework
can be generated with a scenario in the early Universe that is similar to leptogenesis. As succinctly explained in \sectref{sec: Model setup}, we
assume that after inflation there is a thermal bath composed of at least two
Majorana fermions, $N_i$, the fermionic reheaton, $S$, and a complex scalar, $\phi$. The
lepton number asymmetry originates from the decay of the lightest $N_i$ into the reheaton, analogous to the standard leptogenesis mechanism~\cite{Fukugita:1986hr}.\footnote{Unlike the ``Affleck-Dine" inflation type mechanism where the complex scalar (inflaton) carrying lepton number is responsible for leptogenesis~\cite{Affleck:1984fy, Cline:2019fxx, Cline:2020mdt}.} In particular, one can make an analogy with the $N$ being the right-handed neutrino, the $\phi$ being the Higgs, and the $S$ the lepton doublet, but we stress that this is just an analogy and all the fields in the reheaton sector are singlets under all gauge symmetries. 

The decays of the $N_1$ imprint an asymmetry on the remnant population of $S$ which dominates the Universe. 
Later, this asymmetry is transmitted
to a lepton number asymmetry in the various sectors through the reheaton decay and, eventually, it
is converted into baryon asymmetry due to sphaleron processes. 
In what follows, we build a model to explain the production of the lepton asymmetry within our
setup.

\subsection{Particle physics model}
\label{subsec: Toy Model}

We consider a simple sector that contains at least two Majorana fermions, $N_{i}$, and, for simplicity, we take their masses to be hierarchical,  $M_{i-1} \ll M_{i}$. The reheaton, $S$, is a Dirac fermion that couples to the $N_i$ via a Yukawa coupling to a complex scalar $\phi$. The field content and the corresponding charges are given in \tabref{tab:fields-lep}.
The interactions in this setup are given by 
\begin{equation}
    - \mathcal{L}_\text{lepto} = \frac{1}{2} M_{i} N_{i}^2 +  y_{i} \phi^{\dagger} S^{c} N_{i} + \kappa \phi S^{2} + \text{ H.c. } \; , 
    \label{eq:LeptoLag}
\end{equation}
where the subscript $i\geq 2$ denotes the number of Majorana fermions. The Majorana mass term explicitly breaks lepton number and allows a lepton asymmetry to be generated, but all other interactions in the theory conserve lepton number.\footnote{Except the electroweak sphaleron that violates $L$ but conserves $B-L$.}
As we will show, the viable parameter space of interest requires $y\lesssim 0.06$, $\kappa \sim 1$ as well as 
$10^{12}$ GeV $ \lesssim M_1 \lesssim 10^{16}$ GeV and the $S$ at the weak scale. The $\phi$ mass can be anywhere between the weak scale the the scale of the $M_1$.

We take the mass hierarchy to be $M_1 \gg M_\phi \gg m_S$, with $M_1$ the mass of the $N_1$, the lightest $N$ state, so that the decay processes $N_1 \rightarrow S^c \phi$ and $\phi \rightarrow S S$ shown in~\fref{fig:treeLevelN1} are kinematically allowed. At late times compared to the lifetime of the $N$, all that remains is an asymmetric population of $S$ and $S^c$. 
This asymmetry arises due to the interference in the decay of the $N$ between the loop-level diagrams shown in \fref{fig:vertexLoop and self-energy} and the tree-level decay. The standard $CP$ asymmetry is defined as
\begin{equation}
\epsilon \equiv \frac{ \Gamma \left ( N_{i} \rightarrow \phi^\dagger + S \right ) -  \Gamma \left ( N_{i} \rightarrow \phi + S^{c} \right ) }{ \Gamma \left ( N_{i} \rightarrow \phi^\dagger + S \right ) + \Gamma \left ( N_{i} \rightarrow \phi + S^{c} \right ) }. 
\end{equation}
The $CP$-violating parameter $\epsilon$ is given by~\cite{Fukugita:1986hr, Covi:1996wh, Hambye:2003ka, Hambye:2003rt, Buchmuller:2005eh}\footnote{
In our setup, there is an additional one-loop diagram for the $N_1$ decays. Since its matrix element is $\mathcal{M} \propto \abs{\kappa}^{2}$, it does not contribute to the $CP$ asymmetry $\epsilon$. }
 \begin{equation}
 \begin{split}
 \epsilon & = \frac{1}{8 \pi} \sum_{k \neq 1 } \frac{\Im \left ( y_{1}^{\dagger} y_{k}  \right )^{2} }{  y_{1}^{2} } \left ( f(x_{k}) - \frac{M_{1}M_{k}}{M_{k}^{2} - M_{1}^{2}} \right )  \\
 & \sim-\frac{3}{16\pi}\sum_{k\neq1}\frac{1}{\sqrt{x_{k}}}\,I_{1k}  \label{eq:cpAsymmetryApprox} \; ,
 \end{split}
 \end{equation}
 where $x_{k}= \left ( \frac{M_{k}}{M_{1}} \right )^{2}$ , $I_{1k}=\frac{\Im \left ( y_{1}^{\dagger} y_{k}  \right )^{2} }{\left( y^{\dagger} y \right)_{11}}\;$, and 
 \begin{equation}
f(x) = \sqrt{x} \left [ 1 - \left ( 1 + x \right ) \ln \left (  \frac{1+x}{x} \right )  \right ] \xrightarrow{x \gg 1} - \frac{1}{2 \sqrt{x} } \; .
\label{f(x)}
 \end{equation}
\begin{table}[t!]
  \renewcommand{\arraystretch}{1.5}
    \addtolength{\tabcolsep}{5pt} 
    \centering
    \begin{tabular}{| c | c | c |  }
        \hline \hline
  & spin & $L$   \\
        \hline 
$N_i$ & 1/2 & -1    \\ \hline 
$S$ & 1/2 & 1    \\ \hline 
$S^c$ & 1/2 & -1    \\ \hline 
$\phi$ & 0 & -2   \\         \hline \hline
    \end{tabular}
    \caption{Matter content of the leptogenesis sector: the fermionic reheaton ($S$, $S^c$), at least two Majorana fermions ($N_i$) with different masses, and a complex scalar $\phi$.
     Fermions are written in terms of two component (Weyl) spinors. The $L$ represents lepton number which is softly broken by Majorana mass terms for $N_i$.}
    \label{tab:fields-lep}
\end{table}
\begin{figure}[htb!]
\centering
\begin{tikzpicture}
  \begin{feynman}[small]
    \vertex (a) {\(N_{i} \)};
    \vertex [right=of a] (b);
    \vertex [above right=of b] (f1) {\(  \phi \)};
    \vertex [below right=of b] (f2) {\( S^c  \)};
 
    \diagram* {
      (a) -- [black!20!blue, majorana ] (b) ,
      (b) -- [black!20!black, scalar ] (f1),
      (b) -- [black!20!red, fermion] (f2),
    };
  \end{feynman}
\end{tikzpicture} \quad
\begin{tikzpicture}
  \begin{feynman}[small]
    \vertex (a) {\( \phi \)};
    \vertex [right=of a] (b);
    \vertex [above right=of b] (f1) {\(  S \)};
    \vertex [below right=of b] (f2) {\( S \)};
 
    \diagram* {
      (a) -- [black!20!black, scalar ] (b) ,
      (b) -- [black!20!red,  ] (f1),
      (b) -- [black!20!red, ] (f2),
    };
  \end{feynman}
\end{tikzpicture}
\caption{Feynman diagrams for the tree-level decays in our setup.}
\label{fig:treeLevelN1}
\end{figure}
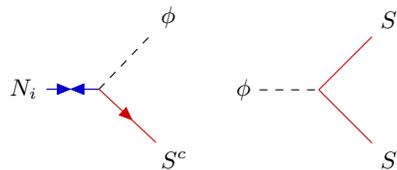
 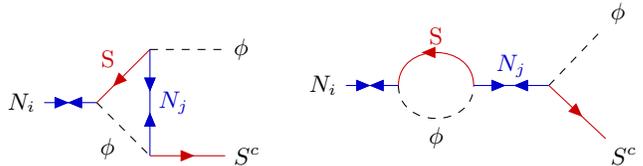
\begin{figure}[t!]
\centering
\begin{tikzpicture}
  \begin{feynman}[small]
    \vertex (a) {\(N_{i} \)};
    \vertex [right=of a] (b) ; 
    \vertex [above right=of b] (f1);
    \vertex [below right=of b] (f2) ;
    
    \vertex [right=of f1] (x1)  {\(  \phi \)}; 
    \vertex [right=of f2] (x2)  {\(  S^c \)};
 
    \diagram* {
      (a) -- [black!20!blue, majorana ] (b) ,
      (b) -- [black!20!red, anti fermion, edge label= S ] (f1),
      (b) -- [black!20!black, scalar, edge label'= $\phi$ ] (f2),
      (f1) -- [black!20!blue, majorana, edge label= $N_{j}$] (f2),
      
      (f1) -- [black!20!black, scalar] (x1),
      (f2) -- [black!20!red, fermion] (x2),
    };
  \end{feynman}
  \end{tikzpicture} \quad  
 \begin{tikzpicture}
  \begin{feynman}[small]
     \vertex (a) {\(N_{i} \)};
    \vertex [right=of a] (b);
     \vertex [right=of b] (m);
      \vertex [right=of m] (n);
    \vertex [above right=of n] (f1) {\(  \phi \)};
    \vertex [below right=of n] (f2) {\( S^c  \)};
 
    \diagram* {
      (a) -- [black!20!blue, majorana ] (b) ,
       (b) -- [black!20!red, anti fermion,half left ,edge label=S] (m) ,
       (b) -- [black!20!black, scalar,half right ,edge label'=$\phi$] (m) ,
        (m) -- [black!20!blue, majorana ,edge label=$N_{j}$] (n),
      (n) -- [black!20!black, scalar ] (f1),
      (n) -- [black!20!red, fermion] (f2),
    };
  \end{feynman} 
 \end{tikzpicture}
 \caption{One-loop diagrams contributing to the CP asymmetry. } \label{fig:vertexLoop and self-energy}
\end{figure}

The generation and evolution of the lepton asymmetry can be described by the usual Boltzmann equations for leptogenesis. In our model, we take into account the decays and inverse decays involving $N_i$, neglecting $2\rightarrow 2$ scattering processes as a first approximation. In the next section, we show an analytical approximation for the lepton number asymmetry that will be transferred to the reheaton.

\subsection{Lepton number asymmetry}
\label{subsec: analytic}

In this section, we first analyze the amount of entropy injected into the thermal bath due to the out-of-equilibrium decay of a massive particle such as the Majorana fermion $N_1$ and the reheaton $S$. The entropy release will dilute the initial lepton asymmetry, affecting the eventual production of baryon asymmetry. For simplicity, suppose that we have a massive unstable particle, $X$, which is nonrelativistic and long lived. In our scenario, the unstable particle $X$ could be the Majorana fermions $N_i$ or the reheaton $S$, and their subsequent decays will produce considerable entropy in our Universe. If $X$ is sufficiently long lived, then it decays while dominating the energy density of the Universe, since $\rho_X \sim a^{-3}$, whereas the energy density of radiation, $\rho_R$, scales with $a^{-4}$, with $a$ being the scale factor of the Universe. Assuming an instantaneous decay, we may estimate the entropy released during the process. The particle $X$ decays when its decay width is of the order of the Hubble parameter, $\Gamma_X \sim H$. The temperature of the thermal bath immediately prior to the decay of $X$, $T_i$, is given by \cite{Kolb:1990vq}
\begin{equation}
\begin{split}
& \Gamma_X^{2} \sim H^{2} = \frac{\rho_X}{3M_{P}^{2} } \simeq  \frac{ M_{X} \; s_i \; Y_{i} }{ 3 \; M_{P}^{2}}    \\
& \Rightarrow  T_i^{3} \simeq \left [ \frac{ 135 \; M_{P}^{2} \;  \Gamma_X^{2}  }{ 2 \pi^{2} \; M_{X}  \; Y_{i} \; g_{\star}   } \right ] \; \label{eq:TIrelation1} \; , \; \\
\end{split}
\end{equation}
where $Y_i\equiv \frac{n_i}{s_i}$ is the initial comoving number density of $X$, with $s_{i}=\frac{2\pi^{2}}{45}\,g_{\star}\,T_{i}^{3}$. Supposing that $X$ decays into relativistic particles that quickly thermalize, the corresponding radiation energy density, $\rho_f$, and temperature, $T_f$, after the $X$ decay are
\begin{equation}
\begin{split}
& \rho_f \simeq \frac{\pi^{2}}{30} g_{\star} T_f^{4} = 3M_P^{2} \; \Gamma_X^{2} 
 \Rightarrow T_{f}^{3} \simeq \left [ \frac{90 \; M_{P}^{2} \; \Gamma_{X}^{2} }{\pi^{2} \; g_{\star} } \right ]^{3/4} \; \label{eq:TFrelation1} \; , \; \\
\end{split}
\end{equation}
where we used the conservation of energy. Hence, the ratio of total comoving entropies is
\begin{equation}
\begin{split}
\delta \equiv \frac{S_f}{S_i} \simeq \frac{T_f^{3}}{T_i^{3}} \simeq  0.78\, \frac{ M_X \; Y_i \; g_{\star}^{1/4}}{ \left ( M_{P} \; \Gamma_X \right )^{1/2} } \; .  \; \label{eq:delta1}   \\ 
\end{split}
\end{equation}
We conducted a careful analysis, by tracking the energy densities for $X$ and radiation with the Boltzmann equations and solving them using numerical methods, which demonstrated that the rough estimate in~\eqnref{eq:delta1} is good, with a difference of $\order{1}$ compared to the numerical value. 

\begin{figure}
     \centering
     \begin{subfigure}[b]{0.44\textwidth}
         \centering
         \includegraphics[width=\textwidth]{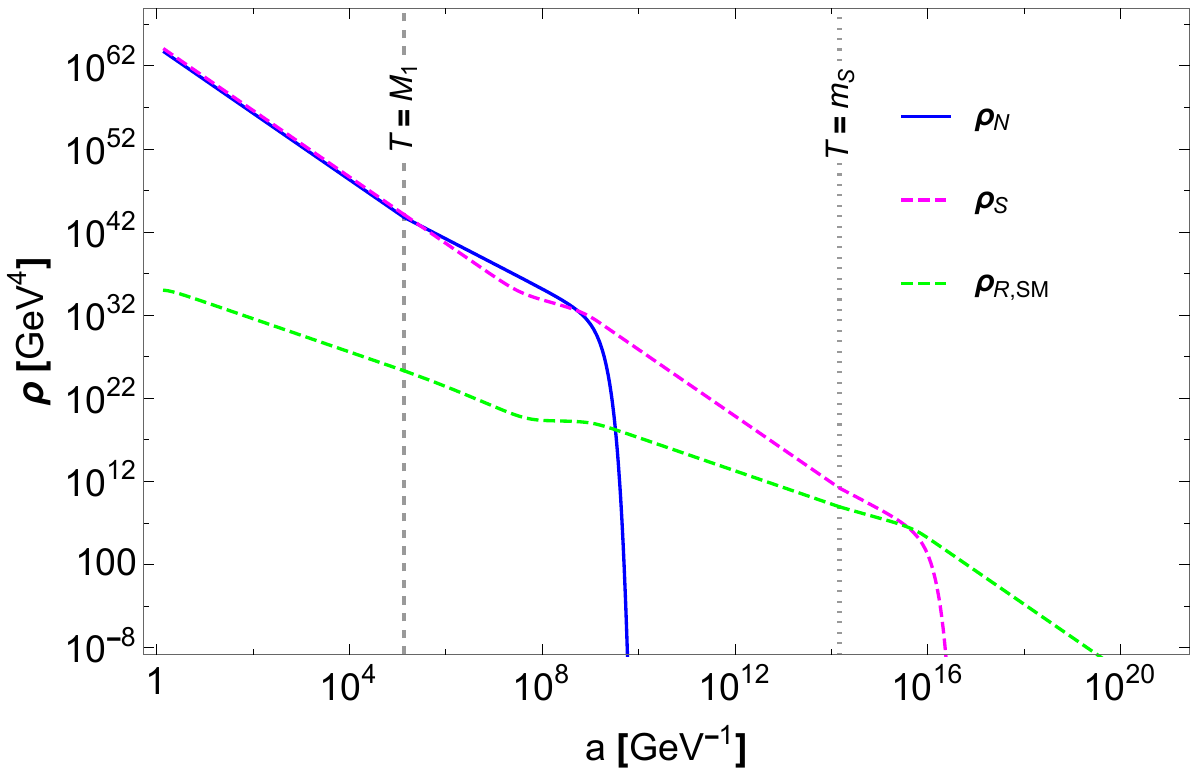}
         \caption{Energy Density }
         \label{fig:EarlyRho}
     \end{subfigure}
     \hfill
     \begin{subfigure}[b]{0.44\textwidth}
         \centering
         \includegraphics[width=\textwidth]{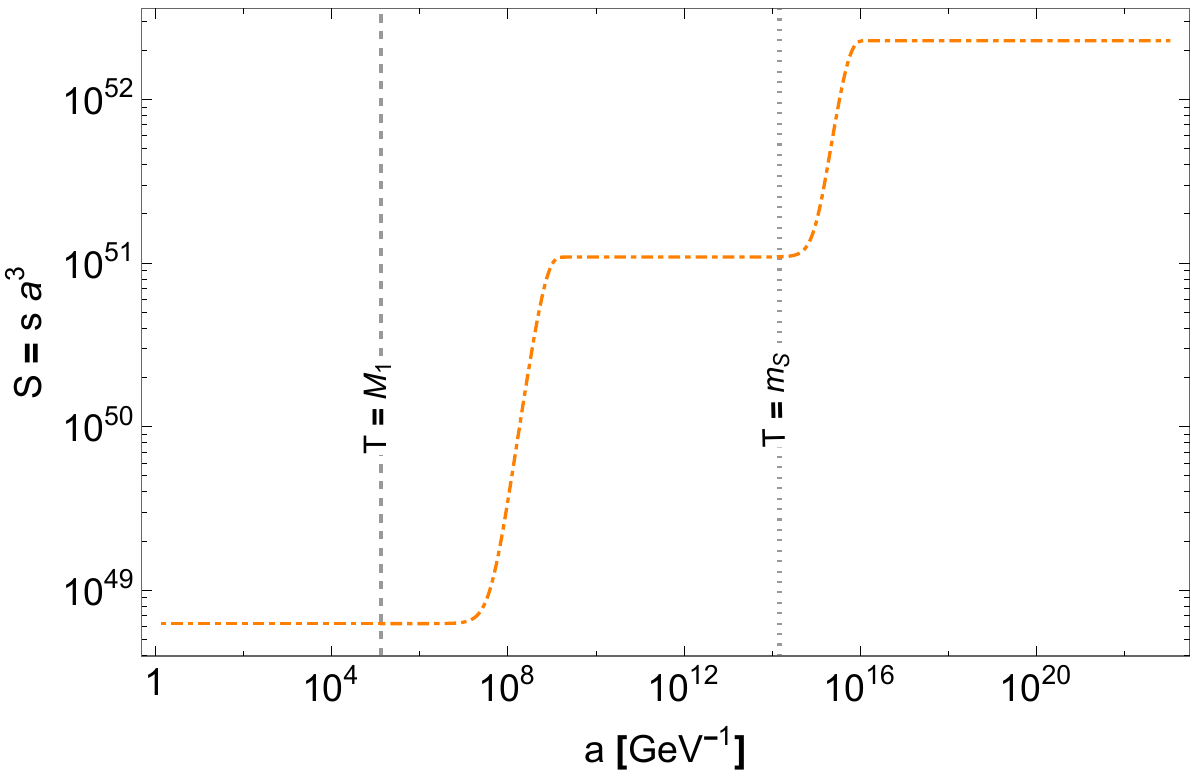}
         \caption{Entropy }
         \label{fig:EarlyS}
     \end{subfigure}
     \hfill
     \begin{subfigure}[b]{0.44\textwidth}
         \centering
         \includegraphics[width=\textwidth]{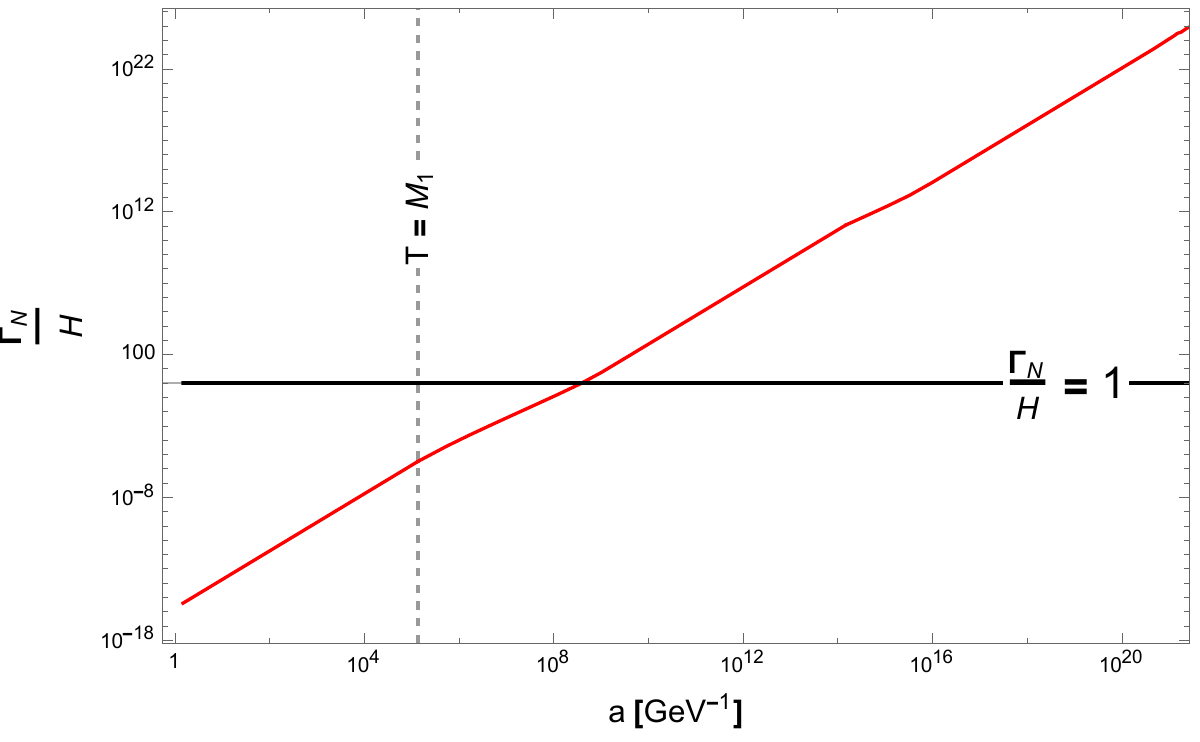}
         \caption{$\frac{\Gamma_N}{H(a) }$ }
         \label{fig:RatioWH}
     \end{subfigure}
\caption{Left upper panel: energy densities of the Majorana fermions, $\rho_{N}$, in solid blue, reheatons, $\rho_S$, in dashed magenta, and the SM thermal bath, $\rho_{\text{R,SM}}$, in dashed green. Right upper panel: the evolution of the entropy per comoving volume as a function of the cosmic scale factor. Bottom panel: the ratio of the $N$ decay width to the Hubble parameter as a function of the scale factor. In all panels, the vertical dashed black line corresponds to the time when $T=M_1$, whereas the black dotted line corresponds to the time when $T=m_{S}$. The benchmark parameters for the three panels are $\Gamma_N=10^{-2}$ GeV, $M_1=10^{11}$ GeV, $\Gamma_S = 10^{-15}$ GeV, $m_S = 5 \times 10^{2}$ GeV, and $a_I=1\; \text{GeV}^{-1}$. 
  }
  \label{fig:BZ Energy Density}
\end{figure}

We now turn to the details of our model. We assume that after inflation there is a thermal bath that includes relativistic $N_1$, $\phi$, and $S$, so the temperature must satisfy $T \gg M_{1}$.\footnote{The couplings we include do not thermalize the $N_i$, but we assume that reheating after inflation or some other mechanism creates a thermal bath. } The total energy density is given by $\rho_I$ and the scale factor at that moment is $a_I=1\; \text{GeV}^{-1}$. 
In addition, we assume that the $N$ SM-like sectors carry negligible energy. 
In our analysis, we do track the evolution of $\phi$; however, $\phi$ is short lived and does not play any relevant role in our analysis.
We then numerically solve the full set of Boltzmann equations for our setup, and the results are illustrated in~\eqnref{fig:BZ Energy Density}. 

The left upper panel in~\fref{fig:BZ Energy Density} shows the energy density evolution of the various particle species present in our Universe. At early times, $T\gg M_1$, the Universe is dominated by a relativistic admixture of $N$, $\phi$, and $S$  with $\rho \propto a^{-4}$. When $T < M_1$, $N_1$ becomes nonrelativistic, behaving like matter ($\rho_N \propto a^{-3}$) and dominating the energy density of the Universe for some period. After the $N_1$ decays are complete, the energy density of the system is stored in the reheatons that take over the evolution of the Universe. The reheaton dominated era will be explored in detail in~\sectref{sec: Reheating Dynamics}. 
As we can see from the right upper panel in~\fref{fig:BZ Energy Density}, the decay of $N_1$ and the $S$ will inject entropy into the SM sector. 
The first increase in the entropy density occurs when $N_1$ decays, while the second one corresponds to the decay of $S$. In the bottom panel, we can observe when $N_1$ departs from thermal equilibrium when $\Gamma_{N}\lesssim H$.

The tree level decay widths of the $N_i$ and the complex scalar, $\phi$, are given by:
\begin{equation}
\begin{split}
  \Gamma_{ N_{i} } & = \frac{ y^{2} }{ 8 \pi } \; M_{i} \; \left ( 1+ \frac{ m_{S}^{2} - m_{\phi}^{2} - 2 m_{S} M_{i}   }{ M_{i}^{2} } \right ) \;  \\
  & \hspace{0.8 cm} \left [ \left (1 - \frac{  m_{S}^{2} + m_{\phi}^{2} }{ M_{i}^{2} } \right )^{2} - \frac{ 4 m_{S}^{2} m_{\phi}^{2} }{ M_{i}^{4} }  \right ]^{1/2} \; , \\ 
  \Gamma_{\phi} & = \frac{ \kappa^{2} }{ 8 \pi } \; m_{\phi} \; \left ( 1- \frac{ 4 m_{S}^{2} }{m_{\phi}^{2}} \right )^{3/2}  \; . 
\end{split}
\label{eq:decay width}
\end{equation}
To quantify departure from the thermal equilibrium of $N_1$, we define the ratio $K$:
\begin{equation}
K=\frac{\Gamma_{N_{1}}}{H\left(M_{1}\right)} \; ,
\label{eq: K}
\end{equation}
where we compare the decay width of the lightest Majorana fermion, $\Gamma_{N_1}$, with the expansion rate of the Universe, given by the Hubble parameter, $H(T)$, at a temperature $T\sim M_{1}$. The decay rate of $N_1$ is given in \eqnref{eq:decay width}, whereas the Hubble parameter for the radiation domination era
at $T=M_1$ is
\begin{equation}
H\left(T_{1}=M_{1}\right)=\sqrt{\frac{\pi^{2}\,g_{\star}\left(T_{1}\right)}{90}}\,\frac{M_{1}^{2}}{M_{P}} \; ,
\label{eq: H at M1}
\end{equation}
where $g_{\star}\left(T_{1}\right)=5.5$  is the total number of relativistic degrees of freedom in $\phi$, $S$, and $S^c$. The $N$ states are nonrelativistic and the other states have not yet been excited in this era.  

As we will see, our model requires a large lepton asymmetry, so we want to be in the weak washout regime, where $K\ll 1$. In order to agree with observations, the Hubble parameter at the end of inflation, $H_I$, must satisfy $H_{I}\lesssim4\times10^{13}\left(\frac{r}{0.032}\right)^{1/2}\mathrm{GeV}$ \cite{Tristram:2021tvh}, with $r$ being the tensor-to-scalar ratio. This imposes an upper bound on the maximum $M_1$ that can be attained. Since $H(T) \lesssim H_I$, $T\lesssim 10^{16}$ GeV and, therefore, the largest mass for the $N_1$ is $M_1\sim10^{16}$ GeV.

The evolution of the yield of $N_{1}$, $Y_{N_{1}}=\frac{n_{N_{1}}}{s}$,
where $n_{N_{1}}$ is the number density of $N_{1}$,
can be tracked by the usual Boltzmann equations \cite{Fong:2012buy}. By defining $z=\frac{M_{1}}{T}$, the equilibrium $N_{1}$ yield is [with $g_{\star}(T)=g_{\star,s}(T)$]
\begin{equation}
Y_{N_{1}}^{\rm eq}\left(z\right)=\frac{90}{7\pi^{4}}\,z^{2}\:\mathcal{K}_{2}\left(z\right),\label{eq: Y_N1 eq} \;
\end{equation}
where $\mathcal{K}_{n}\left(z\right)$ is the $n$th order modified Bessel
function of the second kind. In the weak washout regime, $K\ll1$, the generated lepton asymmetry, $\mathcal{L}_I \equiv Y_{L}-Y_{\bar{L}}$, is~\cite{Fong:2012buy}
\begin{equation}
\mathcal{L}_I  \simeq \epsilon \, Y_{N_1}^{\rm eq} (0) \, \delta_N^{-1}   \simeq - \frac{3}{32 \pi} \, \left ( \frac{81}{256} \right )^{1/4} \, \Delta_1 \, K^{1/2} \, y^{2}   \; , \;
\label{eq: weak wash}
\end{equation}
assuming that $N_1$ has an initial thermal abundance, that is, $Y_{N_1} (z_{i})=Y_{N_1}^{\rm eq} (0)$, where $z_{i}=\frac{M_{1}}{T_{i}}$ and $T_i$ the initial temperature, with $z_i \sim 0$
corresponding to early times. $\Delta_1 \equiv \frac{M_1}{M_2} \ll 1$ and $\delta_N^{-1}$ is the entropy dilution factor in~\eqnref{eq:delta1} for $N_1$ decays.
For simplicity, we consider two Majorana fermions and assume that the imaginary and real parts of the Yukawa type coupling $y_i$ are of the same order, such that 
\begin{equation}
	\begin{split} 
	y^{2} \equiv \Im (y_{1}^{\dagger} y_{2}) \approx \Re (y_{1}^{\dagger} y_{2})  \Rightarrow \frac{\Im \left ( y_{1}^{\dagger} y_{k}  \right )^{2} }{\left( y^{\dagger} y \right)_{11}}= \frac{y^{2}}{2} \; . \;  \; 
	\end{split}
\end{equation}
In this work, we focus on the weak washout regime, since it renders a larger lepton asymmetry compared to other regimes.

\subsection{Favored parameter region}
\label{subsec: Bounds}

We now determine the favored region of parameter space that satisfies all the observational constraints and allows the model to attain the observed value of the SM baryon asymmetry. Since in our scenario there is no direct coupling of the $N_i$ to the SM neutrinos, there are no bounds on their masses $M_i$
(for instance, the Davidson-Ibarra bound \cite{Davidson:2002qv} does not apply here).
Nevertheless, since our ultimate goal is to explain baryogenesis in the SM sector, we have to take into account the initial lepton number asymmetry of the reheaton that is required for successful leptogenesis and the corresponding constraints that it can impose on our model.

To a first approximation, all the asymmetry in the reheaton sector is transferred to the SM, which we can use to set bounds on the lepton asymmetry $\mathcal{L}_I$.
To get a lower bound on $\mathcal{L}_I$, we assume that SM sphaleron is maximally efficient, converting roughly 1/3 of the lepton asymmetry into baryon asymmetry (see \sectref{sec: Baryon Asymm}) and then requiring that all the SM baryon asymmetry comes from this mechanism. The upper bound on $\mathcal{L}_I$ can be obtained using \eqnref{eq: weak wash}. We can take $K=\Delta_1=10^{-1}$, $M_1 = 1 \times 10^{16}$ GeV as the largest values of those parameters. Using \eqnSref{eq:decay width} and \eqref{eq: K} for the width of $N_1$, this leads to an upper limit of $y\simeq 0.06$. Plugging those values in \eqnref{eq: weak wash}, we get
\begin{equation}
  3 \times 10^{-11} \, \lesssim \abs{\mathcal{L}_I}  \lesssim \, 3 \times 10^{-6} \, . \label{eq:InitialLeptonAsymmetryLimit}
\end{equation}
In order to ultimately get a positive baryon number in the Universe, 
${\mathcal{L}}$ 
must be negative.
 
We can also express the weak washout bound ($K \ll 1$) in terms of the particle physics parameters using \eqnref{eq: K}, such that the lower bound on $M_{1}$ is
\begin{equation}
M_{1} \gtrsim (3 \times 10^{11} \text{ GeV}) \left ( \frac{y_{1}^{2}}{10^{-6}} \right ) \, g_{\star}^{-1/2},
\label{eq: M1 upper bound}
\end{equation}
where $y_{1}^{2}=y_{1}^{\dagger} y_{1}$. 
In addition, we must ensure that the $N_1$ mass is larger than the sphaleron freeze-out temperature, $T_{\text{Sp}}\sim100$ GeV \cite{DOnofrio:2014rug}. In what follows, we will explore how the various sectors are reheated.

There are two additional sources of lepton number which could spoil the leptogenesis mechanism. The first is a possible VEV for the $\phi$ field, but we take its mass squared term to be positive which stabilizes it at the origin. The second is the Majorana mass of the $N_i$ which explicitly breaks lepton number and radiatively generates a Majorana mass for the $S$. This must be a collective effect: namely that if either of $\kappa$ or $y_i$ from~\eqnref{eq:LeptoLag} are zero, then one can assign a global symmetry that forbids this Majorana mass. In the limit $M_i \gg m_\phi$, we estimate the size of this mass to be 
\begin{equation}
    m_{S, \; \rm Maj} \sim \frac{y_i^2 \kappa^2}{\left ( 16\pi^2 \right )^2} \frac{m_S^2}{M_i} \, .
    \label{eq:msMaj}
\end{equation}
This mass induces oscillations between $S$ and $S^c$, which would wash out the asymmetry. Therefore, we require this Majorana mass to be smaller than $\Gamma_S$ so that the reheaton decays before oscillating. This constrains the parameter space, but this constraint can easily be accommodated, for example with $y^2\sim 10^{-6}$, $M_1 \sim 10^{11}$ GeV, $m_\phi \sim 10^{9}$ GeV, $m_S\sim 10^2$ GeV, $\kappa \sim 1$. We also require the $\phi$ decay to be fast so that it does not inject entropy, but this is certainly true for $\kappa\sim 1$.

\section{Reheating \emph{N} Sectors }
\label{sec: Reheating Dynamics}

In the previous section, we showed that the initial lepton number asymmetry carried by the reheaton can be generated by a setup that is similar to leptogenesis. Eventually, the decays of  $N_1$ and $\phi$ into $S$ [see \fref{fig:treeLevelN1}] will lead to a Universe dominated by reheatons. According to the N-naturalness framework, the reheaton should couple universally to each sector \cite{Arkani-Hamed:2016rle}. Nevertheless, a large fraction of the energy density of the Universe must be transferred into the SM to reproduce observations. This can be achieved if the width of the reheaton
into each sector decreases as $\left|m_{H}^2\right|$ grows, which will happen if the reheaton mass is such that its decay proceeds mostly through higher dimensional operators. In this section, we explain how the reheating of the several sectors proceeds and show how the temperature evolves in these sectors, which will be important for computing the baryon asymmetries, as we discuss in \sectref{sec: Baryon Asymm}.

\subsection{Reheaton decay}
\label{reheaton decay}

The reheaton decay is mediated by its coupling to the vectorlike lepton doublet shown in \eqnref{eq:lagl4}. As discussed in~\sectref{sec: Model setup}, this will allow the reheaton to have a sufficiently large branching ratio into an exotic sector that we now describe in detail. 
We assume that the vectorlike leptons are heavier than the $S$\footnote{In the SM sector, LHC bounds~\cite{CMS:2019hsm} require the vectorlike leptons to be heavier than $790$ GeV.} and integrating them out leads to a mixing between the neutrinos and $S$. This in turn results in the following terms in the Lagrangian that can lead to possible two- or three-body decays of $S$: 
 \begin{equation}
 \begin{split}
 \mathcal{L}_{\expval{H} \neq 0} & \supset \frac{g}{\sqrt{2} } \frac{\lambda \mu_{L} }{M_{L} m_S } \frac{v_i}{\sqrt{2}} W_{\mu}^{+}  l_i  \sigma^{\mu} {S}^{\dagger} + \\
 &  \frac{\sqrt{g^2 + g'^{2} }}{2} \frac{\lambda \mu_{L} }{M_{L} m_S} \frac{v_i}{\sqrt{2}} Z_{\mu}  \nu_i  \sigma^{\mu}  {S}^\dagger + 
 \frac{\lambda \mu_{L} }{\sqrt{2} M_{L}}  \nu_i  h_i S^c \\ 
& \hspace{4.9 cm}  + \text{H.c.}  ,  \\
\mathcal{L}_{\expval{H} = 0} & \supset  \frac{\lambda \mu_{L} }{M_{L}}   l_i  H_i  S^c + \text{H.c.}\; ,  
 \end{split}
 \label{eq: Effective Lagrangian}
 \end{equation}
where $g$ is the $SU(2)$ gauge coupling, $H_i=(h_i^+ \; h_i^0)^T$, $g'$ is the $U(1)$ gauge coupling, and as above we are using two component Weyl fermions.

If kinetically allowed, $S$ decays dominantly to a Higgs and lepton or to a gauge boson and a lepton. For our choice of parameters, this will happen only in the $i=0$, $i=-1$ and possibly the $i=1$ but in general, the two-body decay widths are given by  
 \begin{equation}
\resizebox{0.5\textwidth}{!}{$
\begin{split}
 \Gamma_{S\rightarrow H_j e_j} & = \left ( \frac{ \lambda^{2} }{16 \pi }  \right ) \; \left ( \frac{ \mu_{L}^{2} m_{S} }{ M_{L}^{2} } \right )  \;  \left ( 1 -  \frac{ m_{H_j}^{2} }{ m_{S}^{2} } \right )^{2} \;, \;\; \; \; j<0 \; , \; \\
  \Gamma_{S\rightarrow h_k \nu_k} & = \left ( \frac{ \lambda^{2} }{64 \pi }  \right ) \; \left ( \frac{ \mu_{L}^{2} m_{S} }{ M_{L}^{2} } \right )  \;  \left ( 1 -  \frac{ m_{H_k}^{2} }{ m_{S}^{2} } \right )^{2} \;, \;\;\;\; k\geq 0 \; , \; \\
 \Gamma_{S\rightarrow W_k e_k+ \;Z_k \nu_k} & = \left ( \frac{ 3\lambda^{2} }{64 \pi }  \right ) \; \left ( \frac{ \mu_{L}^{2} m_{S}  }{ M_{L}^{2} } \right )  \;  \left ( 1 -  \frac{ m_{W_k}^{2} }{ m_{S}^{2} } \right )^{2} \\
 & \hspace{3.4 cm}   \left ( 1 + 2 \frac{ m_{W_k}^{2} }{ m_{S}^{2} } \right )  \;, \;\;\;\; k\geq 0 \; , \; 
\end{split} $}
\label{eq:Partial Widths} 
\end{equation} 
where
$m_{H,k}$ is the mass of the Higgs boson in the $k$th sector, $m_{W,k}$ is the mass of the $W$ boson in the $k$th sector, and the indexes $j$ and $k$ label the exotic and standard sectors, respectively.
In the third line of \eqnref{eq:Partial Widths}, we have assumed that $\frac{m_{W_k}}{m_{S}}\simeq \frac{m_{Z_k}}{m_{S}}$, but the full formula is given with the obvious substitution. We take the $S$ to be heavier than the SM $W$ boson, so the leading decay into the SM sector is $S\rightarrow We$.

\begin{figure*}[t!]
\centering
\begin{minipage}[c]{\textwidth}  \subcaptionbox{ $i=-1$ sector }{
\includegraphics[width=.45\textwidth ]{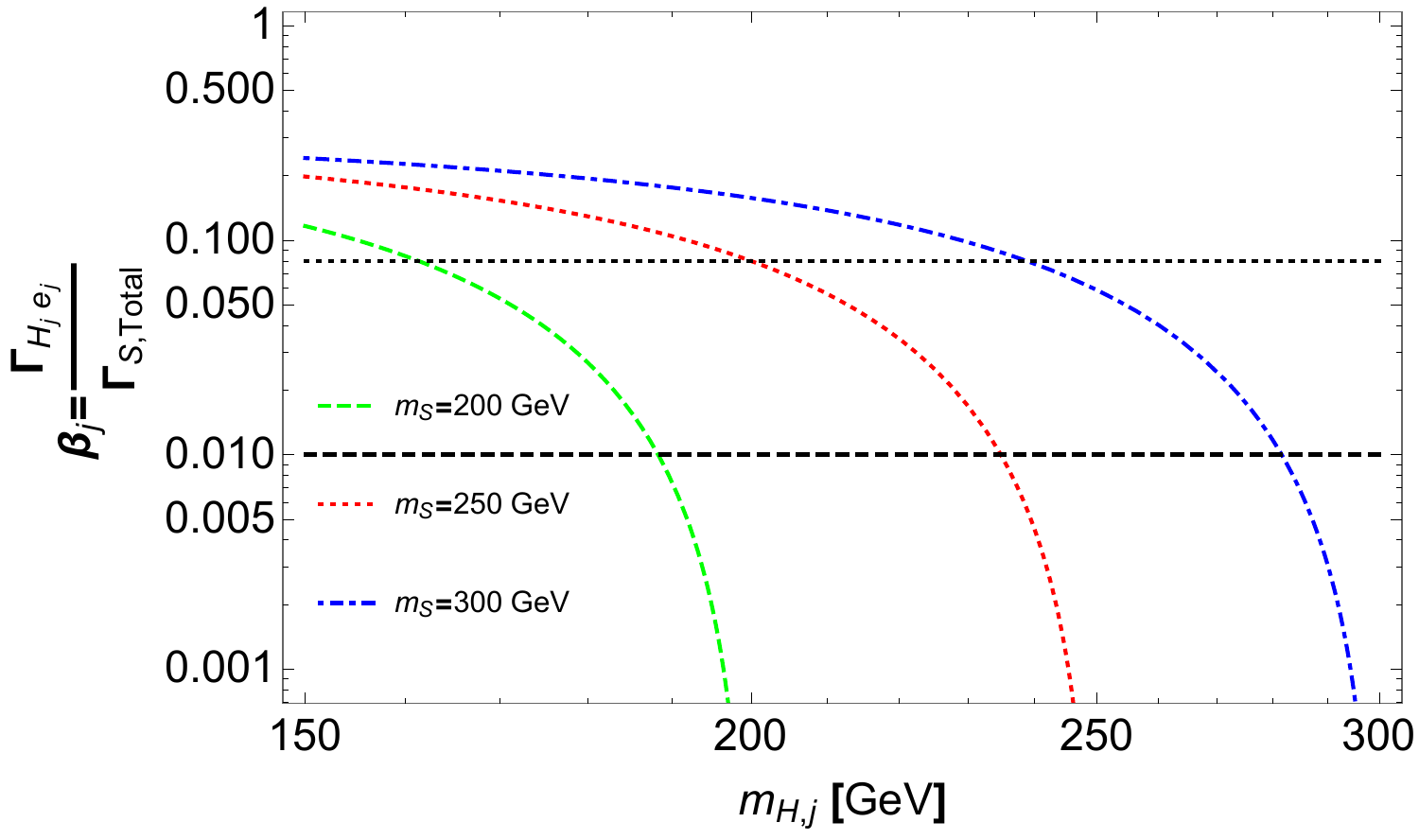} }
\hfill
\subcaptionbox{ $i=+1$ sector }{
\includegraphics[width=.45\textwidth]{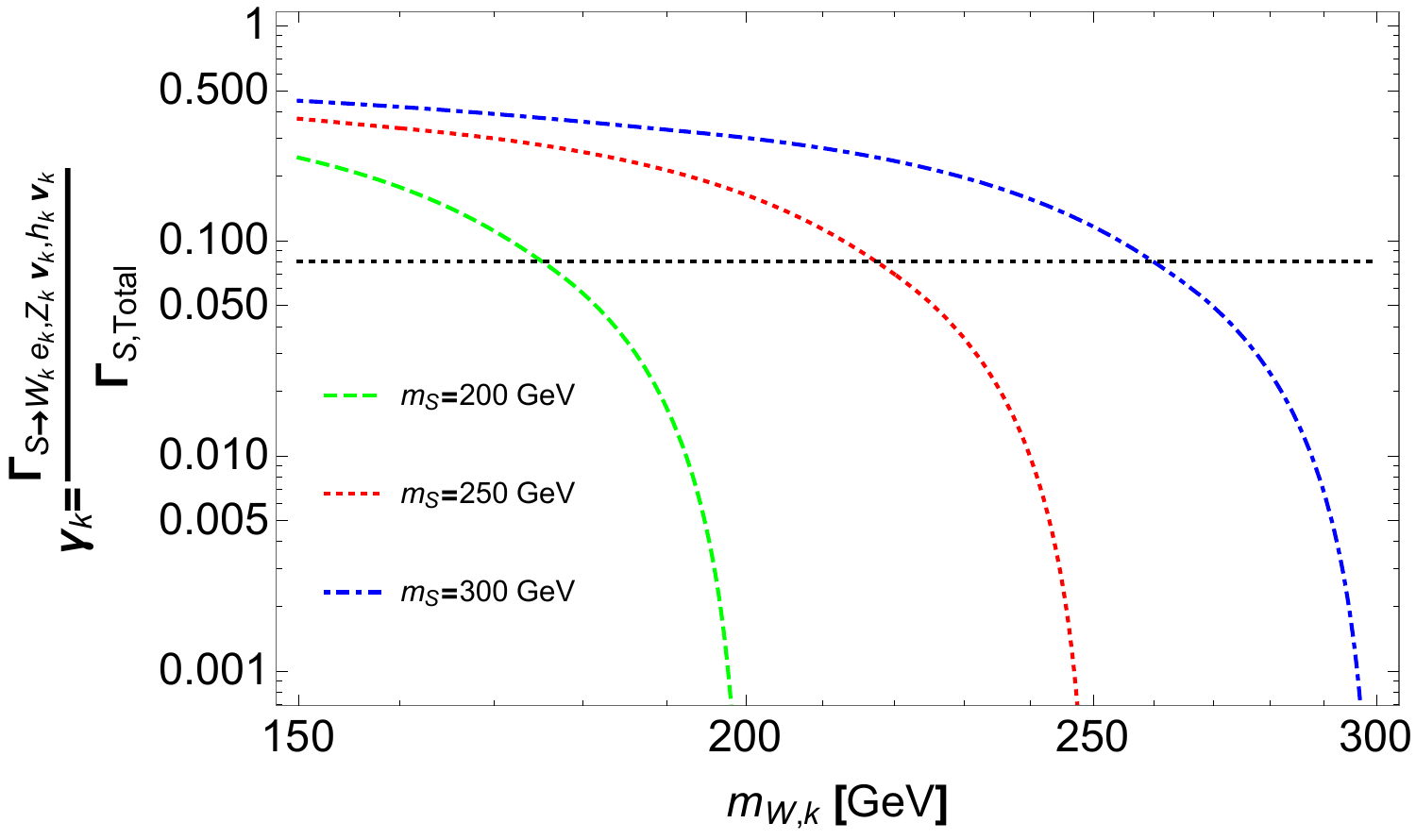}  }
\end{minipage}
\hfill
\caption{The branching ratio of the reheaton into the lowest exotic (left) and standard (right) sectors, as a function of the Higgs and $W$-boson mass, respectively, for different reheaton masses $m_S$ (200 GeV in dashed green, 250 GeV in dotted red, and 300 GeV in dot-dashed blue). The black dotted line corresponds to the reheaton branching ratio of $8\times10^{-2}$, which is the upper bound on the branching ratio from $\Delta N_{\text{eff}} < 0.4$ while the black dashed line corresponds to the reheaton branching ratio of $1\times10^{-2}$. For these plots, we have set
  $\mu_{L}=1$ GeV and $ M_{L}=1$ TeV.
  }
\label{fig:reheaton_width_BR} 
\end{figure*}

If the bosons in a given sector are heavier than the reheaton, then the decay to that sector proceeds via a three-body process, but the scaling is different depending on whether the off-shell boson is a vector or scalar. For a heavy vector ($W/Z$) in a standard sector, the decays are similar to muon decay in the SM, but there is an additional factor of $v_k$ in the operator in \eqnref{eq: Effective Lagrangian}. Therefore the width is given approximately as
\begin{equation}
\Gamma_{S\rightarrow V^*\nu\rightarrow \bar{f} f \nu} \sim \left ( \frac{  \lambda^{2} }{ 3072 \pi^{3}  }  \right )
\frac{ \mu_{L}^{2} m_{S}^{3}  }{ M_{L}^{2} m_{V,k}^{2}} \; ,
\label{eq: 3body}
\end{equation}
where $m_{V,k}$ is the mass of the heavy vector. This is the dominant decay in SM-like sectors where two-body decays are forbidden. Decays via an off-shell Higgs go roughly as
\begin{equation}
\Gamma_{S\rightarrow H^*\nu\rightarrow \bar{f} f \nu} \sim \left ( \frac{  \lambda^{2} }{ 3072 \pi^{3}  }  \right )
\frac{ \mu_{L}^{2} m_{S}^{5}  }{ M_{L}^{2} m_{H,k}^{4}} \; ,
\label{eq:scalar3body}
\end{equation}
where $m_{H,k}$ is the mass of the off-shell Higgs.\footnote{In the fermion reheaton model in~\cite{Arkani-Hamed:2016rle}, there is additional vectorlike field content and the decay to the exotic sector scales as $1/m_H^8$.} For sectors where the two-body decays are kinematically forbidden, the branching ratio to those sectors is expected to be $\lesssim 10^{-5}$.

We denote the branching ratios into the $j$th exotic sector, $\beta_j$, and the branching ratio into the $k$th standard sector, $\gamma_{k}$.
Focusing on sectors where two-body decays are allowed (including the SM sector), we can estimate the branching ratios as:
 \begin{equation}
\begin{split}
\beta_j & \simeq \frac{ \Gamma_{S\rightarrow H_j e_j} }{ \Gamma_{S\rightarrow \text{SM} } } = \left ( \frac{ 4 }{ 3 } \right )  \;  \left ( 1 -  \frac{ m_{H,j}^{2} }{ m_{S}^{2} } \right )^{2} \;  \left ( 1 -  \frac{ m_{W}^{2} }{ m_{S}^{2} } \right )^{-2} \\
& \hspace{5.2 cm}   \left ( 1 + 2 \frac{ m_{W}^{2} }{ m_{S}^{2} } \right )^{-1} \; , \;  \\
\gamma_{k}  & \simeq \frac{ \Gamma_{S\rightarrow W_{k} e_{k} , \;Z_k \nu_k}  }{ \Gamma_{S\rightarrow \text{SM}} } = \left (1 - \frac{m_{W_k}^{2}}{m_{S}^{2}} \right )^{2} \; \left ( 1 + 2 \frac{ m_{W_k}^{2}}{m_{S}^{2}} \right ) \; \\
& \hspace{2.7 cm}   \left ( 1 -  \frac{ m_{W}^{2} }{ m_{S}^{2} } \right )^{-2} \;   \left ( 1 + 2 \frac{ m_{W}^{2} }{ m_{S}^{2} } \right )^{-1} \; , \; \\
\end{split}
\label{branching ratios}
\end{equation}
where we have assumed that the decay to the SM sector dominates, and that the decay to the $h\nu$ in standard sectors is subdominant. We plot the branching ratios for possible two-body decays in \fref{fig:reheaton_width_BR}.
As we will show in \sectref{Neff Constraints}, in order to be consistent with $\Delta N_{\text{eff}}$, we require $\beta_j,\gamma_k \lesssim 0.08$, and in turn the branching ratio to the SM sector will be $\approx 1$, which justifies the above approximation. We will also show in \sectref{sec: Baryon Asymm} that for the dark matter to be saturated by baryons in the exotic sector, we need $\beta_{-1} \gtrsim 0.01$. From \eqnSref{eq:Partial Widths} and \eqref{eq: 3body} as well as \fref{fig:reheaton_width_BR}, the following picture emerges for the spectrum. The reheaton must be heavier than the SM Higgs, so that kinematic suppression of the two-body is negligible and the decay of the reheaton to the SM sector is dominant. The Higgs in the $j=-1$ exotic sector must be approximately degenerate with the reheaton so the decay is somewhat suppressed, but not as small as a natural three-body decay. The lower bound on the $W$ masses in the standard sectors is similar to the reheaton mass to satisfy $\gamma \lesssim 8 \times 10^{-2}$, but the $W/Z/H$ in the first exotic sector can be heavier than the reheaton and decays to that sector will be three-body. Finally, the Higgses in all sectors with $|i|>1$ must be heavier than the reheaton so that the two-body decay is forbidden.

\subsection{Cosmological evolution}
\label{cosmo evo}

We now turn to the cosmological evolution of the reheaton. Let us consider the simplest case where the reheaton decays into two sectors only: the SM one and a hidden sector (that could be an ``exotic sector," with $i<0$, or a ``standard sector," with $i>0$). The generalization for $N$ sectors can be accomplished by replacing $\beta \rightarrow \beta_i$ for the exotic sectors and $\beta \rightarrow \gamma_i$ for standard sectors. The evolution of the energy density of the Universe is given by the following set  of differential equations:
\begin{equation}
	\begin{split}
	& \dot \rho_{S} + 3\left(1+\omega\right) H \rho_{S} = -\, \Gamma_{S}\, \rho_{S}   \; , \\ 
	& \dot \rho_{\text{R,h}} + 4 H \rho_{\text{R,h}} = \beta\, \Gamma_{S}\, \rho_{S} \; ,  \\ 
	& \dot \rho_{\text{R,SM}} + 4 H \rho_{\text{R,SM}} = \left(1- \beta \right )\, \Gamma_{S}\, \rho_{S} \; ,   \\ 
	\end{split}
	\label{eq: BE}
	\end{equation}
where $\beta$ is the fraction of the reheaton that decays into the hidden sector, $\Gamma_{S}$ is the total reheaton's decay width, $ \rho_{S}$, $ \rho_{\rm R,SM} $ and $ \rho_{\rm R,h} $ are the energy density of the reheaton, SM, and hidden sectors, respectively.
The equation of state parameter, $\omega$, is defined by the ratio between the pressure and the energy density of the species (in this case, the reheaton), $p=\omega\,\rho$, where $\omega=0$ if it is nonrelativistic ($T\ll m$), and $\omega=1/3$ if it is relativistic ($T\gg m$).
The Hubble rate, $H$, can be written as
\begin{equation}
 H^{2} = \frac{1 }{3 M_{P}^{2} } \left ( \rho_{S} + \rho_{\text{R,SM}} + \rho_{\text{R,h}} \right)  \; , 
\end{equation}
with $M_{P}\simeq2.43\times10^{18}$ GeV.
The system of equations in \eqnref{eq: BE} is solved numerically with the following set of initial conditions: 
\begin{equation}
\rho_{S} \left(t_{I}\right) =3\,M_{P}^{2}\,\Gamma_{N}^{2} \; , \label{S initial cond1}
\end{equation}
\begin{equation}
\rho_{\text{SM}}\left(t_{I}\right)=0, \quad \rho_{h}\left(t_{I}\right)=0 \; ,
\label{S initial cond2}
\end{equation}
where $t_I$ corresponds to the time where the Majorana fermion $N_1$ has completely decayed. 
We assume that, at this stage, the energy density of the Universe is dominated by the reheaton, $S$, and $\phi$ and $N_i$ have decayed away. As we can see in the numerical analysis shown in~\fref{fig:BZ Energy Density}, \eqnSref{S initial cond1} and~\eqref{S initial cond2} are relatively accurate estimates for the initial conditions for our system when $S$ starts to dominate, since $\rho_{\rm SM}\ll \rho_S$.

The temperature of each sector can be tracked using the relation with the radiation energy density:
\begin{equation}
\rho_{\text{R,i}}  =\frac{\pi^{2}}{30}\,g_{\star,\text{i}}\left(T_{\text{i}}\right)\,T^{4}_{\text{i}} \Rightarrow  T_{\text{i}}  =\left(\frac{30}{\pi^{2}\,g_{\star,\text{i}}\left(T_{\text{i}} \right)}\right)^{1/4}\ \rho_{\text{R,i}}^{1/4} \; .
\label{eq: T evolution}
\end{equation}
These formulas will also apply to the reheaton as long as $T_S > m_S$. Once the temperature of the reheaton equals its mass, we model the equation of state as dropping instantaneously from 1/3 to 0. 
The decay of the reheaton is complete when its decay width is comparable
to the Hubble parameter, $H\sim\Gamma_{S}$. Here, the radiation
energy density starts to dominate the Universe, and the reheating temperature in the SM sector can be estimated as:
\begin{equation}
T_{\text{RH,SM}}=\left(\frac{90}{\pi^{2}\,g_{\star}\left(T\right)}\right)^{1/4}\,\sqrt{\Gamma_{S}\,M_{P}} \, .\label{eq: Trh}
\end{equation}
The reheating temperature is generally defined as the temperature of the thermal bath assuming an instantaneous transfer of the reheaton's energy density into radiation, initiating the radiation-dominated phase of the Universe below which the Universe expands with $T \sim a^{-1}$. However, the reheating phase is not an instantaneous process where the expansion of the Universe is faster compared to the radiation domination and it could attain a temperature much larger than $T_{\rm RH}$. During such period, $t<\Gamma_S^{-1}$, the decays of the reheaton provide a continuous supply of entropy~\cite{Giudice:2000ex}.
For simplicity, in this section, we treat the masses and branching ratios as constant and ignore thermal effects which are important at early times. However, since the reheaton decays mostly after the electroweak scale, the thermal effects that are presented in detail in~\appenref{Thermal Effects} do not change our final result.

In the SM, there is a lower bound on the reheating temperature of $T_{\text{RH,SM}}>4.1$ MeV at $95\%$ confidence level imposed by Big Bang Nucleosynthesis (BBN)~\cite{deSalas:2015glj}. This places a lower limit on the reheaton's width to be roughly $\Gamma_{S}\gtrsim 7\times10^{-24} \text{ GeV }$. This turns out to be a much weaker constraint than the one imposed by requiring the correct baryon asymmetry. Consideration of the latest cosmological bounds on $\Delta N_{\rm eff}$ leads to an upper limit on the reheaton’s width, see~\sectref{sec: Baryon Asymm} for a more detailed discussion.     

\begin{figure}[t]
\begin{centering}
\includegraphics[scale=0.43]{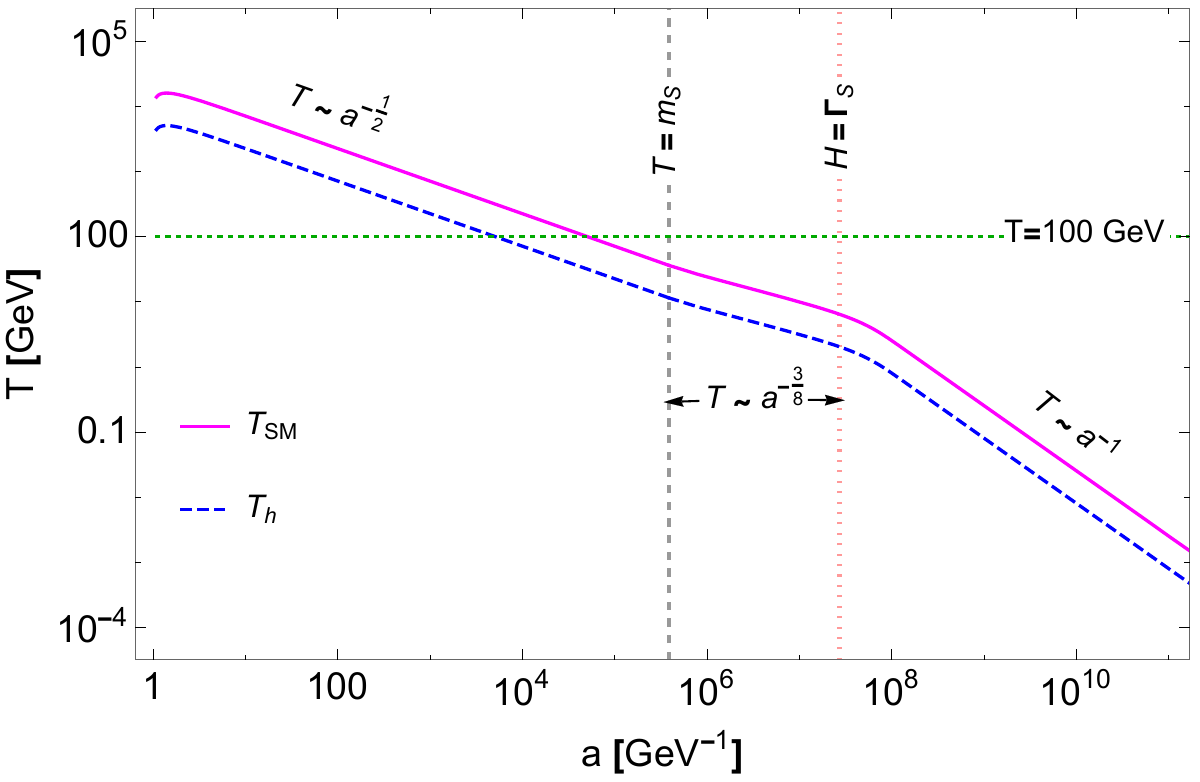}
\par\end{centering}
\caption{The evolution of the temperature in the SM sector (solid line) and a hidden sector (dashed line), as a function of the scale factor, $a$, with $\Gamma_S=10^{-16}$ GeV, $\beta=0.01$, and $\Gamma_N=10^{-2}$ GeV. The green horizontal line corresponds to $T=100$ GeV. On the top of the lines, we write the relation between temperature and the scale factor in each portion of the plot. }
\label{fig: T(a)}
\end{figure}

In order to facilitate numerical computations, we express \eqnref{eq: BE} in terms of dimensionless variables and replace time derivatives with derivatives with respect to the scale factor $a$ \footnote{We use conventions where the scale factor $a$ has dimensions of length.}  (using the definition of the Hubble parameter, $\dd{a}=H\,a\,\dd{t}$) \cite{Giudice:2000ex}. We can then define:
\begin{align}
S & \equiv\begin{cases}
\rho_{S}\,a^{4}, & \mathrm{if}\quad\omega=\frac{1}{3}\\
\frac{\rho_{S}\,a^{3}}{\Lambda}, & \mathrm{if}\quad\omega=0
\end{cases},\label{eq: S dimless}
\end{align}
\begin{equation}
R_{\text{SM/h}}=\rho_{R_{\text{SM/h}}}\,a^{4} \; , \label{eq: R dimless}
\end{equation}
\begin{equation}
A\equiv\frac{a}{a_{I}}=a\,\Lambda \; , \label{a no dim-1}
\end{equation}
where $a_{I}$ is the initial value of the scale factor for the numerical
integration. Since the results cannot depend on the choice of the
scale $a_{I}$, we assume that $a_{I}^{-1}$ is equal to some
$\Lambda$, which we set to be $\Lambda=1$ GeV in our numerical
computation. Thus, with this change of variables,
the initial conditions in \eqnSref{S initial cond1} and \eqref{S initial cond2} become
\begin{align}
S\left(A_{I}\right)\equiv S_{I} & =3\,M_{P}^{2}\,\frac{\Gamma_{N}^{2}}{\Lambda^{4}}\,A_{I}^{3},\quad\mathrm{with\,}A_{I}=1,\nonumber \\
R_{\text{SM}}\left(A_{I}\right)= & 0,\label{eq: initial cond}\\
R_{h}\left(A_{I}\right)= & 0.\nonumber 
\end{align}
As noted above, we model the equation of state $\omega$ as a step function going from $1/3$ to 0 when $T_S = m_S$.
In \fref{fig: T(a)}, we show the temperature evolution of the SM sector (solid line) and a hidden sector (dashed line), as a function of the scale factor, for a fixed reheaton decay width and $\beta$.

It is possible to perform an analytical estimate for  $T_{\text{max}}$ and the dependence of the temperature on the scale factor in the range $T_{\text{RH}}<T<T_{\text{max}}$. At early times ($H\gg \Gamma_S$), we can approximate the Boltzmann equation for $R_{\text{SM/h}}$ by assuming that a large fraction of the energy density is still retained by the reheaton, that is $S\sim S_I$. 
Thus, the equation for $R_{\text{SM/h}}$ can be solved analytically and we are left with
\begin{equation}
\begin{split}
R_{i} & \simeq 
\frac{\sqrt{3}M_P\,\Gamma_S}{\Lambda^2}\, \text{BR}(S\rightarrow i)\, S_I^{1/2} \\
& \hspace{2 cm} \begin{cases}
\frac{1}{2}\,\left(A^{2}-A_{I}^{2}\right), & \mathrm{if}\,\omega=\frac{1}{3}\\
\frac{2}{5}\,\left(A^{5/2}-A_{I}^{5/2}\right), & \mathrm{if}\,\omega=0
\end{cases} \label{eq: R analytic} \; , \;
\end{split}
\end{equation}
where $A_I$ and $S_I$ in the second line are the scale factor ($A_I=a_I\Lambda$) and scaled energy density of the $S$ when the $S$ becomes nonrelativistic.
By substituting~\eqnref{eq: R analytic} in the expression for the temperature in~\eqnref{eq: T evolution}, we can see that
\begin{equation}
T\sim\begin{cases}
a^{-1/2}, & \mathrm{if}\,\omega=\frac{1}{3}\\
a^{-3/8}, & \mathrm{if}\,\omega=0
\end{cases},\label{eq: T a analytic}
\end{equation}
for both SM and hidden sectors, whereas the maximum temperature achieved is
\begin{equation}
T_{\text{max},i}\sim \left( M_{P}^{2}\,\Gamma_{N}\,\text{BR}(S\rightarrow i)\,\Gamma_{S}\right)^{1/4} \; ,
\end{equation}
where $T_{\text{max}} \gtrsim \order{10^{2}\text{ GeV}}$. 
This behavior is confirmed in~\fref{fig: T(a)}. At very early times when the reheaton is relativistic, after attaining a maximum, the temperature decreases with scale factor as $T\sim a^{-1/2}$. Once the reheaton becomes nonrelativistic, the temperature scales as $a^{-3/8}$.  At $T_{\text{RH}}$, which marks the change of the slope in the plot, the decay of the reheaton is complete and there is no more entropy to be injected. From that point on, the usual radiation era commences and the temperature follows $T\sim a^{-1}$ as one would expect. We also see that throughout the evolution, the ratio of the temperature of the SM to the hidden sector remains approximately constant.

In the next section, we will make use of the evolution of the temperature to compute the baryon asymmetry generated in the SM sector but before that let us examine some possible constraints on our model.

\subsection{Constraints from light species}
\label{Neff Constraints}

The existence of $N$ sectors leads to a large number of nearly massless
degrees of freedom, given that all sectors contain photons. The neutrinos will also be relativistic at late times for some sectors. The presence of these extra
relativistic degrees of freedom can have observational effects as they modify the expansion of the Universe, which can lead to changes in predictions for light elements or the cosmic microwave background (CMB).
These contributions are encoded in the effective number of neutrino species, $N_{\text{eff}}$, which, for a completely decoupled hidden
sector can be written as
\cite{Vogel:2013raa, Breitbach:2018ddu}:
\begin{equation}
 N_{\text{eff}} = N_{\text{eff}}^{\text{SM}} + \Delta N_{\text{eff}} = N_{\text{eff}}^{\text{SM}} + \frac{4}{7} \left ( \frac{11}{4} \right )^{4/3} g_{h} \left ( \frac{T_{h}}{T_{\gamma}} \right )^{4}  \; , \;  \label{eq:Decoupled_Sector_Neff}
\end{equation}
where $T_{h} $ and $T_{\gamma}$ are the hidden and SM photon temperatures, respectively. In~\eqnref{eq:Decoupled_Sector_Neff}, $g_{h}$ is the number of effective relativistic degrees of freedom (d.o.f) in the hidden sector and $N_{\text{eff}}^{\text{SM}}$ is the effective d.o.f of the SM neutrinos, $N_{\text{eff}}^{\text{SM}}=3.046$~\cite{Mangano:2005cc}. The current $95\%$ confidence limit constraints from 2018 Planck data, considering the CMB and the baryon acoustic oscillations measurements, lead to $N_{\text{eff}}=2.99^{+0.34}_{-0.33}$~\cite{Planck:2018vyg}. In what follows, the hidden sector parameters (that can be ``standard sectors" with $i>0$, or ``exotic sectors" with $i<0$) are denoted by the subscript $h$. In order to determine the precise contribution to $\Delta N_{\text{eff}}$, we need to track the ratio of temperatures between the hidden and the SM sectors, $\xi \equiv \frac{T_{h}}{T_{\gamma}}$. As we described in \sectref{cosmo evo}, the temperatures of the SM and hidden sectors are, in general, different. Since after $T_{\text{RH}}$ the entropy in each sector is conserved, the following expression holds:
\begin{equation}
\frac{ T_{\text{Dec}}^{i}   }{ T_{\text{RH}}^{i} } = \left [  \frac{ g_{\star}^{i} (T_{\text{RH}}^{i} ) }{ g_{\star}^{i} (T_{\text{Dec}}^{i} ) } \right ]^{1/3} \, \frac{a (T_{\text{RH}}^{i} ) }{ a (T_{\text{Dec}}^{i} )} \; , \label{eq:Conservation_of_Entropy}
\end{equation}
where $T_{\text{Dec}}^{h}$ stands for the temperature of the hidden sector at the time of the CMB. Hence, the ratio between the temperatures of the photon decoupling in the hidden and SM sectors is given by\footnote{The hidden sector quantities are computed at the time of the SM. For example, $T_{\text{Dec}}^{h}$ is the temperature of the hidden sector at the $t_{\text{Dec}}^{\text{SM}}$.}
\begin{equation}
\frac{ T_{\text{Dec}}^{h}   }{ T_{\text{Dec}}^{\text{SM}} } = \frac{T_{\text{RH}}^{h} }{ T_{\text{RH}}^{\text{SM}} } \, \left [  \frac{ g_{\star}^{h} (T_{RH}^{h} ) }{ g_{\star}^{\text{SM}} (T_{\text{RH}}^{\text{SM}} ) } \right ]^{1/3} \, \left [  \frac{ g_{\ast}^{SM} (T_{\text{Dec}}^{\text{SM}} ) }{ g_{\star}^{h} (T_{\text{Dec}}^{h} ) } \right ]^{1/3} \; .
\label{eq:temp_EQ_Ratio}
\end{equation}
As we have discussed in \sectref{reheaton decay}, the reheaton deposits its energy into the various sectors. The ratio between the energy densities of the hidden and the SM sectors is, then,
\begin{equation}
\begin{split}
\frac{ \rho_{h} }{ \rho_{\text{SM}} } = \frac{ g_{\star}^{h} (T_{\text{RH}}^{h} ) }{ g_{\star}^{\text{SM}} (T_{\text{RH}}^{\text{SM}} ) } \, \left ( \frac{ T_{\text{RH}}^{h}  }{ T_{\text{RH}}^{\text{SM}} } \right )^4 \approx \frac{ \Gamma ( S \rightarrow \text{Hidden} ) }{ \Gamma ( S \rightarrow \text{SM} ) } \; .
\label{eq:reheaton_energy_Ratio}
\end{split}
\end{equation}
Using the relations obtained in~\eqnSref{eq:temp_EQ_Ratio} and~\eqref{eq:reheaton_energy_Ratio}, we get
\begin{equation}
\begin{split}
\left( \frac{ T_{\text{Dec}}^{h}   }{ T_{\text{Dec}}^{\text{SM}} } \right )^{4}  & = 
\left [  \frac{ g_{\star}^{h} (T_{\text{RH}}^{h} ) }{ g_{\star}^{\text{\text{SM}} (T_{\text{RH}}^{\text{SM}} ) }} \right ]^{1/3} \left [  \frac{ g_{\star}^{\text{SM}} (T_{\text{Dec}}^{\text{SM}} ) }{ g_{\star}^{h} (T_{\text{Dec}}^{h} ) } \right ]^{4/3} \, \\
& \hspace{2.5 cm} \frac{ \Gamma ( S \rightarrow \text{Hidden} ) }{ \Gamma ( S \rightarrow \text{SM} ) } \;  .
\label{eq:reheaton_energy_Ratio_2}
\end{split}
\end{equation}
Combining \eqnref{eq:reheaton_energy_Ratio_2} with the expression for $\Delta N_{\text{eff}}$ from~\eqnref{eq:Decoupled_Sector_Neff}, the contribution from the additional sectors is:

\begin{equation}
\begin{split}
\Delta N_{\text{eff}} & =\frac{4}{7}\left(\frac{11}{4}\right)^{4/3}\,g_{\star}^{h}(T_{\text{Dec}}^{h})\,\left[\frac{g_{\star}^{h}(T_{\text{RH}}^{h})}{g_{\star}^{\text{SM}}(T_{\text{RH}}^{\text{SM}})}\right]^{1/3}\, \\
& \hspace{0.8 cm} \left[\frac{g_{\star}^{\text{SM}}(T_{\text{Dec}}^{\text{SM}})}{g_{\star}^{h}(T_{\text{Dec}}^{h})}\right]^{4/3}\,\frac{\Gamma(S\rightarrow\text{Hidden})}{\Gamma(S\rightarrow\text{SM})} \; .
 \label{Delta Neff sectors} 
\end{split}
\end{equation}
Let us compute the respective contribution to the exotic and the standard sectors here.

\subsubsection{\textbf{Exotic sectors}}
The computations of $\Delta N_{\text{eff}}$ depend on the particle content of the various sectors and their respective mass spectrum, see~\appenref{App: mass spectrum}. In the exotic sectors, both the photons and neutrinos are relativistic at decoupling. Consequently, the contribution to $\Delta N_{\text{eff}}$ from any given exotic sector ($j<0$) due to the reheaton's decay can be expressed as 
\begin{equation}
\begin{split}
\Delta N_{\text{eff},j}^{\text{Decay}} & \simeq 7.4 \, \left [  \frac{ g_{\star}^{j} (T_{\text{RH}}^{j} ) }{ g_{\star}^{\text{SM}} (T_{\text{RH}}^{\text{SM}} ) } \right ]^{1/3} \, \left ( \frac{\beta_{j}}{1 - \beta -\gamma  } \right ) \; , 
\label{Delta Neff Exotic}
\end{split}
\end{equation}
where $\beta_{j}$ denotes the branching ratio into the exotic sector $j$, $\beta = \sum_j \beta_j$ is the total reheaton's branching ratio into the exotic sectors, and $\gamma = \sum_k \gamma_k$ parametrizes the reheaton's total branching ratio into the standard sectors, $g_{\star}^{j}(T_{\text{RH}}^{j})\sim 102.75$, and we took $g_{\star}^{\text{SM}}(T_{\text{Dec}}^{\text{SM}})\sim 3.36$, and $g_{\star}^{j}(T_{\text{Dec}}^{j})\sim 17.75$.

\begin{figure}[t!]
\begin{centering}
\includegraphics[scale=0.7]{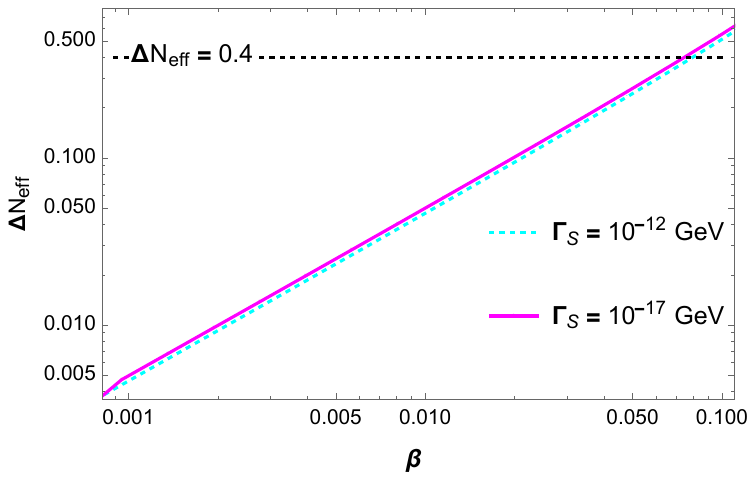}
\par\end{centering}
\caption{The $\Delta N_{\text{eff}}$ as a function of the branching ratio to the lowest exotic sector for two different reheaton widths: $\Gamma_{S}=10^{-12}$ GeV (dashed cyan curve) and $\Gamma_{S}=10^{-17}$ GeV (solid magenta curve). $\Delta N_{\text{eff}}$ changes little when varying the reheaton's decay width.
}
\label{fig:Neff_values}
\end{figure}

The $\Delta N_{\text{eff}}$ contribution is plotted in~\fref{fig:Neff_values} as a function of the fraction of reheatons decaying into the lowest exotic sector, $\beta_{-1}$. We may conclude that $\Delta N_{\text{eff}} \lesssim 0.4$ can be obtained for a branching fraction $\beta_{-1} \lesssim 0.08$, and that changing the reheaton's width for various $\beta$ values alter $\Delta N_{\text{eff}}$ very slightly. Sectors with larger Higgs mass parameters have significantly smaller contributions to $\Delta N_{\text{eff}}$ because the reheaton partial width scales as 
$\Gamma_{m_{H}^{2}>0} \sim 1/m_{h_i}^4$ [see \eqnref{eq:scalar3body}]. For larger $i$, the decay is expected to be three-body and the total contribution goes approximately as $\sim 10^{-5} \sum_j \frac{1}{j^2}$, which is well within the current bounds even for an infinite number of sectors. Hence, for our purpose, the exotic sector contribution is dominated by the sector with the smallest Higgs mass parameter. 

%

\subsubsection{\textbf{Standard sectors photons}}
\label{subsec:standardphotons}

In the standard sectors, the photons are still massless but the neutrinos get mass from electroweak symmetry breaking and are thus expected to be heavier than SM neutrinos.\footnote{An alternative possibility is that neutrino masses are independent of the Higgs VEV, see for example~\cite{Dvali:2016uhn}.} Therefore we consider them separately. For the photons, 
 $\Delta N_{\text{eff}}$  from reheaton decay is given by
\begin{equation}
\begin{split}
\Delta N_{\text{eff},k}^{\text{Decay}}  & \simeq 8.8 \, \left [  \frac{ g_{\star}^{k} (T_{\text{RH}}^{k} ) }{ g_{\star}^{\text{SM}} (T_{\text{RH}}^{\text{SM}} ) } \right ]^{1/3} \, \left ( \frac{\gamma_{k}}{1 - \beta -\gamma  } \right ) \; , 
\label{Delta Neff Standard i}
\end{split}
\end{equation}
where $g_{\star}^{k}(T_{\text{Dec}}^{k})= 2$, which gives $\gamma_1 \lesssim 0.07$.  Unlike the exotic sectors, the dominant reheaton decay is to the gauge bosons and the branching ratio scales as $\gamma_k \propto v_k^{-2} \propto k^{-1}$ [see~\eqnref{eq: 3body}]. Therefore if we sum over all sectors up to $k_{\rm max}$ we get 
\begin{align}
\Delta N_{\rm eff}^{\rm Decay} & \simeq 1.0 \times 10^{-4} \; \left [  \frac{ g_{\star}^{k=1} (T_{\text{RH}}^{k=1} ) }{ g_{\star}^{\text{SM}} (T_{\text{RH}}^{\text{SM}} ) } \right ]^{1/3} \;\left( \frac{\gamma_1}{10^{-5}} \right) \; \nonumber \\
& \hspace{4.95 cm}  \ln(k_{\rm max}) \; . \;
\end{align}
For simplicity, we assumed that $g_{\star,k}$ is constant for all the sectors which will slightly overestimate the size of this contribution. 
Therefore, even for $10^{16}$ sectors, these contributions are unimportant for $\gamma_1 \lesssim 1\times10^{-3}$.

\subsubsection{\textbf{Standard sector neutrinos}}
\label{subsec:standardeutrinos}

As the neutrinos in the standard sectors are heavier than the SM neutrinos, most are nonrelativistic at the time of decoupling and thus are constrained by $\Omega h^2$ rather than $\Delta N_{\text{eff}}$. We will show that the $\Omega h^2$ constraints impose an important limitation on the total number of sectors due to the heavier neutrinos in the standard sectors acting as dark matter. Neutrinos in other sectors can be produced via the decay of the reheaton into the thermal bath of the hidden sectors as described in detail in~\sectref{reheaton decay}. Although, as one ventures to larger $k$ (and thus larger Higgs mass parameter), the branching ratio of the reheaton will be considerably smaller compared to the SM sector, the total contribution to the relic abundance of hidden sector neutrinos becomes significant. 

In order to compute the relic density of neutrinos, one must make an assumption about how the mass of the neutrinos scales with the VEV of the Higgs. We will take the neutrinos to be pure Dirac; they get their mass in the same way as the quarks and charged leptons by marrying a right-handed neutrino field that is different in each sector. The radiative Majorana mass of $S$ from~\eqnref{eq:msMaj} can generate a small Majorana mass for the left-handed neutrinos after electroweak symmetry breaking. This mass is of order 
\begin{equation}
m_{\nu,\text{ Maj}}\sim m_{S,\text{ Maj}} \left( \frac{\lambda \, \mu_L\, v_i}{M_L \,m_S}\right)^2 \; , \;
\end{equation}
which is well below bounds on such a scenario~\cite{deGouvea:2009fp}.
We stress that the $N$ field responsible for leptogenesis introduced in \eqnref{eq:LeptoLag} is not the same as the fermions responsible for neutrino masses. Thus the masses of the neutrinos scale as $m_k \propto v_k$ or 
\begin{equation}
m_k = m_0 \left (  \frac{2k+r}{r} \right )^{1/2} \; , \;
\end{equation}
where $m_0$ is the SM neutrino mass. As noted previously, 
the branching ratio of the reheaton into the $k>1$ sectors scales as $\gamma_k \propto v_{k}^{-2}$ [see~\eqnref{eq: 3body}]. 

The cosmological evolution of the neutrinos also depends on $k$. In the SM, neutrinos decouple from the rest of the SM bath when the weak interactions freeze out at $T_{\rm \nu}\sim $ MeV. Other sectors have lower temperatures and also have heavier $W/Z$ bosons, so the neutrinos freeze out earlier. We can find the temperature of the SM at the time of $k$th sector neutrino decoupling, $T_{\nu,k}$ as follows. We equate the neutrino interaction rate, $\Gamma_k \simeq G_{F,k}^2 T_k^5$, to the Hubble constant, $H\simeq T^2/M_P$. We can use $G_{F,k} \propto G_F (2k+r)^{-1}$, $T_{\rm \nu }^3 \sim  \frac{1}{M_P \; G_{F}^2 }$, and $T_k/T \propto \gamma_k^{1/4}\propto \left(\gamma_1/(2k+r)\right)^{1/4}$ to get
\begin{align}
T_{\nu,k} = T_{\rm \nu}  \; \left ( \frac{2k + r}{r} \right )^{1/2}  \;  \left ( \frac{1 - \beta - \gamma }{\gamma_1} \right )^{1/6} \; \left ( \frac{g_{\star,k}}{g_{\star }} \right )^{1/6} \; , \; \label{eq: neutrino Decoupling T}
\end{align}
where $g_{\star }$ and $g_{\star,k }$ are the number of effective degrees of freedom in the SM and $k$th standard sector, respectively, computed at the time of the SM neutrino decoupling.

The $k$th sector neutrinos are produced by the decay of the reheaton and are reheated to a temperature given by
\begin{align}
&  T_{\rm RH,k} \simeq T_{\rm RH,SM} \; \left [ \frac{2 + r}{2k+r} \right ]^{1/4} \; \left [ \frac{\gamma_1}{ 1- \beta - \gamma} \right ]^{1/4}  \;  \left ( \frac{ g_{\star, \rm RH} }{g_{\star,k, \rm RH}} \right )^{1/4} \; , \;  \label{eq: Reheating temperautre i}
\end{align}
where $T_{\rm RH,SM}$ is the reheating temperature of the SM sector, $g_{\star,\rm RH}$ and $g_{\star,k,\rm RH}$ is the number of effective degrees of freedom in the SM and $k$th standard sector, respectively, at reheating. Therefore only in sectors with $k< k_{\rm max,1} \simeq 2,500$ do the neutrinos achieve equilibrium at all, and then they freeze out when still relativistic. For those sectors, we can estimate the relic density using a standard hot freeze out calculation:
\begin{equation}
\begin{split}
\frac{ \Omega h^2 }{\Omega_{\rm Obs} h^2 } & = \frac{1}{\Omega_{\rm Obs} h^2 } \sum_{k=1}^{k_{\rm max,1}} \Omega_k h^2  \;  \\
\frac{ \Omega h^2 }{\Omega_{\rm Obs} h^2 }  & \simeq \left (5.3 \times 10^{-3} \right )  \; \left (  \frac{m_0}{0.0585 \text{ eV} } \right ) \; \left ( \frac{\gamma_1}{10^{-5}} \right )^{3/4}  \; \\
&  \left ( \frac{k_{\rm max,1}}{2.5 \times 10^{3}} \right )^{3/4} \; \left ( \frac{g_{\star,k}}{75.75} \right ) \; \left (  \frac{g_{\rm eff}}{4.5} \right ) \; \left ( \frac{106.75}{  g_{\star,0} } \right ) \\ 
&\hspace{5 cm} \; \left (  \frac{ 5.5 }{ g_{\star,S} } \right )  \; , \; 
\end{split}
\end{equation}
where we let $r=0.3$, $m_0$ is the mass of the neutrino in the $i=0$ sector, $g_{\rm eff} =\frac{3}{4} g_{f}$ for fermions, and $\Omega_{\rm Obs}h^2 =0.12$ is the current dark matter relic abundance. Moreover, $g_{\star,0}$ and $g_{\star,k}$ are the number of effective degrees of freedom in entropy in the SM and $k$th standard sector respectively computed at the time of the reheaton's decay. On the other hand, $g_{\star,S}$ is the number of effective degrees of freedom in entropy in the SM determined at the time of the neutrino freeze-out. This relic abundance can be compared to that of SM neutrinos. The upper bound on the sum of the masses of neutrinos is $\sum m_\nu \lesssim 0.26$ eV at $95\%$ CI~\cite{Loureiro:2018pdz}. Neutrinos at that mass would contribute 2.4\% of the dark matter budget of the Universe, while the relic abundance from the low $k$ neutrinos is well below that value and thus safe from bounds.

For sectors $k > k_{\rm max, 1}$, the neutrinos never thermalize but are still produced relativistically.\footnote{Reheaton decay to neutrinos is kinematically forbidden for $k \gtrsim 5 \times 10^{22}$, but this is well above the number of sectors allowed.} Therefore, after reheating their number is fixed and can be estimated by finding the number density of the $S$, $n_S$ at the time of its decay when using $H=\Gamma_S$, and then evolving with the expansion of a radiation dominated Universe.  
 The relic abundance can be expressed as
\begin{equation}
\begin{split}
\frac{ \Omega h^2 }{\Omega_{\rm Obs} h^2 }  &  = \frac{ 1 }{\Omega_{\rm Obs} h^2 }   \sum_k  \frac{\rho_{k,\nu} }{\rho_c} h^2 \simeq  \frac{ 1 }{\Omega_{\rm Obs} h^2 }  \sum_k  \frac{  m_k \; n_S (a) }{\rho_c} h^2 \; \\
& \hspace{4 cm}  \left [ \text{ Br } (S \rightarrow 3\nu_k) \; \right ] \;   \\
\frac{ \Omega h^2 }{\Omega_{\rm Obs} h^2 } & \simeq   0.024  \; \left ( \frac{ m_0 }{0.0585 \text{ eV}} \right ) \; \left ( \frac{ \gamma_1 }{10^{-5} } \right )  \; \left ( \frac{ T_{\rm RH,SM} }{10^2 \text{ GeV}} \right ) \; \\
& \hspace{2.5 cm} \left ( \frac{10^2 \text{ GeV}}{ m_{S} } \right ) \; \left( \frac{N_{\rm max}}{1.6\times 10^{{5}}} \right)^{1/2} \; . \; 
 \label{eq: RelicAbundance caseB PureDirac}
\end{split}
\end{equation}
This places an upper bound on the number of sectors in the theory. If we require that these neutrinos contribute less to dark matter than SM neutrinos at their upper limit in mass, then we get $N_{\rm max} \lesssim 1.6\times 10^{5}$. This previous computation gives a conservative bound on the contribution of hidden neutrinos to hot dark matter. Indeed, the precise bound coming from the presence of hot dark matter depends on its free-streaming length. The heavier neutrinos that do not achieve thermal equilibrium have a free-streaming length that can be estimated as
\begin{equation}
\begin{split}
\lambda_{\rm FS} (t) & \equiv r(t) - r(t_i) = \int_{t_i}^{t} \frac{v(t')}{a(t')} \dd{t'} \simeq   2 \frac{t_{\rm NR}}{a_{\rm NR}}   \;  \\
\lambda_{\rm FS} (k)  & \simeq  \left ( 4.5 \times 10^{-2}  \text{ Mpc} \right )   \left ( \frac{ 10^2 \text{ GeV} }{ T_{\rm RH,SM} } \right )  \\
&  \left ( \frac{ m_S   }{ 10^{2} \text{ GeV} }  \right )   \left ( \frac{ 0.0585 \text{ eV} }{ m_0   }   \right )   
\left( \frac{2.5\times 10^3}{k} \right)  \; ,  \; \label{eq: freeStream}
\end{split}
\end{equation}
where $t_{\rm NR}/a_{\rm NR}$ is the time or scale factor where the neutrinos become nonrelativistic. 
Warm dark matter with a free-streaming length shorter than $ 0.015$ Mpc is unconstrained~\cite{DES:2020fxi} by structure formation observation. Hence we expect that sectors with $k > 7.5\times10^{3}$ should behave similar to cold dark matter and not be constrained. Therefore, we can take a more aggressive upper bound on the number of sectors such that the dark matter is saturated by hidden sector neutrinos, given by $ N_{\rm max} \lesssim 3\times 10^8$. In that case, one would require a different parameter space for the dark baryon asymmetry such that it would become a subdominant component of dark matter.  

If we instead took neutrinos to have Majorana masses with $m_k \propto v_k^2$ or $m_{k} = m_0 \; \left ( \frac{2 k + r}{r} \right )$, then the calculation of the relic density goes similarly but the result is very different. The relic density is linear in mass, so each sector gets an additional power of $\sqrt{2k+r}$ compared to the Dirac case. The total contribution to the relic density is then also multiplied by $\sqrt{k_{\text{max}}}$, and these neutrinos reach 2.4\% of the measured dark matter density with only $\sim 250$ sectors.

Another alternative is for the reheaton to couple to the $N$ sectors as in the model presented in~\cite{Arkani-Hamed:2016rle} with the additional vectorlike matter. In that setup, the branching ratio to hidden sectors scales as $1/v_k^8$ rather than $1/v_k^2$ as in \eqnref{eq: 3body}, and thus the upper bound on the number of sectors will be much weaker. While the scenario we present has a bound such that the full hierarchy problem cannot be solved, in this alternative mediation mechanism one could have up to $10^{16}$ and solve the full hierarchy problem. In that case, the phenomenology of the baryon asymmetry and hidden neutron dark matter would be qualitatively similar, as those are dominated by the SM and the exotic sector with the lowest positive Higgs mass squared parameter, while the dominant change is to sectors with large Higgs masses.

Finally, we note that there is another process to produce $k$th sector neutrinos, $\nu \nu \rightarrow \nu_{k} \nu$. In \appenref{FI neutrinos} we show that this process is negligible.

\section{Baryon Asymmetry }
\label{sec: Baryon Asymm}
In this section, we describe how the baryon asymmetry is generated in various sectors. In~\sectref{sec: Lepton asymmetry}, we presented a framework of how the lepton asymmetry can be produced in our model, and in \sectref{sec: Reheating Dynamics} we showed how that lepton asymmetry is distributed to the different sectors. That lepton asymmetry is then reprocessed by the electroweak sphalerons \cite{Manton:1983nd, Klinkhamer:1984di, Kuzmin:1985mm}. In the SM, classically $U(1)_{B+L}$ is an accidental global symmetry but it is violated quantum mechanically due to quantum anomalies
and $SU\left(2\right)_{L}$ field configurations~\cite{Adler:1969gk, Bell:1969ts,PhysRevLett.37.8}.
 The $(B+L)$ violation allows the electroweak sphaleron to partially convert the lepton asymmetry, carried by the reheaton, into the baryon asymmetry distributed across the various sectors since anomalous electroweak processes conserve the difference between baryon ($B$) and lepton numbers ($L$). For the SM field content, assuming the sphaleron interactions are rapid and chemical equilibrium is maintained, the baryon number can be expressed as $B=\frac{28}{79} \left (B-L \right )$~\cite{PhysRevD.42.3344}. However, around the electroweak symmetry breaking phase transition, the sphaleron rate becomes suppressed, complicating the 
relationship between the baryon number and the conserved quantity $(B-L)$ in a manner that will be described here.     

\subsection{Standard sectors}

Mathematically, the sphaleron is a static saddle point solution of the field equations and the \textit{sphaleron rate} per unit time per unit volume, $\Gamma_{\text{diff}}$, is half the Chern-Simons diffusion rate.\footnote{The Chern-Simons diffusion rate is defined in terms of the winding number as
\begin{equation}
\begin{split}
\Gamma_{\text{diff}} \equiv \lim_{V, \; t \to\infty} \frac{ \expval{ \left ( N_{\text{CS}} (t) - N_{\text{CS}} (0)  \right )^{2} } }{V \; t }  \; ,
\end{split}
\end{equation}
where $N_{\text{CS}} (t) - N_{\text{CS}} (0) = \frac{1}{n_G} \left ( B(t) - B(0) \right ) = L_{i}(t) - L_{i}(0)$ and $V$ is the volume of space. 
}  
At high temperatures well above the electroweak scale, the sphaleron occurs via thermal hopping and there is no suppression. 
 The rate can be estimated from lattice simulations~\cite{Bodeker:1999gx, Moore:2000ara}:
\begin{equation}
\Gamma_{ \text{diff} } (T) = \left ( 25.4 \pm 2.0 \right) \alpha_{w}^{5} T^{4},
\end{equation}
where $\alpha_{w} = g^2/(4\pi)$ and $g$ is the $SU(2)$ gauge coupling. In this regime, the rate will always be larger than the Hubble parameter.

At low temperatures, $T\ll 100$ GeV, the transition amplitude between the various vacua is highly suppressed (being proportional to a quantum-tunneling factor $e^{-8\pi^{2}/g^{2}} \sim 10^{-173}$, where $g$ is the $SU(2)_L$ gauge coupling). Here we are interested in the intermediate regime near the temperature of the electroweak phase transition.  In the SM, the phase transition is a crossover~\cite{Kajantie:1996mn, Karsch:1996yh}, and the Higgs VEV is continuous in time/temperature. Defining $T_c$ as the temperature where the Higgs VEV first becomes nonzero, the Higgs VEV and $W$ mass at temperatures below $T_c$ can be written as~\cite{Arnold:1987mh}
\begin{eqnarray}
v^{2} (T) &=& v^{2} (0) - \left [ \frac{1}{2} + \frac{3 \pi \alpha_w}{4 \lambda} \right ]\; T^{2} \nonumber \; , \\
m_{W} \left ( T \right ) &=& m_{W} (0) \left [ 1 - \left ( \frac{T}{T_{c} } \right )^{2} \right ]^{1/2} \; , \; 
\end{eqnarray}
where $\lambda$ is the Higgs quartic coupling and $m_{W} (0)$ is the $W$ boson mass at zero temperature.  
For the SM, $T_c \sim 159$ GeV~\cite{DOnofrio:2014rug}, and the formulas are the same for the standard sector with larger values of $v(0)$ and $T_c$. 

The sphaleron rate for $T < T_{c,i}$ can be estimated as~\cite{Arnold:1987mh}
\begin{equation}
\Gamma_{ \text{diff} } (T) = T^{4} \frac{ \omega_{-} }{ \pi \; v  } \left ( \frac{ v }{ T } \right )^{7} \mathcal{N}_{\text{tr}} \left ( \mathcal{N} \mathcal{V} \right )_{\text{rot}} \ \kappa \ \exp(- \frac{ E_{\text{Sp}} }{ T } ) \; , 
\label{eq:GamDiff}
\end{equation}
where the sphaleron energy $E_{\text{Sp}}\simeq 4 \frac{m_{W} (T) }{ \alpha_{W}}$~\cite{Klinkhamer:1984di, Kunz:1992uh,PhysRevD.40.3463, Carson:1990jm}, $\omega_{-}$ is called the dynamical prefactor, $\abs{ \omega_{-}^{2} }\simeq 2.3 m_{W}^{2}$~\cite{Akiba:1989xu}, $\mathcal{N}_{\text{tr}} \left ( \mathcal{N} \mathcal{V} \right )_{\text{rot}}$ ($\sim 26 \times 5.3\times 10^{3}$) are the normalization factors related to the zero modes of the fluctuation operator around the sphaleron solution~\cite{Arnold:1987mh, Akiba:1989xu, Carson:1989rf} and $\kappa$ is an $\mathcal{O}\left( 1 \right)$ numerical factor~\cite{Carson:1990jm, Baacke:1993aj}. This approximation is valid in the region $m_{W,i} (T) \ll T \lesssim T_{c,i}$.
A simple comparison of the sphaleron rate with the Hubble parameter ($\Gamma_{\text{Sp}} \sim H$) shows that the sphaleron freezes out around $T\sim 131$ GeV and, below this, leptons are no longer converted into baryons. A rapid decrease in the sphaleron rate occurs around the Electroweak (EW) phase transition as it becomes exponentially suppressed when the Higgs acquires a VEV. 

We define the conserved global charge as
\begin{equation}
X \equiv B - L  \; ,
\end{equation}
where $L$ is the total lepton and $B$ is the baryon number ($ B \equiv \frac{ n_{B} - n_{ \bar{B} } }{ s} \; , \  L \equiv \frac{ n_{l} - n_{ \bar{l} } }{ s}$). For the sphaleron 
in chemical equilibrium at a temperature $T$, these numbers are~\cite{Khlebnikov:1996vj}
\begin{equation}
\begin{split}
B_{\text{eq}} & \equiv  \chi \left ( \frac{ v }{ T } \right ) X \; , \quad L_{\text{eq}} \equiv B_{\text{eq}} - X  \; ,  \\
\chi (x) & = \frac{ 4 \left [  5 + 12 n_{G} + 4 n_{G}^{2} + \left( 9+ 6n_{G}\right ) x^{2}   \right ]  }{ 65 + 136 n_{G}  + 44 n_{G}^{2} + \left( 117 + 72 n_{G} \right ) x^{2}   } \; , \label{eq: Beq value}
\end{split}
\end{equation}
where $n_{G}=3$ is the number of generations. In the case of a deviation from the chemical equilibrium, with a given source for leptons $F(t)$, the total lepton and baryon numbers evolve as~\cite{Khlebnikov:1988sr, Khlebnikov:1996vj, Rubakov:1996vz, Burnier:2005hp}
\begin{equation}
		\begin{split}
		\dot B(t) & = - \Upsilon(t) \left [ B(t) + \eta(t)  L (t)  \right ] \;  , \\
		\dot L (t) & = - \Upsilon(t) \left [ B(t) + \eta(t)  L (t)  \right ]  + F (t) \; , 
		\end{split} 
		 \label{eq:baryonEvolution} 
\end{equation}
where
\begin{equation}
\begin{split}
\Upsilon & \equiv n_{G}^{2} \; \rho \left ( \frac{ v }{T}  \right )  \left [ 1 - \chi \left ( \frac{ v }{T}  \right ) \right ]  \frac{ \Gamma_{\text{diff}} \left ( T \right ) }{T^{3}} \; , \; \eta \equiv \frac{  \chi \left ( \frac{ v   }{T}  \right ) }{ 1 -  \chi \left ( \frac{ v }{T}  \right ) } \; , \; \\
\rho \left ( x \right ) & = \frac{3\; \left [ 65 + 136\; n_G + 44\; n_G^{2} + \left (117 + 72\;n_G \right )\;x^2 \right ] }{2\; n_G \left [ 30 + 62\;n_G + 20\; n_G^{2}+ \left (54 + 33\;n_G \right )\;x^2  \right ] } \; . \;
\label{eq: B and L param}
\end{split}
\end{equation}
\begin{figure*}[t!]
\centering
\begin{minipage}[c]{\textwidth}  \subcaptionbox{ $i=0$ and $i=-1$ sector}{
\includegraphics[width=.45\textwidth ]{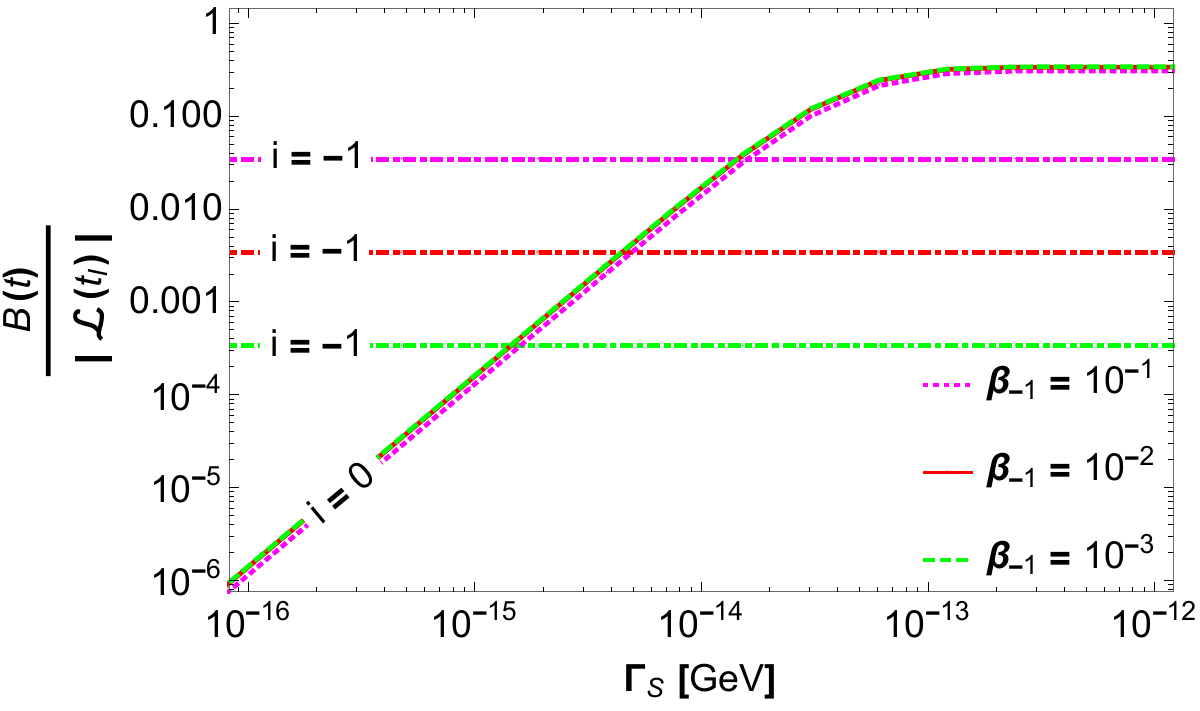} }
\hfill
\subcaptionbox{ $i=+1$ sector}{
\includegraphics[width=.45\textwidth]{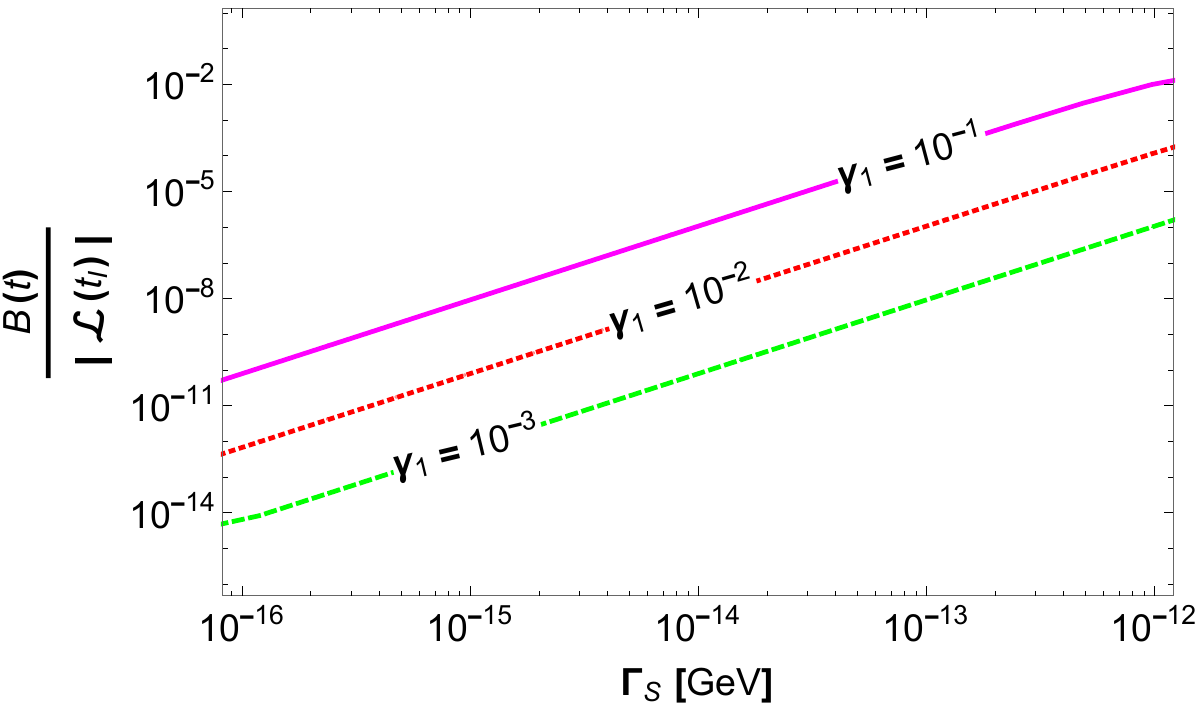}  }
\end{minipage}
\hfill
\caption{ Baryon asymmetry (normalized by the magnitude of the initial lepton asymmetry) as a function of the reheaton's decay width in the SM and the lowest exotic sector (left panel), the lowest standard sector (right panel), with $t_{I} = 10^{-20}$ and $B(t_{I} )= L(t_{I} ) = 0$ evaluated at very late times and the Higgs mass squared parameter in~\eqnref{Higgs mass2 param}: $-(244.8)^{2}\text{ GeV}^{2}$, $-(88.4)^{2}\text{ GeV}^{2}$, $(210.4)^{2}\text{ GeV}^{2}$ corresponding to $i=+1$, $0$, and $-1$ sectors respectively. We consider three different branching ratios for each panel: $\beta_{-1}=10^{-3}$ in dashed green, $\beta_{-1}=10^{-2}$ in solid red, and $\beta_{-1}=10^{-1}$ in dashed magenta in the left panel, and $\gamma_{1}=10^{-3}$ in dashed green, $\gamma_{1}=10^{-2}$ in solid red, and $\gamma_{1}=10^{-1}$ in dashed magenta in the right panel. Note that the vertical axes on the right panel is very small compared to the left panel. On the left panel, the horizontal dot-dashed lines correspond to the $i=-1$ sector.}
  \label{fig:Bnorm_widths_timei0i1}
\end{figure*}

To obtain the baryon asymmetry in the various sectors, we need to know the lepton source function $F(t)$. In our model, the total lepton asymmetry is carried by the reheaton (S) at early times and can be tracked using Boltzmann equations, assuming the decay of the reheaton into two particles that rapidly reach thermal equilibrium with each other. The decay products of the reheaton are SM particles and we have explored the branching ratios in~\sectref{sec: Reheating Dynamics}. At an early epoch, reheatons will be in thermal equilibrium with each other, so the evolution of their number density can be written as  
\begin{equation}
		\begin{split}
		\dot n_{S}  + 3 H n_{S} &  = -  \expval{\Gamma_{S} } \left ( n_{S} - n_{S}^{\text{eq}} \right ) \; ,  \\
		\dv{ Y_{S} }{t} & =  - \expval{\Gamma_{S} }  \left (  Y_{S} - Y^{\text{eq}}_{S}   \right ) \; , \label{eq:yield_evolution} \\
		\end{split}
\end{equation}
where $Y_S \equiv \frac{n_S}{s}$ is the comoving number density, $\expval{\Gamma_{S} } = \frac{ K_{1} \left( \frac{m_{S}}{T} \right ) }{ K_{2} \left( \frac{m_{S}}{T} \right ) }  \Gamma_{S} $, $K_{1} \left( \frac{m_{S}}{T} \right )$, and $K_{2} \left( \frac{m_{S}}{T} \right )$ are the modified Bessel functions of order 1 and 2 respectively. Defining $\mathcal{L} \equiv Y_{S} - \bar{Y}_{S} $, we can subtract the equations for $S$ and $\bar{S}$
\begin{equation}
\begin{split}
 \dv{\mathcal{L}}{t}  =  - \expval{\Gamma_{S} }  \mathcal{L} \; ,  \label{eq:yieldAssymetry}
\end{split}
\end{equation}
with the boundary condition $\mathcal{L} \left ( t_I \right ) \equiv \mathcal{L}_I$, and $\mathcal{L}_I$ satisfying the bound in~\eqnref{eq:InitialLeptonAsymmetryLimit}.
Thus, the lepton number source function can be deduced from~\eqnSref{eq:baryonEvolution} and~\eqref{eq:yieldAssymetry}, yielding
\begin{equation}
\begin{split}
		F(t)_{0} & =  \expval{\Gamma_{S} } (1-\beta-\gamma) \; \mathcal{L}  \; , \quad  F(t)_{j} =  \expval{\Gamma_{S} } \beta_j \; \mathcal{L} \; , \\
	        F(t)_{k}  &=  \expval{\Gamma_{S} } \gamma_k \; \mathcal{L} \;
		\; .
\end{split}
 \label{eq:sourcefunction}
\end{equation}
As before $\beta$ ($\gamma$) is the sum of the branching ratios to all exotic (standard) sectors. Using the source function in~\eqnref{eq:sourcefunction}, one can compute the baryon asymmetry as a function of time for a given reheaton width. We are interested in the late time behavior of the baryon asymmetry. In \fref{fig:Bnorm_widths_timei0i1}, we represent the BAU as a function of the reheaton width, for the SM ($i=0$) and $i=1$ sectors, displayed in the left and right panels respectively, for different branching ratios ($\beta_{-1}=10^{-3}$ in dashed green, $\beta_{-1}=10^{-2}$ in solid red, and $\beta_{-1}=10^{-1}$ in dashed magenta). The horizontal dot-dashed lines correspond to the lowest exotic sector ($i=-1$) with magenta, red, and green denoting $\beta_{-1}=10^{-1}$, $\beta_{-1}=10^{-2}$, and $\beta_{-1}=10^{-3}$, respectively. The SM curve is proportional to $1-\beta-\gamma \approx 1$, while the $i=1$ curves are proportional to $\gamma_1$.
We can clearly observe that the conversion of lepton into baryon asymmetry is less efficient for sectors $i \geq 1$ rather than for $i=0$. This is primarily due to the fact that $m_{W,i} > m_{W,0} $, $v_{i} > v_{0}$, and $T_{\text{RH},i} < T_{\text{RH},0} $ for $i \geq 1$. The modification in these parameters changes the sphaleron rate which, in turn, alters the baryon asymmetry that is produced.

We can give an approximate analytical expression for the final baryon asymmetry as follows. If we first ignore the sphaleron process, then the lepton asymmetry as a function of time is given by solving~\eqnref{eq:baryonEvolution}:
\begin{equation}
    L(t) \simeq \int^t dt F(t) \simeq \Gamma_S\,  t \, \mathcal{L}_I \, ,
    \label{eq:lt}
\end{equation}
using~\eqnref{eq:sourcefunction} but ignoring thermal effects. This of course, only applies for $t \ll \Gamma_S^{-1}$, while for $t \gg \Gamma_S^{-1}$, we have $L(t) \simeq \mathcal{L}_I$.
There will also be a branching ratio factor which will differ for each sector. 

We can now model the sphaleron process, and in particular the function $\Upsilon(t)$ as a step function which shuts off when the temperature of a given sector is below some sphaleron temperature $T_\text{Sp}$. In the limit where the $B \ll L$, then the baryon asymmetry can be approximated as only sourced by the lepton asymmetry. From~\eqnref{eq:baryonEvolution}, the baryon asymmetry can then be approximated as 
\begin{equation}
\begin{split}
    B & \simeq -\int^t dt \, \Upsilon(t) \,\eta(t) \, L(t) \simeq -\frac{\eta}{1+\eta}\int^{t_\text{Sp}} dt F(t) \\
   B & \simeq -\Gamma_S\, t_\text{Sp} \mathcal{L}_I \; , \;
\end{split} 
    \label{eq:Bapprox}
\end{equation}
where $t_\text{Sp}$ is the time at which the sphaleron becomes inactive, and we have assumed $\Gamma_S \,t_\text{Sp} \ll 1$. The sphaleron becomes inactive when the temperature of the bath is of order the electroweak phase transition temperature. Using the fact that $\rho \propto T^4$ and that in a radiation dominated Universe $t \propto a^2$, we can plug into~\eqnref{eq: R analytic} to get
\begin{equation}
    t_\text{Sp} \propto \frac{\Gamma_S}{T_\text{Sp}^4} \; .
\end{equation}
The scaling is the same if the sphaleron freezes out when the reheaton is nonrelativistic. Therefore, if the sphaleron freezes out well before the reheaton finished decaying, then the baryon asymmetry is proportional to $\Gamma_S^2$. On the other hand, if the sphaleron is in equilibrium after most $S$ decays, then the last approximation in \eqnref{eq:Bapprox} will just give $\mathcal{L}$, and the baryon number will be independent of $\Gamma_S$. In our convention, it is necessary for the Universe's lepton asymmetry to be negative, i.e., $\mathcal{L}_I < 0$, in order to produce a positive baryon number.

Putting the branching ratio and entropy dilution factors back in, and doing a numerical fit to get the overall coefficient, we get
 \begin{equation}
  \resizebox{0.5\textwidth}{!}{$
\begin{split}
\frac{B_{j}}{ \mathcal{L}_{I}} \simeq \begin{dcases} 
\left(\frac{\Gamma_S}{5.4\times 10^{-14} \text{ GeV}} \right)^2 \frac{\eta}{1+\eta} \;
\gamma_1 \; \delta_S^{-1} \; ,  & \text{ For  } j =1  \\  
\left(\frac{\Gamma_S}{4\times 10^{-14} \text{ GeV}} \right)^2 \frac{\eta}{1+\eta} \;
\left (1- \beta -\gamma \right ) \; \delta_S^{-1} \; ,  & \text{ For  } j =0  \\ 
\frac{\eta}{1+\eta}  \; \beta_{-1} \; \delta_S^{-1} \; , & \text{ For } j = -1  \end{dcases} \label{eq:Byield} \; ,
\end{split} $ }
\end{equation}
where $\mathcal{L}_{I}$ is the initial lepton asymmetry with $\abs{\mathcal{L}_{I}} \lesssim 3 \times 10^{-6}$, as obtained in \eqnref{eq:InitialLeptonAsymmetryLimit} and $\delta_S^{-1}$ is the entropy dilution factor in \eqnref{eq:delta1} from the reheaton decay. 
The upper bound on the initial lepton asymmetry (see \eqnref{eq:InitialLeptonAsymmetryLimit})places a lower bound on the reheaton width while using the current $\Delta N_{\text{eff}}\simeq 0.4$ measurements~\cite{Planck:2018vyg} imposes an upper bound on the width. Thus, the reheaton's width lies in the following window,
\begin{equation}
\begin{split}
6.2 \times 10^{-16} \text{ GeV} \lesssim  \Gamma_{S} & \lesssim 1.4\times10^{-14} \,\mathrm{GeV.}
\end{split}
\label{eq: reheaton range}
\end{equation}
For the exotic sector ($j=-1$), the sphaleron is in equilibrium much later, as described in detail below, and thus  $\Gamma_S \,t_\text{Sp} \gtrsim 1$ so the baryon asymmetry is independent of $\Gamma_S$.

\subsection{Exotic sectors}
\label{sec:exotic asymmetry}

The baryon asymmetry generated in the exotic sectors, $i<0$, are dramatically different from the $i>0$ sectors due to a combination of factors: (1) 
 electroweak symmetry will be broken by the QCD confinement at a much lower temperature around $T\approx 89$ MeV~\cite{PhysRevD.20.2619}; (2) the reheating temperatures in those sectors are large compared to the masses of the particles that live there; (3) the baryon asymmetry is generated very efficiently by the sphalerons due to the low electroweak scale, permitting the sphalerons to be active much longer. 
In these sectors, the phase transition is also qualitatively different because there are six massless quarks, so it is expected to be first order~\cite{PhysRevD.29.338,PhysRevD.45.466}. Therefore we model the phase transition as a step function:
\begin{equation}
v \rightarrow \Lambda_{\text{QCD}} = \begin{dcases} 0 \; , &  T > T_c  \\ 89 \text{ MeV} \; , &  T \lesssim T_c  \end{dcases} \; \label{exotic sector condition}.
\end{equation}
The calculated baryon asymmetry in the exotic sectors does not depend on how exactly the phase transition is modelled. 

One may worry that there will be baryon or lepton number violation during this first order phase transition. This turns out not to be a problem because there is not enough $CP$ violation for such processes. The $CP$ violation can be written in terms of the Jarlskog invariant~\cite{PhysRevLett.55.1039, Jarlskog:1985cw, PhysRevD.50.774} and goes like $\sim \prod \Delta m_q^{2}/T_c^{12}$. Since the quarks are much lighter than in the SM sector [see~\eqnref{eq:exotic_fermionMasses}] while the phase transition temperature changes little, this dimensionless quantity is $\sim 10^{-103}$, so these processes during the phase transition are negligible.

Because the phase transition in the exotic sector is much lower than the reheating temperature from reheaton decays, the sphaleron reaches equilibrium and the baryon asymmetry can be parametrically larger than in the SM sector, even though the reheating temperature is lower. 
The behavior of the baryon asymmetry in the exotic sector is illustrated in the left panel in~\fref{fig:Bnorm_widths_timei0i1} by the horizontal dot-dashed lines, for different branching ratios ($\beta_{-1}=10^{-3}$ in green, $\beta_{-1}=10^{-2}$ in red and $\beta_{-1}=10^{-1}$ in magenta).  We can see that in the $i=-1$ sector the baryon asymmetry is much larger compared to the $i=1$ case, as shown in the right panel of~\fref{fig:Bnorm_widths_timei0i1}, in particular, for smaller reheaton's widths. This enhancement of the baryon asymmetry in the exotic sector is due to the sphaleron being active well below the SM electroweak scale for a longer duration. Thus in this model, the dark matter can be the lightest baryon of the exotic sectors, dominated by $i=-1$. The baryon asymmetry in the $i=-1$ sector is independent of the reheaton decay width because the reheating temperature is well above the sphaleron freeze-out temperature so the sphaleron reaches equilibrium for all the allowed values of $\Gamma_S$.

\begin{figure}[t]
\begin{centering}
\includegraphics[scale=0.4]{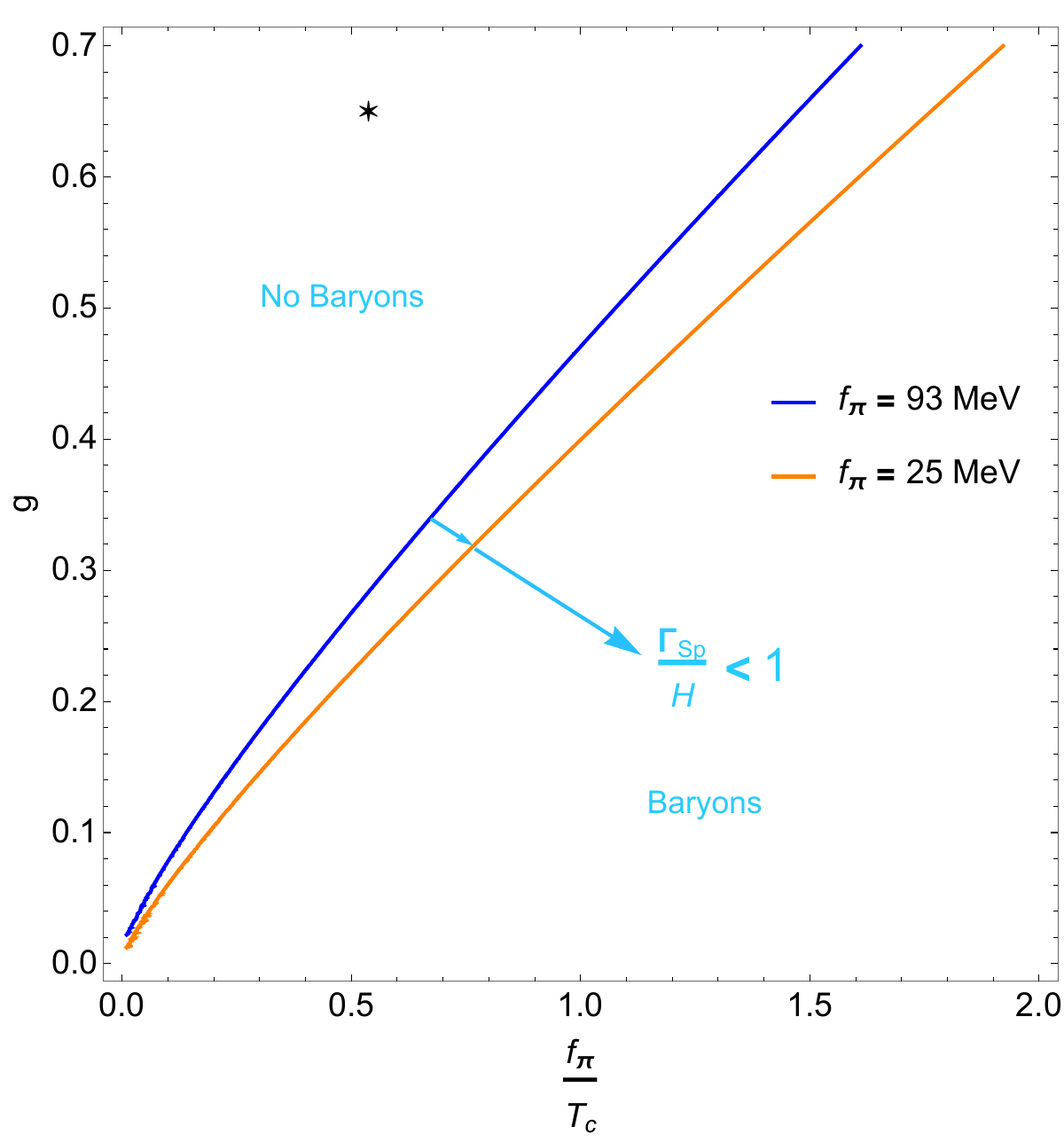}
\par\end{centering}
\caption{The $SU(2)_L$ gauge coupling constant, $g$, as a function of the ratio of pion decay constant over temperature, $\frac{f_\pi}{T_c}$, with contours corresponding to $\Gamma_{\text{Sp}}(T_c) = H(T_c)$ for some fixed pion decay constants ($f_\pi=93$ MeV in solid blue, $f_\pi=25$ MeV in solid orange). The black star represents the SM value (with $g\simeq 0.65$ and $\frac{f_{\pi,0}}{T_{c,0}}\simeq 0.54$). In the region below the curves, baryon washout is frozen out and the hidden sector is a viable dark matter model.  }
\label{fig:ratioSphaleronHubble}
\end{figure}

It was argued in~\cite{arkani2005predictive} that in exotic sectors there will be no late time baryon asymmetry. This is because, after the QCD/EW phase transition, the lightest baryon is much heavier than the temperature, so the sphaleron reaction is strongly biased toward baryons' destruction. The rate of $B$ violation, in~\eqnref{eq: Beq value}, can be approximated as:
\begin{equation}
 \frac{1}{B}\dv{B}{t} \simeq - \Gamma_{\text{Sp}}(T) \; ,
\end{equation}
where $\Gamma_{\text{Sp}}(T) \sim \Gamma_{\text{diff}}/T^{3}$ from \eqnref{eq:GamDiff},
with $m_{W,i}^{2} \left ( T_i \right ) =\left ( \frac{g}{2} \right )^{2} f_{\pi}^{2}$~\cite{PhysRevD.20.2619} with $f_\pi$ being the pion decay constant, and the VEV replaced as in \eqnref{exotic sector condition}. 
Right after the phase transition, the sphaleron is exponentially suppressed, $\Gamma_{\text{Sp}} \sim e^{-f_\pi/(g T_c)}$, but the question remains if it is smaller than the Hubble rate which is Planck suppressed. One could estimate $T_c$ by rescaling the SM value by the ratio $\Lambda_{\text{QCD},i}/  \Lambda_{\text{QCD},0} $ to compute the sphaleron, and by that method, one confirms the conclusions of~\cite{arkani2005predictive} that the sphaleron is fast enough to wash out any baryon asymmetry. We note, however, that the phase transition of the exotic sector is qualitatively different from that of the SM, and there is no reason to expect the nonperturbative parameters of $f_\pi$ and $T_c$ to be the same as the SM. Furthermore, the sphaleron is exponentially sensitive to that ratio, so small changes can alter the qualitative picture. If  $f_{\pi}/T_c \gtrsim 1.5$, then the sphaleron rate after the phase transition is always smaller than Hubble and there is no washout. This ratio can ultimately only be calculated with nonperturbative methods such as the lattice. The critical temperature of the QCD phase transition of the SM (considering two flavors only),  determined using QCD lattice simulations, is $T_{c,0} \simeq \left ( 172 \pm 3 \right ) \text{ MeV}$~\cite{Karsch:2000kv, Fodor:2001pe}. However, its precise value for a larger number of flavors is fairly ambiguous. In the literature, the value varies significantly so there is a large uncertainty associated with the six flavors $T_{c}$~\cite{Braun:2006jd, Miura:2012zqa, Lombardo:2014fea, Dini:2021hug, Kotov:2021hri}.

We also note that if $SU(2)_L$ gauge coupling $g$ in the $i=-1$ sector is smaller than the SM value of $g$, then the sphaleron rate is also suppressed. While the simplest N-naturalness setup has $g$ equal in all sectors, this is not necessary for the solution to the hierarchy problem. Therefore, the exotic sectors can more easily suppress this freeze-out process and accommodate dark matter if $g$ is reduced. These results are summarized in~\fref{fig:ratioSphaleronHubble} where we show contours for $\Gamma_{\text{Sp}}(T_c) = H(T_c)$ in the $f_{\pi}/T_c$ vs.~$g$ plane. The region below the curve will not have any baryon washout.

 %
\begin{figure*}[t!]
\centering
\begin{minipage}[c]{\textwidth}  \subcaptionbox{ $i=-1$ sector }{
\includegraphics[width=.45\textwidth ]{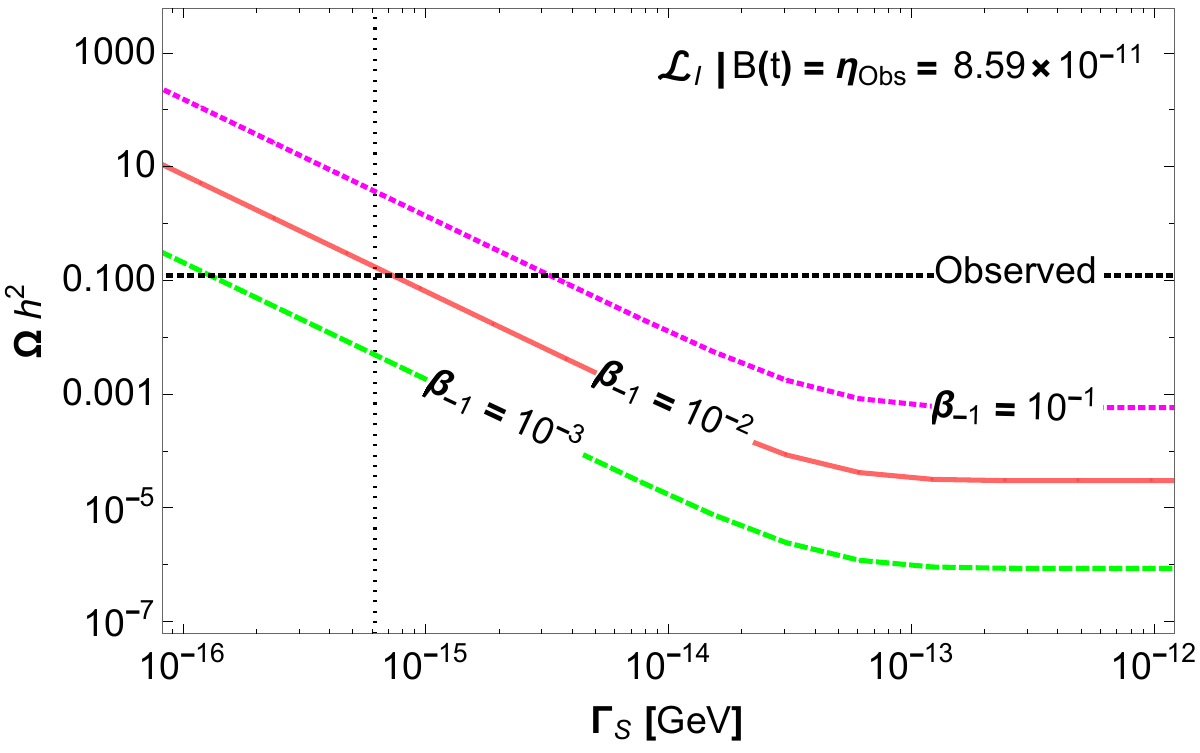} }
\hfill
\subcaptionbox{ $i=+1$ sector}{
\includegraphics[width=.45\textwidth]{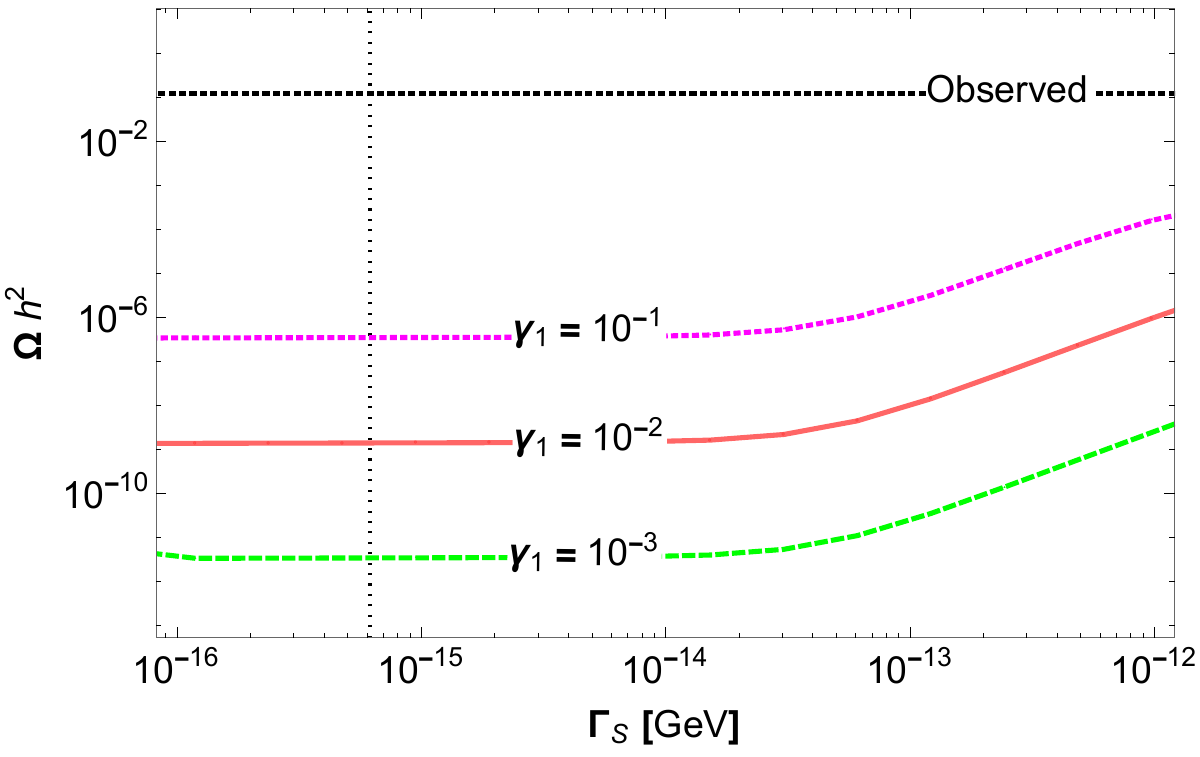}  }
\end{minipage}
\hfill
\caption{The relic density of the DM as a function of the reheaton's width for various $\beta$ in the $i=-1$ (left panel) and $i=1$ sector (right panel). The black dotted horizontal line represents the experimentally observed dark matter relic abundance and the black dotted vertical line indicates the decay width value taking into account the upper bound on initial lepton asymmetry in~\eqnref{eq:InitialLeptonAsymmetryLimit}.
  }
  \label{fig:DM_sectori-1i1}
\end{figure*}

\subsection{Dark baryon relic abundance} 
\label{sec:DM_abundance}
The exotic sectors can have large baryon asymmetries, so the lightest baryon in the exotic sectors is a stable dark matter candidate that can provide the observed relic abundance. 
From observations~\cite{Planck:2018vyg}, the current DM relic abundance is:
\begin{align}
\Omega h^2 = 0.120 \pm 0.001 \; , \quad \text{ and  } h=0.673 \; .
\end{align}
We focus on the dark neutron\footnote{The neutron is expected to be lighter than the proton in these sectors, making the proton unstable. In the SM, the proton is lighter than the neutron mainly due to the down quark outweighing the up quark. However, in the exotic sectors, the $u$ and $d$ (and the electron) are nearly massless, and the dominant contribution to the proton-neutron mass splitting is the electromagnetic self-energy that raises the mass of the proton. The decay width of the dark proton $\Gamma \propto G_{F,i}^2  \abs{m_{n,i} - m_{p,i} }^5$ similar to the SM beta decay. The lifetime of the dark proton in exotic sectors will be much shorter than that of the SM neutron because $G_{F,i} \gg G_{F,\text{SM}}$ and because there is no phase space suppression due the electron mass.\label{foot:proton}} of the $i=-1$ to serve as the dominant DM candidate, and we assume that the warm and cold neutrinos described in~\sectref{subsec:standardeutrinos} are subdominant.
The masses of the nucleons in these sectors, $m_{N,i}$, are linearly proportional to the confinement scale (neglecting the constituent mass effects)~\cite{Gasser:1982ap} and can be written as:
\begin{equation}
m_{N,i} \simeq m_{N,0} \frac{ \Lambda_{\text{QCD},i} }{ \Lambda_{\text{QCD},0} },
\label{DN mass}
\end{equation}
where the subscript $0$ denotes the usual SM sector. Using $B_{i} = \frac{n_{b_i} - n_{\bar{b_i}}}{s_i}$ 
as the dark neutron yield, the dark matter abundance is given by:
\begin{align}
\Omega h^{2} & \equiv\frac{\rho_{i}}{\rho_{c,0}}\nonumber \\
 & \simeq\frac{m_{N,i}\;B_{i}\;s_{0}}{\rho_{c,0}/h^{2}}\;\mathcal{R}\label{eq: DM1} \; ,
\end{align}
where $\rho_{c}/h^{2}=8.09\times10^{-47} \text{ GeV}^{4}$ is the critical density of the Universe, $s_{0} $ is the entropy density of the SM sector today and the parameter $\mathcal{R}$ is the ratio of the entropy densities of $i=-1$ over $i=0$ sector at late times ($t_{f}$):
\begin{equation}
\mathcal{R} \equiv \frac{ s_{i=-1}}{ s_{i=0}}= \frac{ g_{\star,i=-1} (t_{f} ) }{ g_{\star,i=0} (t_{f} ) } \; \left [  \frac{  T_{i=-1} (t_{f}) }{T_{i=0}  (t_{f}) } \right ]^{3}. \label{eq: Rdefination} 
\end{equation}
Using~\eqnSref{eq:Byield},~\eqref{DN mass},~\eqref{eq: Rdefination}, and the relation $\frac{T_{-1}}{T_{0}} \simeq \left ( \frac{\beta_{-1}}{1-\beta-\gamma} \right )^{1/4}$, the DM relic abundance can be written as
\begin{equation}
\begin{split}
    \Omega h^{2}   \simeq  & 0.12  \; \left ( \frac{1}{1-\beta-\gamma} \right )^{7/4} \; \left ( \frac{\beta_{-1}}{10^{-2}} \right )^{7/4} \; \\
    & \left ( \frac{B_{0}}{8.59 \times 10^{-11} } \right ) \;   \left ( \frac{ 2.8 \times 10^{-16} \text{ GeV} }{\Gamma_S} \right )^{2} \; .  \label{eq:relicAbundance}   
\end{split}
\end{equation}
The $\Gamma_S$ dependence comes from~\eqnref{eq:Byield} using the fact that $\Omega h^2 \propto B_{-1}\propto {\mathcal L}_I \propto \Gamma_S^{-2}$ where the last proportionality is for fixed $B_0$.

In \fref{fig:DM_sectori-1i1}, we show the parameter space for DM in two sectors: $i=-1$ (left panel) and $i=+1$ (right panel). In both panels, the horizontal black dashed line is the observed DM abundance, whereas
the vertical black dotted line corresponds to the upper bound on the initial lepton asymmetry in~\eqnref{eq:InitialLeptonAsymmetryLimit}. The upper bound on the initial lepton asymmetry gives a lower bound on $\Gamma_S$ using the relation in~\eqnref{eq:Byield} and requiring the observed baryon asymmetry of the Universe (and the correct relic abundance for DM). In the left panel, we represent the DM relic abundance provided by our model, for three different branching ratios:  $\beta_{-1}=10^{-3}$ in dashed green, $\beta_{-1}=10^{-2}$ in solid red and $\beta_{-1}=10^{-1}$ in dashed magenta. We can see that, in the $i=-1$ sector, there is a region of the parameter space that yields the observed DM abundance. For small $\Gamma_S$, the relic abundance is well approximated by~\eqnref{eq:relicAbundance}. At large $\Gamma_S$, the reheaton fully decays before the sphaleron goes out of equilibrium, so further raising $\Gamma_S$ no longer has any effect. The change in slope occurs where the $S$ reheating temperature is comparable to the sphaleron decoupling temperature, $T_{\rm Sp}$ which occurs for $\Gamma_S \sim 5 T_{\rm Sp}^2/M_P\sim 10^{-13}$ GeV.  

In the right panel, we plot the DM abundance in the $i=+1$ sector for three different branching ratios as well: $\gamma_{1}=10^{-3}$ in dashed green, $\gamma_{1}=10^{-2}$ in solid red, and $\gamma_{1}=10^{-1}$ in dashed magenta in the right panel. As expected, we can see that there are not enough baryons in the standard sectors ($i>0$) to accommodate a DM candidate.

\begin{figure}[h!]
\begin{centering}
\includegraphics[scale=0.55]{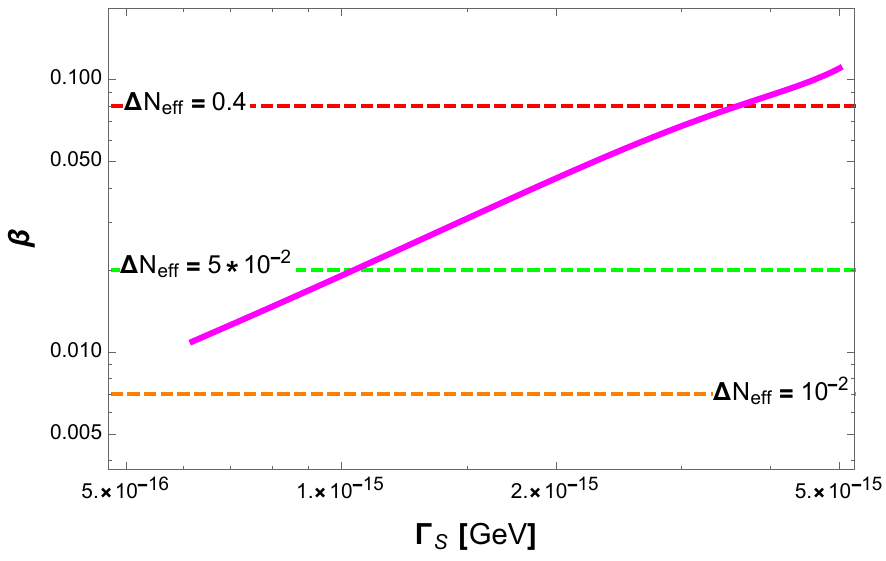}
\par\end{centering}
\caption{ Contours of $\Delta N_{\text{eff}} $ in  $\Gamma_S-\beta$ plane, with the solid magenta curve providing the observed DM abundance, the observed baryon asymmetry and satisfying the theoretical upper bound on the initial lepton asymmetry [$\abs{\mathcal{L}_{I}} \leq 3 \times 10^{-6} $, computed in \eqnref{eq:InitialLeptonAsymmetryLimit}]. The horizontal axis only represents the allowed range for $\Gamma_S$, as given in \eqnref{eq: reheaton range}. 
}
\label{fig:beta_Reheaton_countour}
\end{figure}

In \fref{fig:beta_Reheaton_countour}, we represent the viable parameter space of our model (solid magenta curve) that can account for the observed DM relic abundance (in the $i=-1$ sector) and the observed baryon asymmetry (in the $i=0$ sector), and it includes the theoretical upper bound on the initial lepton asymmetry $\abs{\mathcal{L}_{I}} \leq 3 \times 10^{-6} $, in the plane of the branching ratio into the exotic sector, $\beta$, as a function of the reheaton width, $\Gamma_S$. The horizontal axis only shows the allowed range for $\Gamma_S$, as computed in \eqnref{eq: reheaton range}. The dashed lines correspond to different contours of $\Delta N_{\text{eff}}$, with $\Delta N_{\text{eff}}=10^{-2}$ in orange, $\Delta N_{\text{eff}}=5\times 10^{-2}$ in green and $\Delta N_{\text{eff}}=0.4$ in red. 
The minimum values for our parameters in $\Gamma_S-\beta$ plane are approximately $\beta=0.011$ and $\Gamma_S=6.2 \times10^{-16}$ GeV, yielding $\Delta N_{\text{eff}}\simeq 0.048$.
CMB-S4 is expected to constrain $\Delta N_{\text{eff}} \le 0.06$ at the $95\%$ confidence level~\cite{ abazajian2019cmb}, allowing our model to be potentially probed by the future experiment.

\section{Dark Matter Phenomenology }
\label{sec: DM}
In this section, we study the possibility of large DM bound states. Dark matter is made up of dark neutrons in the $i=-1$ exotic sector. The dark protons in those sectors will be slightly heavier (see footnote~\ref{foot:proton}), but have a short lifetime which can be estimated as
\begin{align}
\tau_{p,j} \sim \tau_{n,\mathrm{SM}} 
\left( \frac{G_{F,\mathrm SM} }{G_{F,j}} \right)^2
\sim (10^3 \,s) \left ( \frac{m_{W,j}}{m_{W,0}} \right )^4 \sim 10^{-11} \, s \, ,
\end{align}
where we have used the estimate for the exotic sector $W$-boson mass for exotic sector from~\eqnref{eq:wmassesSpectrum}. We have also assumed that $m_p-m_n$ in the exotic sector is comparable to $m_n-m_p-m_e$ in the SM. Therefore, when the Universe is cool enough that nuclear bound states can form (dark BBN), the dark baryon population is all neutrons. 

SM neutrons have significant self-interactions, and these are potentially even larger in the exotic sector since the pions are lighter than those of the SM. The dark neutrons can also form dark nuclei~\cite{Samuel:1999am,Hardy:2014mqa}, which turns out to be the dominant state of dark matter in this model. This allows constraints from self-interacting dark matter to be evaded. Finally, we briefly comment on the possibility of dark quark nuggets~\cite{PhysRevD.30.272}, and show that these nuggets evaporate rapidly.

\subsection{Dark neutrons self-interaction}
Before considering dark matter composite states, we first consider observational constraints on free dark neutrons if they are all of the dark matter.
The limit on dark matter self-interaction cross section coming from the Bullet Cluster is~\cite{Randall_2008}
\begin{equation}
\ensuremath{\frac{\sigma_{NN}}{m_{N}}<1.25\,\text{cm}^{2}\,\text{g}^{-1}}.\label{eq: Bullet Cluster constraint}
\end{equation}
For a $1$ GeV DM neutron, this bound limits the cross section to be
\begin{equation}
\ensuremath{\sigma_{NN}<2.2\times10^{-24}\,\text{cm}^{2}}. \label{eq: Bullet Cluster final}
\end{equation}
In the SM, the total cross section of neutrons scattered by protons at low energy is $\sigma_{pn}\sim 2.04 \times 10^{-23} \, \text{cm}^{2}$~\cite{PhysRev.76.1744, PhysRevC.3.1886,1955RSPSA.230...19S} and, by isospin symmetry, (neglecting the $p$ and $n$ mass difference), the following approximation is valid: $\sigma_{nn}\simeq \sigma_{pn}$. Applying  naive dimensional analysis, the scattering cross section in the exotic sector is
\begin{equation}
\sigma \sim \sigma_{nn}\; \left (  \frac{  \Lambda_{\text{QCD},0} }{  \Lambda_{\text{QCD},i} } \right)^{2} \simeq 2.84 \times 10^{-22} \, \text{cm}^{2}. \label{eq: cross section bound}
\end{equation}
From \eqnref{eq: cross section bound}, we may conclude that the 1 GeV neutron DM is inconsistent with the stringent bound from the Bullet Cluster given in \eqnref{eq: Bullet Cluster final}. In addition, the mass of the neutron in the exotic sector is smaller than that of the SM one by a factor of $\Lambda_{\text{QCD},i} / \Lambda_{\text{QCD},0}\sim 0.27$, which makes the upper bound even stronger on the self-interaction. In order to satisfy \eqnref{eq: Bullet Cluster final}, the DM mass is required to be $m_{\text{DM}} \gtrsim 1.3 \times 10^{2}$ GeV.
An alternative way to evade the Bullet Cluster bound is raising the confinement scale of the exotic sectors. Assuming the mass and cross section can be scaled using naive dimensional analysis, the confinement scale must be
\begin{align}
\frac{\sigma_{nn}}{m_{n,0}}\;\left(\frac{\Lambda_{\text{QCD},0}}{\Lambda_{\text{QCD},i}}\right)^{3} & <1.25\,\text{cm}^{2}\,\text{g}^{-1}\nonumber \\
  \Rightarrow\Lambda_{\text{QCD},i} & \gtrsim 700\text{ MeV }.\label{eq: QCD exotic BC}
\end{align}
This can be achieved by assuming the $SU(3)$ gauge coupling of the $i=-1$ sector is a bit larger than the SM at high energy. As noted in~\sectref{sec: Baryon Asymm}, different gauge couplings in different sectors are consistent with the N-naturalness paradigm. We also note that the Standard Model nucleon scattering cross section on which these estimates are based is unnaturally large. It is in fact much larger than the geometric cross section, a feature that is not expected to be generic~\cite{Kaplan:1998tg,Kaplan:1998we,Bedaque:2002mn,Hall:2019rld}; the nucleons of the exotic sectors might very well have a smaller scattering cross section. However in either case, as we will show next, we expect the dark matter to form large bound states which can easily evade the bounds.

\subsection{Nuclear dark matter}
\label{sec: Nuclear DM}

In this section, we argue that composite DM, made of stable dark nuclei (DN) of large dark nucleon number, can evade the constraints from the Bullet Cluster on the DM self-interactions and provide a viable DM candidate~\cite{Krnjaic:2014xza, Hardy:2014mqa}. As the detailed mechanism is explained in~\cite{Hardy:2014mqa}, in this work, we will just point out some of the main steps to get the observed DM abundance and the maximum size of these DN. In the SM, the main bottleneck that prohibits the synthesis of nuclei is the substantial binding energy per nucleon of helium-4, $\sim 7$ MeV, relative to the following smaller nuclei.\footnote{During SM BBN, almost all the nucleons present wind up in hydrogen and helium-4 while a small fraction leads to the synthesis $4<A<8$. There are subsequent bottlenecks post helium-4, such as $^{12}$C, where the binding energy per nucleon exceeds helium-4. However, there are no bottlenecks in the formation of large dark nuclei in the exotic sectors.  } 
As the Coulomb repulsion term is absent in this scenario, unlike in the SM, there can be stable DN up to a large dark nucleon number. DM will thus be produced at low temperatures given that, in this case, the energy term dominates over the binding energy, which favors the generation of bound states, and it will be built up by aggregation, as fusion processes dominate over dissociations and fissions in the low temperature regime \cite{Hardy:2014mqa}.

Our composite dark nucleons can be made to satisfy all the assumptions stated in~\cite{Hardy:2014mqa} that are necessary for the formation of the large dark nucleon (A-DN). By assuming the cross section to be geometric\footnote{As mention previously, in the SM the cross section of neutrons scattering is larger than the geometric cross section as a consequence of an accidental cancellations in the effective field theory~\cite{Kaplan:1998tg,Kaplan:1998we,Bedaque:2002mn,Hall:2019rld}.}, the self-scattering cross section of the A-DN over its mass can be expressed as
\begin{equation}
\begin{split}
\frac{\sigma_{AA}}{m_{A}}  & =\left(\frac{4\pi}{m_{1}\;A}\right)^{1/3}\left(\frac{3}{\rho_{-1}}\right)^{2/3}    \\
 & \simeq 4.4 \text{ cm}^{2}\text{ g}^{-1}\left(\frac{252\text{ MeV }}{A\;m_{1}}\right)^{1/3}\left(\frac{1 \; \text{MeV}\;\text{fm}^{-3}}{\rho_{-1}}\right)^{2/3} ,
 \end{split}
 \label{sigma A over mA}
\end{equation}
where $A$ is a dark nucleon number (analogous to the mass number in usual atoms), $m_1$ is the mass of a single nucleon, and $\rho_{-1}$ is the nucleon internal mass-energy density in the closest exotic sector to the SM which we find by rescaling from a typical Standard Model value: $\rho_{-1} = \rho_0 \; \left (  \frac{ \Lambda_{\text{QCD,-1}} }{\Lambda_{\text{QCD,0}}} \right )^4$, with $\rho_0 \sim 0.2\;\text{GeV}\;\text{fm}^{-3}$~\cite{Andronic:2014zha}. 
This will evade the Bullet Cluster bounds if the typical $A$ is larger than $\sim 43$. 
Also, due to the slight differences in our parameters compared to the benchmark points employed in~\cite{Hardy:2014mqa}, our DN could be orders of magnitude larger. 

To finish this section, we can determine the maximum size of the DN due to the aggregation process. We should consider two regimes: the case where the last fusions to freeze-out are those between large DN ( ``large" + ``large" fusions) and the scenario where the last fusions to freeze-out are between small and large DN (``small" + ``large" fusions)~\cite{Hardy:2014mqa}.    
In the case of ``large" + ``large" fusions, we assume that there is a peaked DNs mass distribution where almost all of the mass lies in the large A-DN. We can estimate when the freeze-out occurs by comparing the rate of 
the thermally average cross sections of fusions times the A-DN number density, with the Hubble parameter. The rate of the fusion over the Hubble parameter can be expressed as
\begin{equation}
\begin{split}
\frac{\Gamma}{H} & =\frac{\left\langle \sigma v\right\rangle n_{A}}{H}  \\
 & \simeq5.5\times10^{9}
 \left(\frac{\text{1 MeV}\;\text{fm}^{-3}}{\rho_{-1}}\right)^{2/3} 
  \left(\frac{m_{1}}{252\text{ MeV}}\right)^{-5/6}  \\
 & \left(\frac{g_{\star}}{10.75}\right)^{1/2} \left(\frac{T}{\text{1 MeV}}\right)^{3/2} \left(\frac{\beta_{-1}}{1-\beta - \gamma}\right)^{1/8}
 A^{-5/6} ,
 \end{split}
\label{eq:LargeLargeDN}
\end{equation}
 where we have used $m_{N,-1}=m_{N,0} \, \frac{\Lambda_{\text{QCD,-1}}}{\Lambda_{\text{QCD,0}}} \simeq 251.9$ MeV with $\Lambda_{\text{QCD,-1}} = 89$ MeV, $\Lambda_{\text{QCD,0}} = 332$ MeV and $m_{N,0}=939.6$ MeV, the approximations $\frac{T_{0}}{T_{i}}  \simeq \left ( \frac{1-\beta-\gamma}{\beta} \right )^{1/4} \xrightarrow{\beta =\gamma =0.01} 3.15$. 
Thus, the maximum mass obtained from~\eqnref{eq:LargeLargeDN} is
\begin{align}
M_{\text{max}} & =A_{\text{max}}\;m_{1}\nonumber  \\
 & \simeq\left(5.5\times10^{9}\right)^{6/5}\;\left(\frac{\beta_{-1}}{1-\beta- \gamma}\right)^{3/20}\;m_{1}\nonumber  \\
 & \simeq 6.2 \times 10^{10}\text{ GeV } \; . \;
 \label{max mass large large}
\end{align}
In the ``small" + ``large" fusions regime, the number density of small DN, $k$-DN, with $k<A$, may be larger than in the ``large"+``large" scenario, implying that the number of fusion processes can be larger as well. In addition, the velocities of ``k-DN" in thermal equilibrium are smaller than the ones of ``A-DN", which contributes to a size enhancement.  Assuming there is a sufficient population of small DNs, and it remains long enough, we can compute their rate of fusion to be
\begin{align}
\Gamma & \sim\left\langle \sigma v\right\rangle _{k+A}\;n_{k}\;\frac{k}{A}\nonumber \\
 & \sim\frac{1}{4}\;\delta\;\sigma_{1}\;v_{1}\;k^{-1/2}\;A^{-1/3}\;n_{0} \; ,\label{eq: Gamma small large}
\end{align}
where $v_1$ is the velocity of one nucleon, $n_{0}$ is the total DN number density, $\left\langle \sigma v\right\rangle _{k+A}=\frac{\delta}{4}\,\sigma_{1}\,v_{1}\,k^{-1/2}\,A^{2/3}$ \cite{Hardy:2014mqa} and $\frac{1}{4} \,\delta$ is a suppression factor for ``small"+``large" cross section relative to the geometric limit. This leads to a similar result as before but with $A^{-5/6} \rightarrow \frac{1}{4} \; \delta \; k^{-1/2} \; A^{-1/3}$. Thus, the fusion rate over the Hubble parameter for the ``small"+``large" regime is
\begin{equation}
\begin{split}
\frac{\Gamma}{H} & \simeq5.5\times10^{9}\left(\frac{\text{1 MeV}\;\text{fm}^{-3}}{\rho_{-1}}\right)^{2/3}\left(\frac{g_{\star}}{10.75}\right)^{1/2}\; \left(\frac{\beta_{-1}}{1-\beta - \gamma}\right)^{1/8}\\
 & \left(\frac{m_{1}}{252\text{ MeV}}\right)^{-5/6}\left(\frac{T}{\text{1 MeV}}\right)^{3/2} \left(\frac{1}{4}\;\delta\;k^{-1/2}\;A^{-1/3}\right) \; . 
\end{split}
\label{eq: gamma H small large}
\end{equation}
Similarly, the maximum DN synthesized would be, assuming that there is no suppression,
\begin{equation}
\begin{split}
M_{\text{max}} & =A_{\text{max}}\;m_{1}  \\
 & \simeq\left(\frac{5.5\times10^{9}}{4}\right)^{3}\;\left(\frac{m_{1}}{252\text{ MeV }}\right)^{-5/2}\; \\
 & \hspace{2 cm} \left(\frac{\beta_{-1}}{1-\beta-\gamma}\right)^{3/8}\;m_{1}  \\
 & \simeq 1.2\times10^{26}\text{ GeV ,}
 \end{split}
 \label{eq: Mmax small large}
\end{equation}
where in the last step we set $\beta=\gamma=0.01$.

\subsection{Comment on the formation of dark quark nuggets}
\label{DQN}

In addition, we study the possible formation of quark nuggets in the exotic sectors. The hypothesis of quark nuggets proposed by Witten~\cite{PhysRevD.30.272} is that after the QCD phase transition, there will be stable nuggets that remain in the quark phase and have very high baryon density. The formation of such objects requires a first order QCD phase transition, which does occur in exotic sectors~\cite{PhysRevD.29.338} with six nearly massless quark flavors. 
Studies of the six-flavor quark matter (6FQM) quark nuggets have been conducted in~\cite{Bai:2018vik, PhysRevD.99.055047}. One process not considered in those scenarios is the evaporation of a pion and a lepton from the quark nugget. Such a process conserves all gauge quantum numbers but it does change $B-L$ and decreases the lepton number of the nugget. In \appenref{Dark Nugget}, we show that such processes are quite fast and thus render the nugget lifetime much too short to play any role in dark matter phenomenology.

\section{Summary}
\label{conclusions}

Models with multiple SM-like hidden sectors may have interesting phenomenology. On the one hand, if the Higgs mass varies across sectors, these models can potentially solve the electroweak hierarchy problem~\cite{PhysRevD.80.055001, Arkani-Hamed:2016rle}. On the other hand, if the other sectors have similar temperatures as the Standard Model, then such a scenario is immediately excluded by cosmological observations such as $\Delta N_{\text{eff}}$, and novel cosmological histories are required.

Focusing on the reheaton scenario of~\cite{Arkani-Hamed:2016rle}, the cosmological history requires a reheating temperature of the order of the weak scale. In such a setup, the creation of the baryon asymmetry of the Universe and dark matter are open problems. In this work, we solve them. Although the strong self-interaction rules out dark neutrons of the exotic sectors as the dark matter candidate, the unique structure of the exotic sectors naturally allows the formation of large dark nuclei~\cite{Krnjaic:2014xza,Hardy:2014mqa}, which easily evade the self-interaction bounds from the Bullet Cluster. Dark matter will then form large neutral dark nuclei, which can have interesting observable consequences. 
We assume that the reheaton, which dominates the energy of the Universe at early times, is a fermion and that the population of reheatons develops an asymmetry from out-of-equilibrium decays, analogous to leptogenesis. The decay of the asymmetric population of reheatons reheats the SM and hidden sectors and imparts the asymmetry into the lepton number of those sectors. Part of the lepton number asymmetry is then processed into a  baryon number asymmetry via the electroweak sphaleron. 

The constraint coming from $\Delta N_{\text{eff}}$ requires the temperatures of all the hidden sectors to be significantly lower than the temperature of the SM sector. Naively, this would mean that the energy density of dark baryons should be less than that of visible baryons, in contradiction with observation. This is resolved by the fact that the dark matter lives in an exotic sector where the electroweak symmetry is not broken by a Higgs. 
In such a sector, the QCD confinement at a scale around 100 MeV leads to electroweak symmetry breaking, which in turn means that the electroweak sphaleron is active down to a much lower temperature. These sectors have light photons and neutrinos, and the requirement of achieving the correct baryon and dark matter density predicts $\Delta N_{\text{eff}} \gtrsim 0.05$, which may be detectable in the next generation of CMB experiments~\cite{abazajian2019cmb}. We must stress that the standard sectors have a dark baryon density substantially smaller than that of the visible sector,
which precludes the possibility of having a dark matter candidate in those sectors.

\section{Acknowledgments}

We would like to thank Yang Bai for useful discussions. C.C.~is supported by the Generalitat Valenciana Excellence grant PROMETEO-2019-083, by the Spanish Ministerio de Ciencia e Innovacion PID2020-113644GB-I00, the European Union’s Horizon 2020 research and innovation programme under the Marie Sk\l{}odowska-Curie Grant Agreement No. 860881-HIDDeN, and partially by the FCT project Grant No. CERN/FIS-PAR/0027/2021, and was supported by the Arthur B. McDonald Canadian Astroparticle Physics Research Institute. H.E., T.G., and D.S.~are supported in part by the Natural Science and Engineering Council of Canada (NSERC). 


\appendix

\section{MASS SPECTRUM OF \emph{N} SECTORS}
\label{App: mass spectrum}
In this section, we study the mass spectrum of the $N$ sectors. The ``standard sectors," with $m_{H,k}^{2} < 0$, corresponding to $k \geq 0$, exhibit electroweak symmetry breaking just like the SM case, with the exception that the VEV of the Higgs is given by
\begin{equation}
 v_{k}^{2}= - \frac{ \left ( m_{H}^{2} \right )_{k}  }{ \lambda } = v_{0}^{2} \;  \mathcal{C}_k \; , 
 \label{Higgs VEV}
 \end{equation}
 where $\mathcal{C}_k= \frac{2\,k + r }{r}$ and the parameter $r$ indicates the spacing between sectors, with $r=1$ corresponding to uniform spacing and $r<1$ corresponding to a large splitting between our sector and the next one~\cite{Arkani-Hamed:2016rle}. The sector with the smallest absolute value of $m_H^2$, $k=0$, corresponds to our SM sector, with $v_{0}=246 $ GeV and $\left ( m_{H}^{2} \right )_{0} = - \frac{\Lambda^{2}}{N} \; r \simeq -\left (88.4 \text{ GeV} \right )^2$. In these sectors, the electroweak symmetry is broken by the Higgs
 VEV given in \eqnref{Higgs VEV} and, consequently, the masses of the particles
 (both fermions and gauge bosons) will increase proportionally to $\sqrt{k}$.
 In particular, for $k>10^{8}$, the quarks are heavier than their
 corresponding QCD scales, meaning that those heavy sectors will not
 feature baryons~\cite{Arkani-Hamed:2016rle}.

 On the other hand, ``exotic" sectors, with $m_{H}^{2} > 0$ and corresponding to $i < 0$, are radically different from the SM~\cite{PhysRevD.20.2619,Samuel:1999am,PhysRevD.79.096002}. Since $m_{H}^{2} > 0$, these sectors do not acquire a VEV for the Higgs and the electroweak symmetry is broken at low scales by the phase transition from free quarks to confinement at the QCD scale, $\Lambda_{\text{QCD}}$~\cite{PhysRevD.20.2619}. In these sectors, fermions masses, $m_{f,i}$, are obtained from the four-fermion interactions with the Higgs being integrated out:
 \begin{equation}
 m_{f,j} \sim y_{f} y_{t} \frac{ \Lambda_{\text{QCD}}^{3} }{\left ( m_{H}^{2} \right )_{j} } = - \frac{ \sqrt{2}  y_{t}  }{ \lambda } \left (  \frac{ \Lambda_{\text{QCD}} }{ v_{0} } \right )^{3}  \frac{ m_{f,0} }{ \mathcal{C}_i } \lesssim 100 \text{ eV } \; , \label{eq:exotic_fermionMasses}
 \end{equation}
 where $y_{t}$ and $y_{f}$ are the top and the fermion $f$ Yukawa couplings, respectively, and $m_{f,0}$ is the mass of the given fermion in the SM. The gauge bosons receive masses when $SU(3)$ confines and their mass is given by~\cite{PhysRevD.20.2619}
 \begin{equation}
 m_{W}^{2} = \left ( \frac{g}{2} \right )^{2} f_{\pi}^{2}  \; , \; m_{Z}^{2} = \left ( \frac{g + g^{\prime} }{2} \right )^{2} f_{\pi}^{2} \; , \label{eq:wmassesSpectrum}
 \end{equation}
 where $f_{\pi}$ is the pion decay constant, $g$ and $g^{\prime}$ are the $SU(2)$ and $U(1)$ gauge coupling, respectively. As all six flavor quarks are lighter than the $SU(3)$ confinement, 
there will be many more light hadrons than in the SM~\cite{Samuel:1999am}. The spontaneous breaking of $SU(6)\times SU(6)\rightarrow SU(6)$ results in 35 pseudo-Goldstone bosons, three of which are absorbed to become the longitudinal polarization components of the $W$ and $Z$ bosons, and the remaining ones are analogous to the SM pions. The masses of the pions in the various sectors through the QCD phase transition can be obtained by applying the well-known Gell-Mann-Oakes-Renner relation~\cite{GellMann:1968rz, Schwartz:2013pla}:
 \begin{equation}
 m_{\pi}^{2} = \frac{V^{3}}{ f_{\pi}^{2} } \left ( m_{u} + m_{d} \right )  \; ,
 \end{equation}
 where $V \sim \Lambda_{\text{QCD}}$. Assuming $V^{3} / f_{\pi}^{2} \sim V^{3}_{i} / f_{\pi,i}^{2} $, the pion's masses in the ``standard" and ``exotic" sectors are:
 \begin{equation}
 m_{\pi , i }^{2} \simeq \begin{dcases}
   \mathcal{C}_i^{1/2} \; m_{\pi}^{2}  & \; , \; \text{Standard}\;  \\
   \frac{ \left ( m_{a,i} + m_{b,i} \right ) }{ \left( m_{u} + m_{d} \right ) }m_{\pi}^{2} & \; , \;  \text{Exotic}   
   \end{dcases} \; ,
  \label{eq:pion_mass}
 \end{equation}
 where $a$ and $b$ denote the flavor of the component quark given in \eqnref{eq:exotic_fermionMasses}, and $m_\pi$ is the experimentally measured pion mass. We have ignored corrections due to the changes in running couplings induced by different quark masses, although those effects are detailed in~\cite{PhysRevD.101.095016}.

With the spectrum, we can compute the effective number of relativistic d.o.f., $g_{\star}(T)$, for each sector. 
The $i=0$ sector 
has $g_{\star}=106.75$ at $T \gtrsim 100$ GeV. As the temperature decreases, the various particle species become nonrelativistic and they need to be removed from the total $g_{\star}$ value. The top quark is the first particle to decouple at 
$T \sim \frac{1}{6} m_t$, reducing the number of relativistic d.o.f to $g_{\star}=96.25$ and, similarly, the rest of the particles above QCD scale follows. There is a significant drop in $g_{\star}$ when the QCD phase transition occurs, at $T\sim 200$ MeV, with quarks and gluons confined into hadrons, and the only particles being left are three pions, electrons, muons, neutrinos, and photons, resulting in total $g_{\star} = 17.25$. These particles, with the exception of photons, will also eventually become nonrelativistic and decouple as the temperature drops. The story of the relativistic d.o.f is analogous in the $i>0$ sectors with different particle species removed from the total $g_{\star}(T)$ counting at different times due to the distinct mass spectrum and temperature of those sectors. In the SM sector, photons decouple and form the cosmic microwave background at $T_0 \sim 0.32$ eV, so that the total relativistic d.o.f can be computed as $g_{*,\text{Dec}}^{0} = 2 + \frac{7}{8} (2\times 3) \times \left( \frac{4}{11} \right)^{4/3}\simeq 3.36$ where the last factor is due to photons being reheated relatively to neutrinos. The standard sectors are colder than $i=0$, so like the SM the only relativistic particles are only the photons and neutrinos giving $g_{*,\text{Dec}}^{k} \simeq 3.36$.  

The number of relativistic degrees of freedom in the exotic sectors is of course different because of the altered spectrum. First, Electroweak Symmetry Breaking (EWSB) is triggered by the QCD scale at $T\simeq 89$ MeV, above which $g_{\star} = 102.75$, with only the complex Higgs doublet integrated out. After the QCD phase transition, we need to remove the contribution due to all the quarks and gluons being trapped inside hadrons. Moreover, we need to take into account the presence of the 35 pseudo-Goldstone bosons from the spontaneous breaking of $SU(6)\times SU(6)\rightarrow SU(6)$ symmetry, thus reducing the total d.o.f to $g_{\star}=58.75$.
Next, at a temperature approximately $T\sim \frac{1}{6} m_{W,i}$, gauge bosons $W^{\pm}, Z$ annihilates so $g_{\star}= 58.75 - 9 = 49.75$. The rest of the particle species annihilates analogously to the SM as the temperature drops over time.  
The temperature of the exotic sector at the time of the photon decoupling can be computed using the relation
$T_{j}/T_{0} \simeq \left ( \frac{\beta_j}{1-\beta - \gamma} \right )^{1/4}$. Using $T_{\text{Dec}}^0 \sim 0.32$ eV and the most optimistic value of $\beta \sim 10^{-2}$ leads to $T_{\text{Dec}}^j \sim 0.1$ eV, which is larger than $\frac{1}{6}m_{\tau,-1} \simeq 0.03$ eV. Therefore, the total number of relativistic d.o.f, at the time of the CMB, in the first exotic sector is $g_{*,\text{Dec}}^{-1} = 17.75$.

\section{DARK NEUTRINO FREEZE-IN}
\label{FI neutrinos}

Given the coupling between the leptons and the reheaton, electroweak symmetry breaking induces mixing between the SM neutrinos and hidden sector neutrinos in standard sectors. This in turn can mediate the production of hidden sector neutrinos from the SM neutrino bath. In this Appendix, we show that this process gives a negligible contribution to the neutrino energy density.

\begin{figure}[h]
\begin{centering}
\includegraphics[scale=0.3]{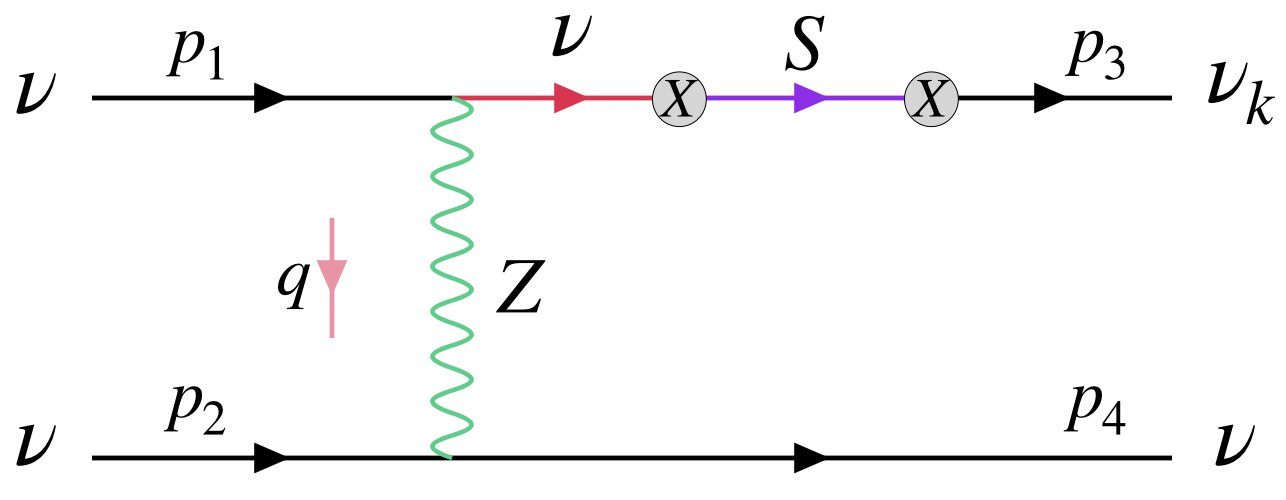}
\par\end{centering}
\caption{The Feynman diagram contributing to the SM neutrino freeze-in production in the $k$th sector. }
\label{fig:FIProcess}
\end{figure}

We can begin with the model where the reheaton couples directly to the leptons via the coupling
\begin{equation}
   -\mathcal{L}_l \supset \lambda S^c \sum_k l_k H_k \; ,
\end{equation}
where $l_k$ is one of the SM-like lepton doublets in the $k$th sector.\footnote{SM flavor indices are ignored for simplicity.} The neutrino freeze-in process is then\footnote{Since the population of the $\nu$ and $\bar{\nu}$ is roughly symmetric, there is also a contribution from the $\nu \bar{\nu}$ initiated process for the freeze-in our model. These additional contributions are approximately of the same size as $\nu \nu$ initiated process.} 
\begin{equation}
    \nu\nu \rightarrow \nu_{k} \nu,
\end{equation}
where $\nu$ corresponds to the SM neutrinos and $\nu_k$ to the neutrinos in the $k$th sector. A representative Feynman diagram for this process is shown in  \fref{fig:FIProcess}.

This process proceeds through a heavy mediator; however, the dominant freeze-in occurs at low energy. 
The cross section of this process in the low energy limit ($E \ll M_Z , m_{S}$), is given by
\begin{align}
       \sigma & \simeq 
       \left(\frac{\lambda^2 \, m_0^2 \, v_0 \,v_k}{(m_k^2-m_0^2)\,m_S^2}\right)^2
       \left( \frac{g}{2 \cos(\theta_W)}\right)^4
       \frac{ E^2 }{64 \pi  M_Z^4 } \; , \;
\label{eq:FIcrossSection Lmodel}
\end{align}
where $E$ is SM neutrino’s energy in the Centre of Mass (CM) frame, $g$ is the $SU(2)$ gauge coupling, and $\theta_W$ the Weinberg angle, $M_{Z}$ is the $Z$ boson mass, $m_{k}\equiv m_{\nu_{k}}$ is the $k$th sector neutrinos mass, $m_{0}$ is the mass of the SM neutrinos, $v_{0}$ is the SM Higgs VEV and $v_{k} =  v_{0} \left [  \frac{ 2 k + r }{ r } \right ]^{1/2}$ is the Higgs VEV in the $k$th sector. 
The above approximation is relatively accurate for our purpose since we are interested in the allowed reheaton widths, given in~\eqnref{eq: reheaton range}, corresponding to $T_{\rm RH,SM} \lesssim 100$ GeV.

The term in the first parentheses in~\eqnref{eq:FIcrossSection Lmodel} is the mixing angle between the SM and $k$th sector neutrinos, and it is very small for all sectors. Therefore, this process is never in equilibrium so the abundance can be computed as a freeze-in process. We estimate the late time yield of this process to be
\begin{equation}
\begin{split}
Y_{\infty}  \simeq \frac{\pi\, g^4 \,\lambda^4\, m_0^4 \,v_0^2\, v_k^2 }{276480 \cos^4\theta_W \, M_Z^4\; m_S^4} \; \frac{ g_{\star,s} \; T_{\text{RH,SM}}^3}{ m_k^2 \; H(m_{k})} \,  Y_{\rm 0,EQ}^{2}  \; 
\; , \;
\label{Yield limits}
\end{split} 
\end{equation}
where $H(m_k)$ is Hubble when the SM bath has a temperature of $m_k$, and $Y_{\rm 0,EQ} \simeq 0.2$ is the equilibrium yield of neutrinos. We have also taken $m_k \gg m_0$. We can then compare the energy density in neutrinos from this freeze-in process to that of neutrinos coming from the reheaton decay for a given sector $k$:
\begin{equation}
\begin{split}
\frac{\rho_{\rm k, FI}}{ \rho_{\rm k, Decay }} & \simeq \left [ 5.8 \times 10^{-30}  \right ] \; \left ( \frac{g_{\star,k}}{75.75} \right ) \; \left ( \frac{106.75}{g_{\star}} \right )^{1/2} \;   \\
     &   \left ( \frac{10^{-5}}{\gamma_1} \right)^{1/4} \; \left ( \frac{200 \text{ GeV}}{m_S} \right)^{6}  \left ( \frac{\Gamma_S}{10^{-15} \text{ GeV }} \right )^2 \; \\
     & \left ( \frac{m_0}{0.0585 \text{ eV} } \right) \; \left ( \frac{T_{\rm RH,SM}}{100 \text{ GeV}} \right)^3 \; \left [ \frac{r}{2k+r} \right ]^{1/4} \; , \;
  \end{split}   
  \label{eq: lmodel ratio}
\end{equation}
where we have used $\gamma_k = \gamma_1 \; \frac{ (2 + r )}{ (2 k + r ) }$, $m_k = m_0 \left ( \frac{2k+r}{r} \right )^{1/2}$, $v_k = v_0 \left (  \frac{2k+r}{r} \right )^{1/2}$, and $\Gamma_S \simeq \frac{\lambda^2 \; m_S}{64 \pi}$. Therefore we see that we can ignore this contribution in the $\ell$ model. In the model presented here, the cross section, of this process is further suppressed by a factor of $(\mu/M_L)^4$, so it can be safely ignored.

%
\begin{figure*}[t!]
\centering
\begin{minipage}[c]{\textwidth} \subcaptionbox{ Energy density }{
\includegraphics[width=.45\textwidth ]{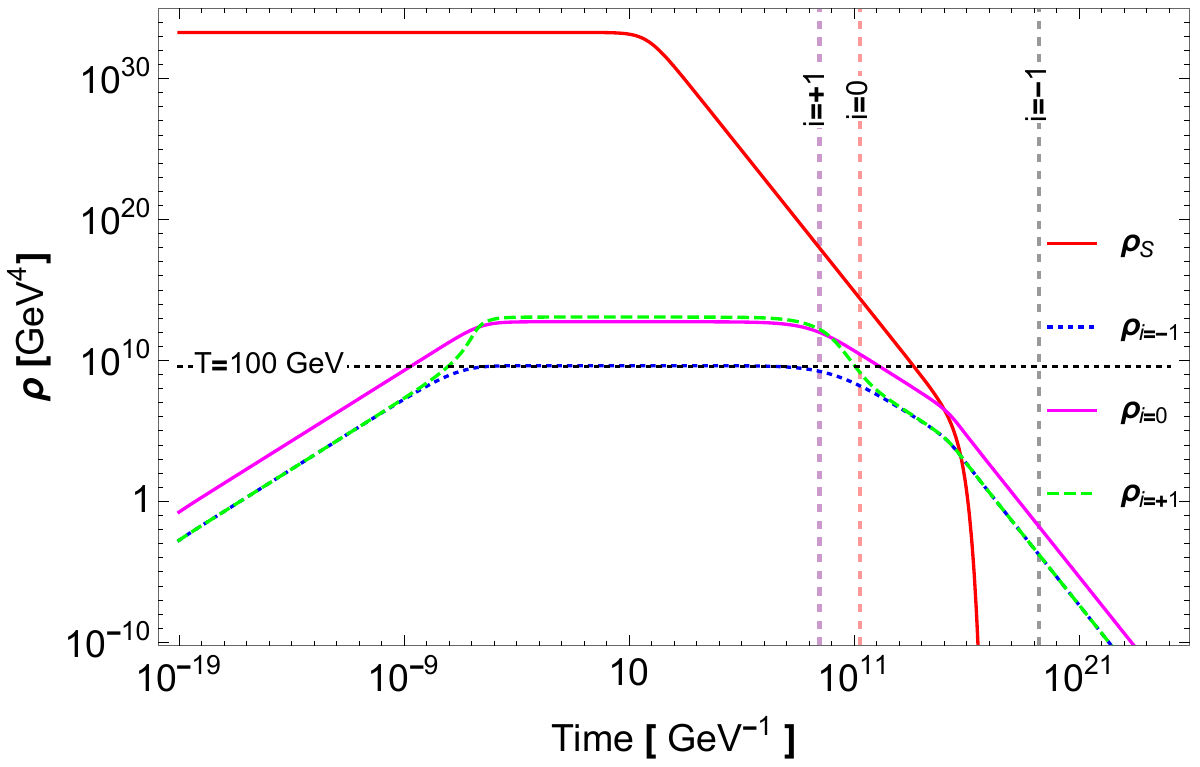} }
\hfill
 \subcaptionbox{ Ratios of energy densities  }{
\includegraphics[width=.45\textwidth]{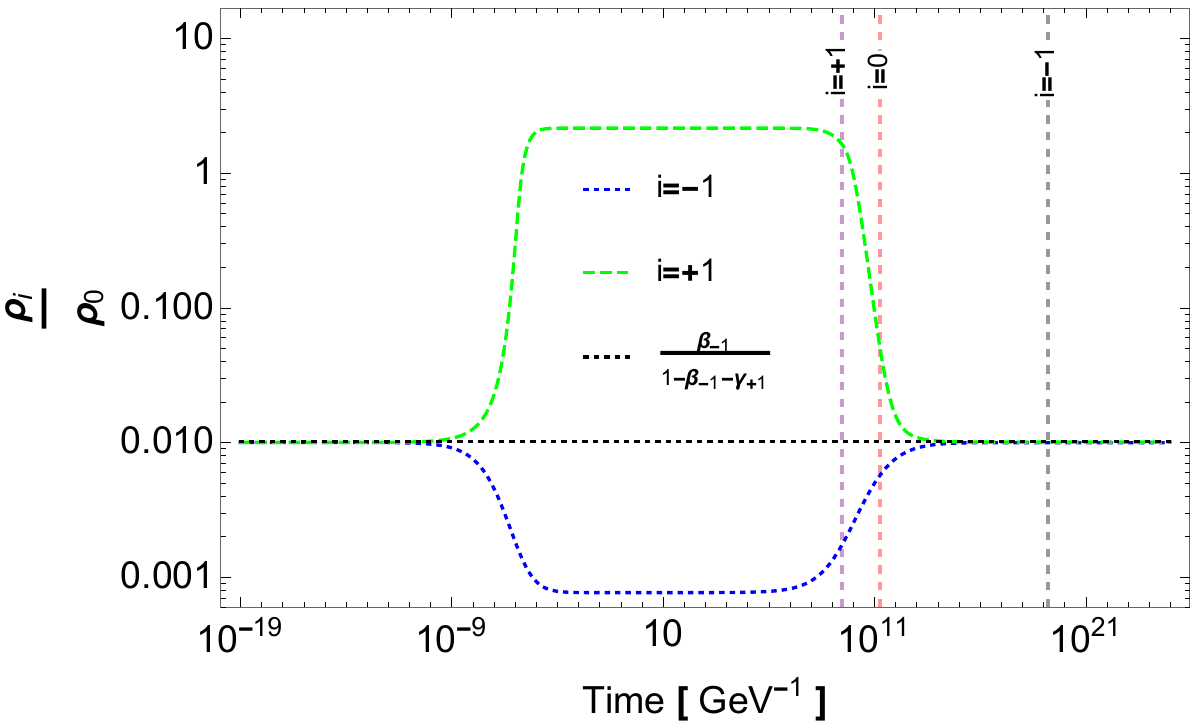} }
\hfill
 \subcaptionbox{ Partial width }{
\includegraphics[width=.45\textwidth]{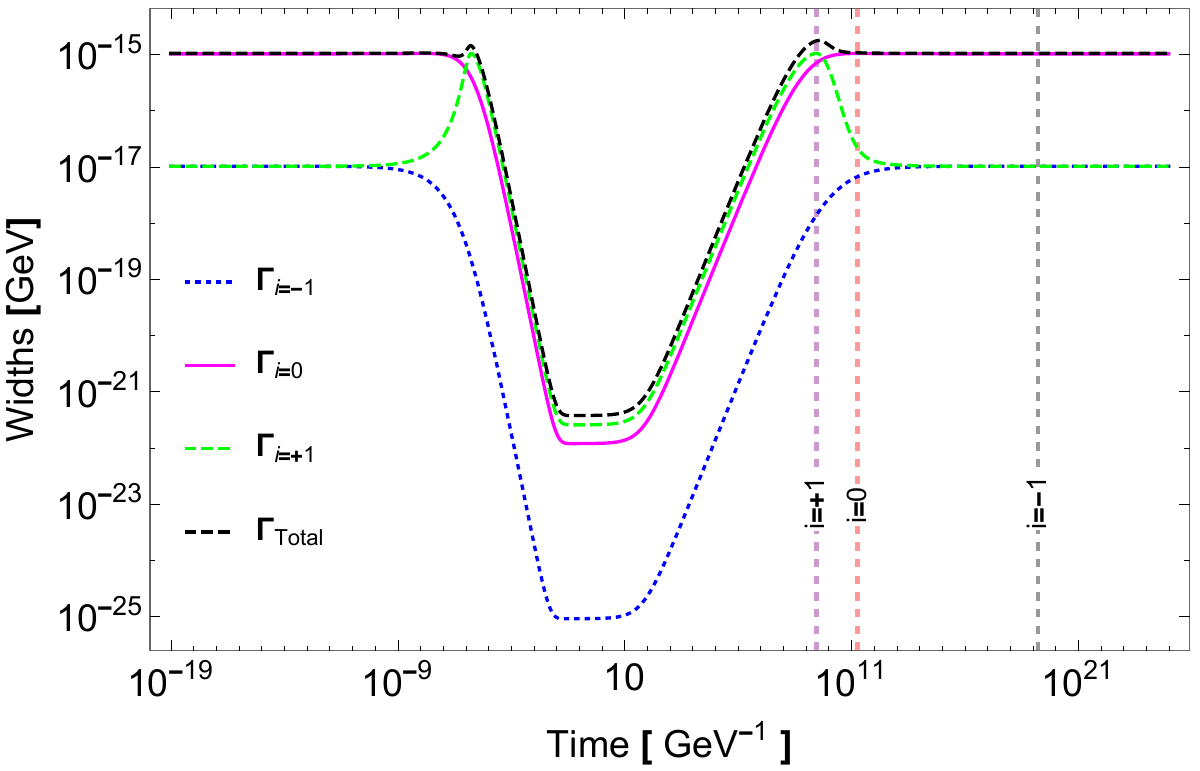} }
\hfill
 \subcaptionbox{ Baryon asymmetry ratio }{
\includegraphics[width=.45\textwidth]{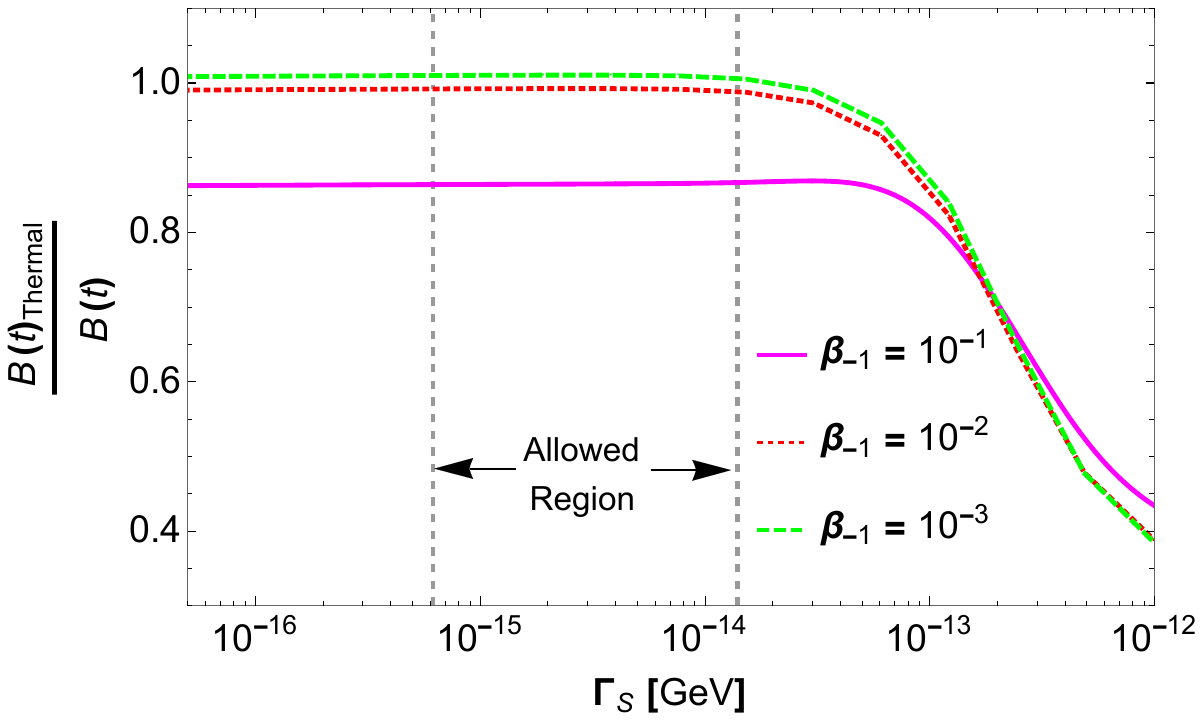} }
\end{minipage}
\hfill
\caption{ The evolution of the energy densities (a), ratios of the energy densities (b), and the partial widths (c) for sectors $i=-1,0,1$ as the Universe cools down over time. The ratio of the baryon asymmetry, with the case with thermal corrections over the case without, is computed at very late time for the $i=0$ sector (d) as a function of the reheaton's width for some fixed zero temperature branching ratios. The vertical dashed lines set the crossover temperature of the electroweak phase transitions for the different sectors ($i=-1,0,1$). The benchmark branching ratios are $\beta_{-1}=\gamma_{+1}= 10^{-2}$ at $T=0$, $m_S=300$ GeV and $\Gamma_S (T=0) \simeq 10^{-15}$ GeV. In these curves, the initial conditions are: $\rho_i \left ( t_I \right ) = 0$ and $\rho_{S} \left(t_{I}\right) =3\,M_{P}^{2}\,\Gamma_{N}^{2}$ with $t_I=10^{-20} \text{ GeV}^{-1}$ and $\Gamma_N = 10^{-2}$ GeV. The enhancement, for $i=+1$, in the ratio of the energy densities in (b) can be justified by the rise in the partial decay width (c) near the EW phase transition where the $W$-boson thermal mass approaches zero. 
  }
  \label{fig:Thermal Effects}
\end{figure*}

\section{EFFECTS OF THERMAL CORRECTIONS }
\label{Thermal Effects}
In this section, we study the thermal effects on the various quantities relevant to baryogenesis and dark matter in the N-naturalness framework, in particular, whether the thermal corrections to the Higgs and $W$-boson masses of the different sectors are relevant in the partial decay widths in~\eqnref{eq:Partial Widths}. The temperature-dependent masses are~\cite{Katz:2014bha, Arnold:1987mh}
 \begin{equation}
 \resizebox{0.5\textwidth}{!}{$
\begin{split}
m_{W,i}^{2} \left ( T_i \right )  & =  \begin{dcases}
 m_{H,i}^{2}(T_i)  =    \frac{y_t^2}{4} \left (  T_i^2 - T_{c,i}^2  \right )  \; , \; & \text{for  } T_i > T_{c,i}  \; , \;  \\
  m_{W,i}^{2} (0) \left [1 - \left ( \frac{T_i}{T_{c,i}} \right )^{2} \right ]  \; , \; & \text{for  } T_i < T_{c,i} \; , \; \\
\end{dcases} \; , \;  \\
m_{H,j}^{2} \left ( T_j \right )  & = m_{H,j}^{2} (0) + \frac{y_t^{2}}{4} T_j^{2}  \; , \; \quad  \forall \; T_j  \; , \; 
\end{split} $}
 \label{eq:Thermal Masses}
\end{equation}
where $y_t$ is the top Yukawa coupling and $T_{c,i}$ is the phase transition temperature of the $i$th sector, $m_{H,j} (0)$ and $m_{W,i}(0) $ are the zero temperature Higgs and $W$ boson masses, respectively. 
If thermal masses are sufficiently large that the two-body decays ($S\rightarrow H_j e_j$ and $S\rightarrow W_{k} e_{k}$) are kinematically forbidden, then the reheaton's decay is three-body. The decay $S\rightarrow t \bar{b} e $, in the limit of $m_{H} \gg m_{S} $ is given as
\begin{equation}
\Gamma_{S\rightarrow t \bar{b} e} \simeq \left ( \frac{ N_{c} y_b^{2}  \lambda^{2} }{ 3072 \pi^{3}  }  \right )\frac{ \mu_{L}^{2} m_{S}^{5}  }{ M_{L}^{2} m_{H}^{4}} \; ,
\end{equation}
where $N_c=3$ is the number of colors and $y_b$ is the bottom quark Yukawa coupling. Taking into account the temperature dependence of the various particle masses, the evolution of the energy densities in the different sectors can be determined by solving~\eqnref{eq: BE} with the appropriate temperature-dependent widths and branching ratios.

The behavior of the energy densities and the partial decay widths with thermal corrections taken into account are shown in~\fref{fig:Thermal Effects} as a function of time. We can compare this to the results in the left panel of~\fref{fig:BZ Energy Density}, which shows the dynamics ignoring thermal effects. In~\fref{fig:Thermal Effects}(a) we can observe that, as before, the energy density of the Universe is dominated by the reheaton for times $t \lesssim \Gamma_S^{-1}$.\footnote{The evolution here begins after the $N_1$ has fully decayed.} The initial condition is that the $N$ sectors are not populated, so early decays of the reheaton quickly heat these sectors up. Once these sectors reach the electroweak temperature, they undergo a phase transition from the broken to the unbroken phase. This in turn reduces the reheaton width which slows down the decays of the reheaton. This can be seen as the plateau in the top left panel of~\fref{fig:Thermal Effects}. It can also be seen as the decrease of partial widths in the bottom left panel. The expansion of the Universe then cools the sectors and they undergo another phase transition back to the electroweak broken phase, and the width increases again. This interesting double phase transition behavior could potentially have interesting observable consequences which we leave to future work.

In the top right panel of~\fref{fig:Thermal Effects}, we show the ratio of the energy density of the hidden sector to the SM sector as a function of time. Without thermal effects, this ratio will be equal to the ratio of branching ratios, and that is the result before the first phase transition. After the first phase transition, the $W$ mass in the $i=1$ reduces and the branching ratio increases, so the energy density in that sector increases relative to the others. After the second phase transition, the ratio of energy densities is restored to the naive branching ratio prediction. This shows that thermal effects are unimportant, and this is ultimately because most of the decays of the $S$ happen at late times, so the complicated dynamics at earlier times do not make a large impact. 

This conclusion can be confirmed in the bottom right panel of~\fref{fig:Thermal Effects}, which shows the ratio of late time baryon asymmetry with and without thermal effects. We see that this ratio is very close to one as long as $\beta$ or $\Gamma_S$ are not too large, which is the case in the allowed region of $6.2 \times 10^{-16} \text{ GeV} \lesssim  \Gamma_{S}  \lesssim 1.4\times10^{-14} \,\mathrm{GeV}$ and $0.01 \lesssim \beta \lesssim 0.08$.

\section{DARK QUARK NUGGETS}
\label{Dark Nugget}

In this section, we will lay out some requisite details on the possible formation of quark nuggets in the exotic sector ($i=-1$) of our model. The formation of dark quark nuggets can be another interesting possibility in the scope of macroscopic dark matter. These objects are primarily composed of quarks and their formation requires first-order QCD phase transition. This hypothesis was first proposed by Witten in Ref.~\cite{PhysRevD.30.272}, and has gained some attention lately~\cite{Bai:2018vik, PhysRevD.99.055047}.

Lattice studies have shown that the SM QCD transition is a continuous crossover, meaning that SM physics alone cannot form quark nuggets in the early Universe \cite{Fodor:2001pe}. However, it was also shown that the phase transition in QCD-like gauge theories (that are not the SM one) is first order if the number of light quarks below the confinement scale is $N_{f} \geq 3$~\cite{PhysRevD.29.338,PhysRevD.45.466}. In~\cite{Bai:2018vik, PhysRevD.99.055047}, the authors study the formation of six quark flavor matter (6FQM) quark nuggets, arguing that, in addition to a first-order phase transition, it is also necessary for a nonzero baryon number in the dark sector. This is exactly the setup of the exotic sectors, with the $i=-1$ sector having the largest baryon asymmetry.

Here we study a possible evaporation process that was not discussed in~\cite{Bai:2018vik, PhysRevD.99.055047}, which is the emission of a pion and a lepton, or in terms of $SU(2)\times U(1)$ states,  $q_{L} \overline{u}_{R} \overline{\ell}_{L}$ and  $d_{R} \overline{u}_{R} \overline{e}_{R}$. This emission conserves all gauge quantum numbers, but it does change the lepton number and the $B-L$ charge of the quark nugget. 

\begin{figure}[t!]
\begin{centering}
\includegraphics[scale=0.42]{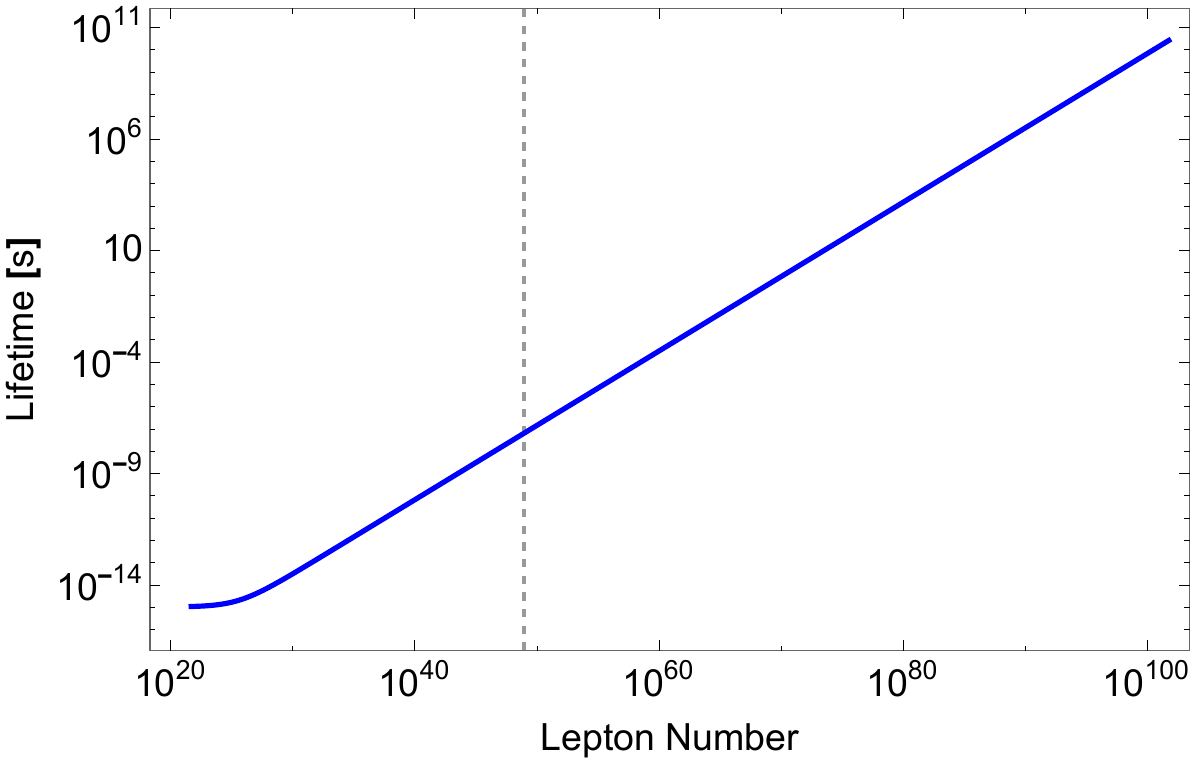}
\par\end{centering}
\caption{ The lifetime of the nugget as a function of the lepton number, $N \left ( t \right )$, for $B^{1/4} = 150 \text{ MeV}$, $g_{\star} = 49.75$, and $g_{\nu} = 6 $. As an example, for a nugget with a lifetime of approximately $1$ ns corresponds to $N \sim 10^{43}$. This does not take into account (re)absorption. The vertical dashed black line corresponds to $N=10^{49}$ which is the maximum number of leptons in a nugget determined by the simple Hubble patch argument. 
}
\label{fig:minimuLtillToday}
\end{figure}

The total number of leptons, in the 6FQM, contained in the nugget is~\cite{Bai:2018vik}
\begin{equation}
\begin{split}
N_{\text{Initial L}} & \equiv n_{L} \; V_{\text{QN}} \\
& = - \frac{55}{2 \pi^{2}} \left ( \frac{8 \pi^{2}}{415} \; B \right )^{3/4} \; \frac{4}{3} \; \pi \; R_{\text{QN}}^{3} \; ,
\end{split}
\label{eq: N leptons QN}
\end{equation}
where $B$ is the MIT bag constant, $V_{\text{QN}}$ is the quark  nugget volume and $R_{\text{QN}}$ is its radius. The MIT bag constant is not well known; nevertheless, our results for the nugget's lifetime will be relatively insensitive to any variation in $B$.
Then, the rate of the nugget's evaporation can be estimated following the technique in~\cite{Madsen:1986jg}, as
\begin{align}
\frac{dN_{\nu}}{dt} & =n_{\nu}\cdot v_{\nu}\,A\nonumber \\
 & =\left(\frac{g_{\nu}}{6\pi^{2}}T^{3}\right)\;\left[\left(\frac{\mu_{\nu}}{T}\right)^{3}+\pi^{2}\frac{\mu_{\nu}}{T}\right]\;4\pi\;R_{\text{QN}}^{2} \; ,
\label{eq: dN nugget dt}
\end{align}
assuming that it is proportional to the flux of neutrinos from the nugget, $n_{\nu}\,v_{\nu}$, times the nugget's surface area, $A$. In~\eqnref{eq: dN nugget dt}, $\mu_{\nu}$ is the chemical potential of neutrinos, and we assumed that the neutrinos are relativistic, i.e., $v_{\nu} \sim c$. Using $\mu_{\nu}=-3 \; \left ( \frac{8\pi^{2}}{415}\;B \right )^{1/4}$, which is obtained by assuming an equilibrium is maintained between the QN and the surrounding $\Delta P_{\text{Particles}} = \Delta P_{\text{Vacuum}}$, the expression above becomes
\begin{equation}
\begin{split}
\frac{dN_{\nu}}{dt}= & 2g_{\nu}T^{2}\left(\frac{\pi}{36}\right)^{2/3}\left(\frac{8\pi^{2}}{415}\right)^{-1/4}B^{-1/4} \\
& \left[\pi+\frac{9}{T^{2}}\left(\frac{8}{415}\right)^{1/2}B^{1/2}\right]N_{\nu}^{+2/3} \; .
\end{split}
\label{eq: dN nugget dT}
\end{equation}
By writing $ \dv{N_{\nu}}{T}$ in terms of time using $T^{2}\simeq \left( \frac{90}{\pi^{2}\;g_{\star}} \right)^{1/2} \frac{M_P}{2\;t}$, and integrating \eqnref{eq: dN nugget dT} over time, we get:
\begin{equation}
\begin{split}
\Delta N\left(t\right)^{1/3} = & \frac{2g_{\nu}}{3} \left(\frac{\pi}{36}\right)^{2/3}\left(\frac{8\pi^{2}}{415}\right)^{-1/4} B^{-1/4}   \\
 & \Bigg[\left(\frac{45}{2\;g_{\star}}\right)^{1/2}\; M_P \;\ln(\frac{t}{t_{C}})+ \\
 & 9\left(\frac{8}{415}\right)^{1/2}B^{1/2}\left(t - t_C \right)\Bigg] \; , 
\end{split}
\label{eq: Delta N nugget}
\end{equation}
where $\Delta N^{1/3}  \left (t \right) \equiv N  \left (t \right)^{1/3} - N  \left (t_{C}\right)^{1/3}$ and $t_{C}$ is the time of the phase transition where
the nugget forms ($T\simeq \Lambda_{\text{QCD,i}} \simeq 89$ MeV).
A plot of the nugget's lifetime as a function of $N \left ( t \right )$ is shown in~\fref{fig:minimuLtillToday}.
We can see that for a nugget with a lifetime of about $1$ second requires $N \sim 10^{63}$. To understand the order magnitude of the lepton number computed in~\eqnref{eq: Delta N nugget}, one might inquire, what is the typical number of baryons contained in one Hubble patch? 
\begin{equation}
\begin{split}
 N_{B} (T) & = \frac{4}{3} \pi \;  R^{3} \;  s(t) \; \eta_{\rm Obs} = \frac{8}{135} \left ( 90 \right )^{3/2} \frac{\eta_{\rm Obs}}{g_{\star}^{1/2}} \left ( \frac{M_{P}}{T}\right)^{3} \; , \;  \\
N_{B} (T)  & \simeq 9 \times 10^{48} \left (  \frac{100 \text{ MeV}}{T}  \right )^{3}  \; , 
\end{split}
\end{equation}
where we used $R = \frac{1}{H} \; , \; \eta_{\rm Obs}=8.59 \times 10^{-11}$, and $g_{\star} = 49.75$. Consequently, assuming the lepton number to be the same order magnitude as the baryon number, the maximum lepton number in a single patch can be approximately $\sim 10^{49}$ which corresponds to the nugget's lifetime of about $100$ nanoseconds and it is represented by the vertical dashed black line in~\fref{fig:minimuLtillToday}.   

\begin{figure}[t!]
\begin{centering}
\includegraphics[scale=0.65]{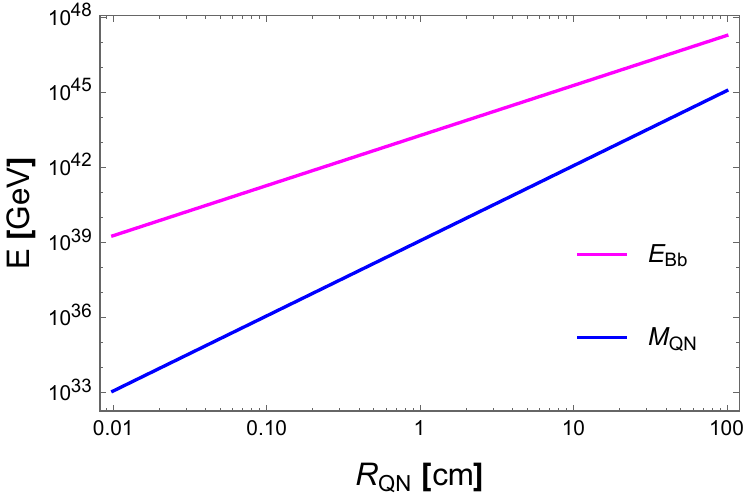}
\par\end{centering}
\caption{ Energy loss due to neutrino emission and the total mass of the quark nugget as a function of its radius in the blackbody treatment of neutrino evaporation. In the above curves, we set $B^{1/4} = 150 \text{ MeV},\;  T_{i} = 100 \text{ MeV}, \text{ and } T_{f} = 1 \text{ MeV } $.   }
\label{fig:QN_E}
\end{figure}

Another approach to model the nugget dynamics is to treat it as a blackbody. We compute the total energy loss of the quark nugget and compare it to the total energy contained in the nugget. Inside the nuggets, electroweak symmetry is unbroken and, therefore, leptons and quarks are massless. Considering that the quark nugget behaves like a blackbody is reasonable since the mean free path of the neutrino is small compared to the typical $R_{\text{QN}}$.
This way, in principle, all the neutrinos falling on it will be absorbed and the quark nugget will also emit neutrinos with a thermal spectrum. The typical power for a given surface area can thus be written using the Stefan-Boltzmann law:
\begin{equation}
P = 4 \pi \sigma \; \epsilon \;  R^{2}  \; T^{4}  \; ,
\end{equation}
 where $\epsilon$ is the emissivity ($\epsilon=1$ for idealized blackbody) and the Stefan-Boltzmann constant is $\sigma = 5.6705 \times 10^{-8} \; W/m^{2} K^{4} = \frac{\pi^{2}}{60}$ (in terms of fundamental constants). Then, the total amount of energy expended over a period $\Delta t = t_{f} - t_{i}$ is
 \begin{equation}
  \resizebox{0.5\textwidth}{!}{$
 \begin{split}
 \Delta E_{Bb} & = \int_{t_{i}}^{t_{f}} P \dd{t}=  \left ( 360 \right )^{1/2} \; M_{P} \; \sigma \; \epsilon \; R^{2}_{\text{QN}} \left [ \frac{T_{i }^{2}}{g_{\ast,i}^{1/2}} - \frac{T_{f }^{2}}{g_{\ast,f}^{1/2}}  \right ]  \; . 
 \end{split} $}
 \end{equation}
On the other hand, using the MIT bag model, the energy density of the quark nugget can be expressed in terms of the Bag constant $B$ as (in the massless particle regime):
\begin{equation}
\rho_{\text{QN}} = \sum_{i} \rho_{i} + B = 3 \sum_{i} P_{i} + B = 4 B.
\end{equation}
Hence, the total energy contained inside the QN can be expressed as
\begin{equation}
M_{\text{QN}} =  \frac{16}{3} \pi \; R_{\text{QN}}^{3} \; B.
\end{equation}
 In the SM, the typical value for the Bag parameter is $B^{1/4} \simeq 150 \text{ MeV}$. In~\fref{fig:QN_E}, we compare the total energy of the nugget with the energy loss due to neutrino emissions, approximating the nugget as a blackbody, and we see that the nugget is completely depleted by evaporation. In our analysis, we have neglected any (re)absorption of leptons from the environment.  

To conclude, we may estimate the consequences of the shut off of the sphaleron process once the size of the nugget becomes smaller than the size of the sphaleron.
For the classical Yang-Mills theory, the sphaleron size is~\cite{Moore:1999fs}:
\begin{equation}
R_{\text{Sp}} \sim \frac{5}{g^{2} \; T } \simeq \begin{dcases} 0.12 \text{ GeV}^{-1} \; ,  &  \text{ For } T=100  \text{ GeV} \\ 120 \text{ GeV}^{-1} \; ,  &  \text{ For } T=100  \text{ MeV}  \end{dcases} \; ,
\end{equation}

whereas its energy is
\begin{equation}
E_{\text{Sp}} \sim \frac{4 \; m_{\text{W}}}{\alpha_{\text{W}} } \simeq \begin{dcases} 9.6 \text{ TeV} \; ,  &  \text{ For } m_{\text{W}}=80.4  \text{ GeV} \\ 3.6 \text{ GeV} \; ,  &  \text{ For } m_{\text{W}}=30.3  \text{ MeV}  \end{dcases} \; .
\end{equation}
We may then conclude that the sphaleron interactions in the nugget never shut off since $R_{\text{QN}} \gg R_{\text{Sp}} $ (for a typical value $R_{\text{QN}} = 1 \text{ mm } = 5.06 \times 10^{11} \text{ GeV}^{-1}$).

From these estimates, we see that if the dark quark nugget does form in this scenario, its lifetime will be much too small to affect the dark matter phenomenology.


\clearpage
\newpage
\appendix
\bibliography{BaryogenesisPRD}

\end{document}